\documentclass[lettersize,journal]{IEEEtran}
\usepackage{amsmath,amsfonts}
\usepackage{algorithmic}
\usepackage{algorithm}
\usepackage{array}
\usepackage[caption=false,font=normalsize,labelfont=sf,textfont=sf]{subfig}
\usepackage{textcomp}
\usepackage{stfloats}
\usepackage{url}
\usepackage{verbatim}
\usepackage{graphicx}
\usepackage{cite}
\usepackage[colorlinks=true,citecolor=blue]{hyperref}

\usepackage{amssymb}
\usepackage{pifont}
\newcommand{\cmark}{\checkmark}
\newcommand{\xmark}{\ding{55}}

\usepackage{wasysym}

\usepackage{tabularx}
\usepackage{colortbl}
\usepackage[utf8]{inputenc}
\usepackage{amssymb}
\usepackage{float}

\usepackage{listings}  
\usepackage{xcolor}    
\usepackage{orcidlink}

\lstdefinelanguage{yaml}{
    keywords={true,false,null,y,n},
    keywordstyle=\color{blue},
    basicstyle=\ttfamily,
    sensitive=false,
    comment=[l]{\#},
    morecomment=[s]{/*}{*/},
    morestring=[b]",
    morestring=[b]'
}

\definecolor{codegreen}{rgb}{0,0.6,0}
\definecolor{codegray}{rgb}{0.5,0.5,0.5}
\definecolor{codepurple}{rgb}{0.58,0,0.82}
\definecolor{backcolour}{rgb}{0.96,0.96,0.96}

\lstdefinelanguage{yaml}{
  keywords={true,false,null,y,n},
  keywordstyle=\color{blue}\bfseries,
  basicstyle=\ttfamily\small,
  sensitive=false,
  comment=[l]{\#},
  commentstyle=\color{codegreen},
  morestring=[b]",
  morestring=[b]',
  stringstyle=\color{codepurple},
}

\lstdefinestyle{mystyle}{
    commentstyle=\color{codegreen},
    keywordstyle=\color{blue},
    stringstyle=\color{codepurple},
    basicstyle=\ttfamily\small,
    breaklines=true,
    captionpos=b,
    keepspaces=true,
    frame=single,
    showspaces=false,
    showstringspaces=false,
    showtabs=false,
    tabsize=2
}

\begin{document}

\title{Predictive Autoscaling in Cloud-Native and Federated Cloud-Edge Computing Environments: A Taxonomy and Future Directions}

\author{Bablu Kumar\orcidlink{0009-0007-6337-3700}, Anshul Verma\orcidlink{0000-0003-2597-2458}, and Rajkumar Buyya\orcidlink{0000-0001-9754-6496}
    \thanks{This work was supported in part by the Institutions of Eminence (IoE) Scheme of Banaras Hindu University, Varanasi, India under Dev. Scheme No. 6031 and and an Australian Research Council (ARC) Discovery Project grant.}
    
    \thanks{Bablu kumar, Anshul Verma, and Rajkumar Buyya are with the Quantum Cloud Computing and Distributed Systems (qCLOUDS) Laboratory, School of Computing and Information Systems, The University of Melbourne, Australia, as well as Bablu Kumar and Anshul Verma are with the Department of Computer Science, Banaras Hindu University, Varanasi, India.   Email: bablu.kumar.1@student.unimelb.edu.au, anshul.verma@unimelb.edu.au, rbuyya@unimelb.edu.au.}}

\markboth{}%
{}

\IEEEpubid{}

\maketitle

\begin{abstract}
Autoscaling is a key capability in cloud-native systems, where dynamic workloads, heterogeneous environments, and latency-sensitive applications require efficient and adaptive resource management. Traditional reactive approaches based on fixed thresholds often respond too late, leading to resource imbalance, performance degradation, and unstable scaling behavior. Recent advances in predictive models, Kubernetes Custom Resource Definitions (CRDs), Monitor-Analyse-Plan-Execute (MAPE) based control loops, and federated learning (FL) have enabled more proactive and autonomous autoscaling strategies. This paper presents a structured review of these developments. It first introduces a taxonomy of autoscaling techniques based on triggers, targets, prediction models, and evaluation metrics. It then examines predictive autoscaling approaches and CRD-based mechanisms, including Kubernetes operators and reconciliation workflows. Further, it analyses autoscaling in federated learning environments, highlighting reactive and proactive strategies alongside privacy-preserving techniques and container-level isolation. The paper also discusses drift-aware and uncertainty-aware autoscaling, incorporating concepts such as the Autoscaling Drift Index (ADI), feedback-driven correction, and stability control for heterogeneous workloads. Finally, it outlines open challenges and future research directions, providing a foundation for next-generation intelligent predictive autoscaling in cloud-edge environments.
\end{abstract}


\section{Introduction} \label{sec1}
\IEEEPARstart{C}{loud-native} computing has emerged as the dominant paradigm for deploying modern applications across scalable, distributed, and highly elastic infrastructures by fully leveraging the capabilities of cloud environments \cite{TARI2024100650, bittencourt2025computing}. Unlike traditional monolithic systems, cloud-native applications adopt a modular and flexible microservice-based architecture  \cite{al2024containerized, 8125550} and each microservice operates as an independent unit and is typically deployed inside containers, which offer strong isolation, lightweight execution, and efficient resource management \cite{di2018migrating, deng2024cloud}. Built on microservices packaged as containers orchestrated through pods and Kubernetes \cite{kubernetesOverview, kumar2026critical} provide scaling efficiency \cite{jeong2025autoscaling} and autoscaling, the capability of a system to automatically adjust computing resources, and play a crucial role in ensuring performance continuity. 

Most traditional autoscaling solutions are based on reactive mechanisms, such as triggering scaling operations based on thresholds of resource metrics such as CPU or memory usage \cite{kubernetesToolsMonitoring}. Although simple and widely used, it has its limitations: it reacts only once performance is degraded, which means resource allocation is delayed, under-provisioning, or over-provisioning occurs. Despite being simple and popular, reactive scaling has its drawbacks: it only kicks in when performance is degraded, which delays or wastes resources. With increasing unpredictability, bursts, and latency requirements of loads, it becomes challenging to maintain consistent performance using reactive strategies, particularly in distributed cloud-edge environments \cite{10025811,10335918,zhai2025energy}. Intelligent resource management plays a vital role in an evolving landscape where users are mobile, services are interactive and real-time, apps are distributed, and devices are heterogeneous \cite{furnadzhiev2025efficient,kubernetesAutoscalingWorkloads}.

Recent improvements have turned autoscaling from a purely reactive process to an intelligent, predictive, and context-aware process known as proactive autoscaling~\cite{kumar2024optimizing}. The Monitor-Analyse-Plan-Execute (MAPE) framework gives a systematic approach to include workload forecasting, policy optimization and autonomous decision making in cloud native systems \cite{kumar2025optimizing}. Recently, several deep learning models, such as LSTM, GRU, Bi-LSTM, CNN-LSTM, Transformers, etc., have been successfully adopted for predicting highly nonlinear and time-dependent workload patterns, including long-sequence models like Informer-like models and multivariate Transformers, etc.~\cite{kumar2024optimizing,kumar2025optimizing}. These predictive models can be used for proactive autoscaling, which allocates resources ahead of time based on expected demand, leading to greater efficiency, responsiveness, and cost-effectiveness.

At the same time, Kubernetes~\cite{kubernetesOverview} has evolved beyond its default autoscalers, the Horizontal Pod Autoscaler (HPA)~\cite{van2025properties,hpaDocs2023}, Vertical Pod Autoscaler (VPA)~\cite{pham2024elastic}, and Cluster Autoscaler~\cite{kubernetesAutoscalingWorkloads}, by supporting advanced customization through custom resource definitions (CRDs)~\cite{kumar2024optimizing}, operator-based automation,  coroutine-based scheduling, and reconciliation loops. These mechanisms can be integrated directly into the orchestration layer, allowing for the seamless incorporation of predictive models, decision policies, and adaptive controllers, thus creating intelligent, autonomous, and fully customizable autoscaling pipelines~\cite{carrion2022kubernetes}.

With often demanding and unpredictable workloads, it is important to minimise training overheads and to ensure that data is handled securely and efficiently. Federated learning (FL) combined with differential privacy (DP) can be used to facilitate secure distributed learning in the cloud–edge systems~\cite{pham2024elastic, zhao2022edge, 9060868}. The unique characteristics of FL workloads, such as varying client participation, diverse data distributions, and a wide range of data types and sources, make them inherently dynamic in terms of their computational and communication requirements~\cite{zhai2025energy, albaseer2023data}. To maintain convergence, reduce stragglers, and stabilize training, there are proactive autoscaling frameworks like FedScaleEdge and FedALoRA~\cite{yi2025fedalora} and FedInv~\cite{liu2025fedinv} and Dogani's work on proactive autoscaling frameworks~\cite{dogani2024proactive}. However, new challenges arise in FL, such as autoscaling drift, prediction misalignment, and runtime uncertainty~\cite{ye2024openfedllm}. In the context of a dynamic cloud–edge environment, addressing these needs necessitates drift-aware mechanisms, like the Autoscaling Drift Index (ADI), uncertainty-aware correction loops, and stability controllers for a robust performance.

Although these advances have been made, there are still a number of research gaps that need to be addressed. As the data grows more complex, such as multimodal, the federated learning paradigm~\cite{ye2024openfedllm} and large-scale distributed inference, as well as real-time edge analytics, must also be supported by autoscaling. Furthermore, a few new requirements like uncertainty-aware orchestration~\cite{zhai2025energy, kumar2024optimizing}, fairness, privacy-preserving optimization, explainable autoscaling, and cross-layer coordination between network–edge–cloud systems are not well explored. Next-generation intelligent predictive autoscaling must include a common framework that encompass core autoscaling mechanisms, predictive intelligence, and Kubernetes-based automation and FL dynamics~\cite{kumar2025optimizing,kumar2025multivariate}.

This paper offers a comprehensive and unified review of the cloud-native autoscaling, Kubernetes orchestration, MAPE-based adaptive frameworks, predictive modeling, proactive scaling algorithms, drift-aware FL controller, and CRD-based automation. The work brings together core concepts, categorizes the most recent research, evaluates performance metrics and use case, and summarizes the latest trends in research, providing a baseline resource for both researchers and practitioners and system designers designing intelligent, autonomous and adaptive auto-scaling systems in cloud-native and federated cloud–edge environments. The major contributions of this work are outlined below:

\begin{itemize}
    \item This work introduces a unified overview that combines cloud-native architecture, autoscaling fundamentals, and MAPE-guided adaptive control into a single conceptual foundation.

    \item This paper proposes a well-structured four-dimensional taxonomy that covers scaling triggers, scaling targets, prediction models, and evaluation dimensions.
    
    \item This paper classifies intelligent predictive autoscaling approaches based on predictive autoscaling models and CRDs, including Proactive long-sequence forecasting along with Deployments: Docker, Kubernetes, and Cloud Native Stacks such as CRDs, operators, and reconciliation loops that enable autonomous resource management.
    
    \item We investigate federated learning (FL)-specific autoscaling challenges, proactive scaling frameworks, differential privacy (DP) implications, and Kubernetes-based FL orchestration mechanisms such as the KubeFlower operator, container Isolation and Multi-Region FL autoscaling.
    
    \item This work explains autoscaling drift phenomena, introduces the Autoscaling Drift Index (ADI), and analyzes correction loops and the FRSC to achieve stable and robust scaling in heterogeneous, privacy-preserving FL environments.
    
   \item This paper also identifies open challenges and future research directions in predictive autoscaling for cloud-native and federated cloud–edge environments.
\end{itemize}

This paper is guided by the following research questions:

\begin{itemize}
   \item RQ1: How do modern predictive models, including statistical forecasting, neural networks, Transformer-based architectures, and multivariate predictors, enhance proactive autoscaling accuracy and responsiveness?

    \item RQ2: How can Kubernetes mechanisms such as CRDs, operators, reconciliation loops, and runtime extensions be leveraged to design autonomous and intelligent predictive autoscaling pipelines?

    \item RQ3: What unique challenges arise when autoscaling federated learning workloads, and how do proactive FL frameworks, DP constraints, and operator-based orchestration techniques address these challenges?

    \item RQ4: What is autoscaling drift, what factors cause it, and how can uncertainty-aware correction loops and stability controllers improve robustness in dynamic and heterogeneous systems?
\end{itemize} 

\begin{figure*}[!ht]
    \centering
    \includegraphics[width=0.90\linewidth]{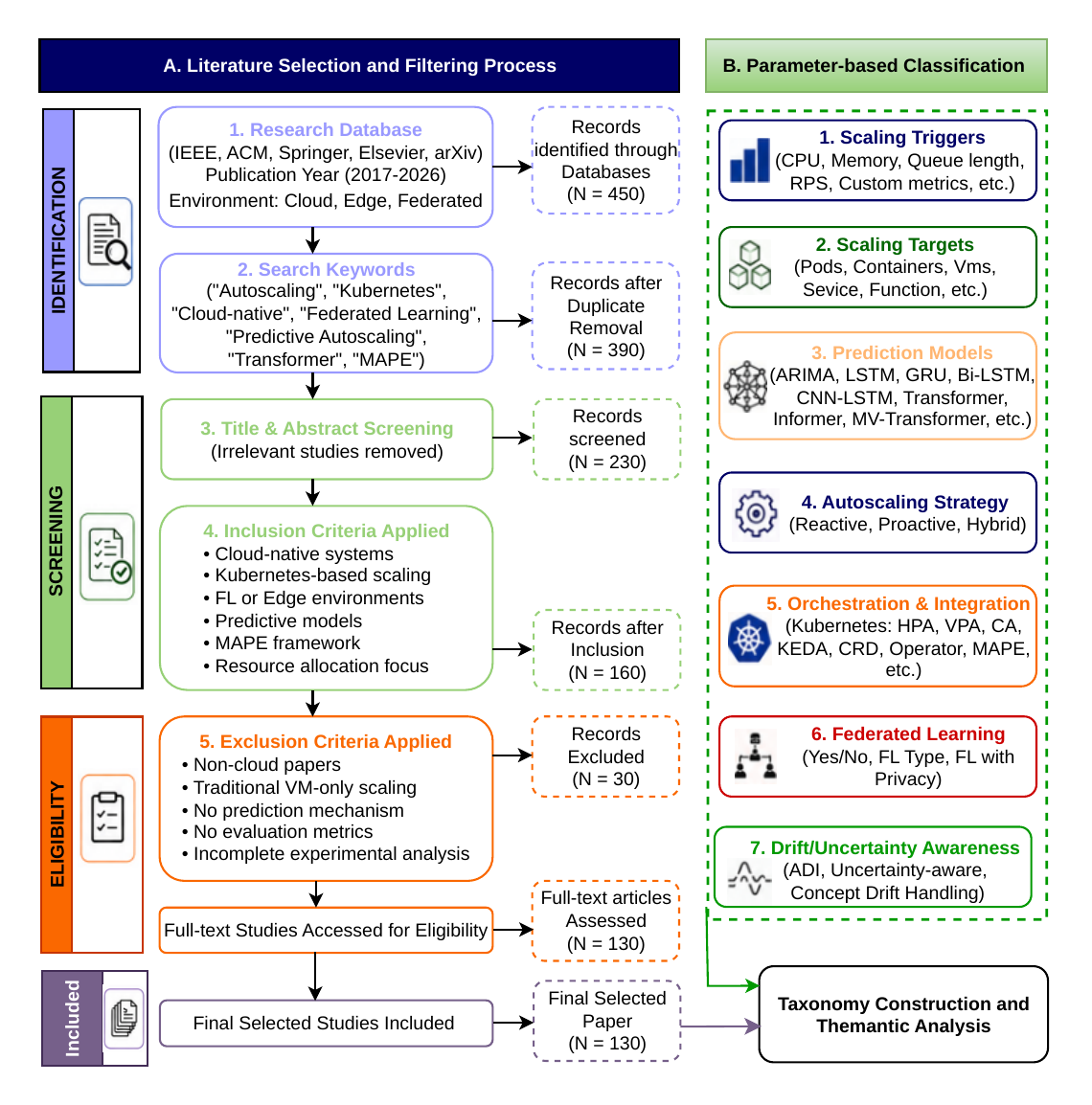}
    \caption{Systematic literature selection and filtering methodology for intelligent predictive autoscaling studies in cloud-native and federated cloud environments.}
    \label{fig:SLR}
\end{figure*}

\subsection{Systmetic Literature Review Process} 
The literature review process was conducted using a systematic and parameter-based filtering methodology to identify high-quality and relevant studies related to intelligent predictive autoscaling in cloud-native and federated cloud environments. As illustrated in the literature selection workflow in Figure \ref{fig:SLR}, research articles were collected from major scientific databases including IEEE Xplore, ACM Digital Library, SpringerLink, Elsevier (ScienceDirect), arXiv, and Google Scholar using search keywords such as “Autoscaling,” “Kubernetes,” “Cloud-native,” “Federated Learning,” “Predictive Autoscaling,” “MAPE,” “Transformer,” and “Drift-aware.” Initially, a large collection of studies was obtained, after which duplicate records were removed to improve dataset consistency and avoid redundancy. Subsequently, title and abstract screening was performed to eliminate irrelevant studies not directly related to autoscaling, cloud-native orchestration, federated learning, or predictive resource management.

After the preliminary screening stage, inclusion and exclusion criteria were systematically applied to refine the final set of studies. The inclusion criteria focused on studies associated with cloud-native autoscaling environments, Kubernetes-based orchestration, predictive or AI/ML-driven autoscaling mechanisms, federated learning or cloud–edge systems, MAPE-guided decision frameworks, and performance evaluations using standard metrics. In contrast, exclusion criteria removed non-cloud or traditional VM-only systems, studies without autoscaling or prediction mechanisms, papers lacking experimental evaluation, and duplicate or short editorial works without significant technical contributions. Following eligibility assessment through full-text analysis, the final selected papers were retained for a comprehensive review and structured taxonomy construction.

Furthermore, the selected studies were classified according to multiple parameter dimensions to support a detailed comparative analysis and taxonomy development. These parameters included scaling triggers, scaling targets, prediction models, deployment environments, federated learning support, autoscaling strategies, drift and uncertainty awareness, orchestration mechanisms, evaluation metrics, and datasets or benchmarks. The classification framework enabled the systematic organization of existing literature into categories such as reactive, proactive, and hybrid autoscaling; Kubernetes mechanisms including HPA, VPA, CA, KEDA, CRDs, and Operators; and predictive models such as ARIMA, LSTM, Bi-LSTM, Transformer, Informer, and MV-Transformer. Finally, filtered and classified studies formed the basis for taxonomy construction and thematic analysis, enabling a comprehensive understanding of intelligent predictive autoscaling techniques in modern cloud-native and federated cloud–edge computing environments.

\begin{figure*}[!ht]
    \centering
    \includegraphics[width=1.0\linewidth]{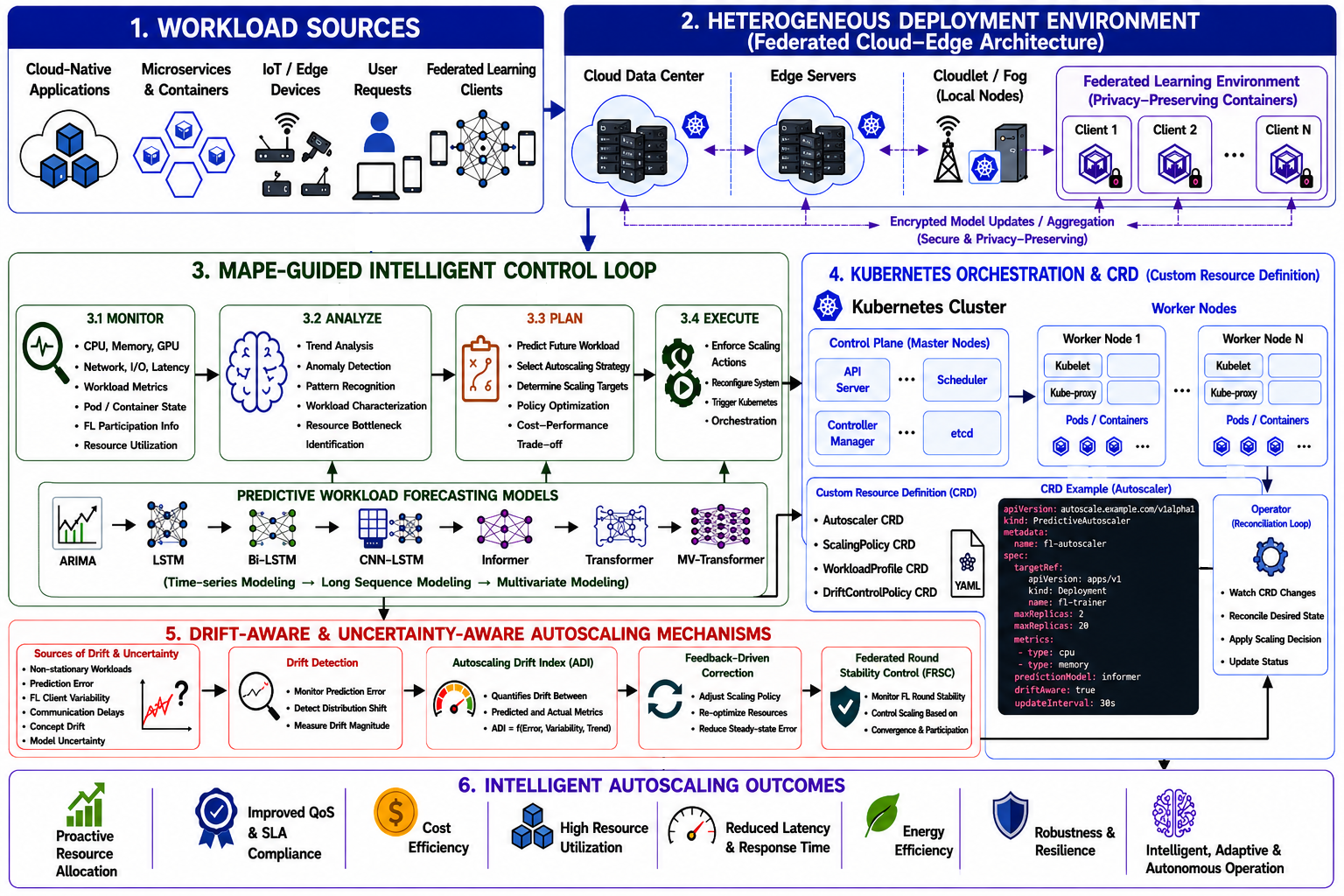}
    \caption{Unified Intelligent predictive Autoscaling Framework for Cloud-Native and Federated Cloud-Edge Environments}
    \label{fig:unififedframework}
\end{figure*}

\subsection{Unified Intelligent Predictive Autoscaling Framework}
This Figure \ref{fig:unififedframework} presents a unified intelligent predictive autoscaling framework for cloud-native and federated cloud–edge environments. The workflow begins with diverse workload sources including cloud-native applications, microservices, IoT devices, user requests, and federated learning clients. These workloads are deployed across heterogeneous infrastructures consisting of cloud data centers, edge servers, fog nodes, and privacy-preserving federated environments.

The collected workload and system metrics are processed through a MAPE-guided intelligent control loop containing monitoring, analysis, planning, and execution phases. The framework integrates predictive workload forecasting models such as ARIMA, LSTM, Bi-LSTM, CNN-LSTM, Informer, and Transformer models to estimate future resource demand and support proactive autoscaling decisions.

The predicted scaling decisions are enforced through Kubernetes-native orchestration mechanisms using Custom Resource Definitions (CRDs), Operators, master-worker node architecture, and reconciliation loops for automated resource management. Since federated cloud-edge environments introduce workload heterogeneity, communication delays, and model uncertainty, the framework further incorporates drift-aware and uncertainty-aware autoscaling mechanisms including drift detection, Autoscaling Drift Index (ADI), feedback-driven correction, and Federated Round Stability Control (FRSC).

Finally, the integrated framework enables intelligent predictive autoscaling outcomes such as proactive resource allocation, improved QoS and SLA compliance, cost efficiency, high resource utilization, reduced latency, energy efficiency, robustness, and adaptive autonomous cloud-edge operation.

\subsection{Workflow and Conceptual Connections}
\begin{figure*}[!ht]
    \centering
    \includegraphics[width=0.9\linewidth]{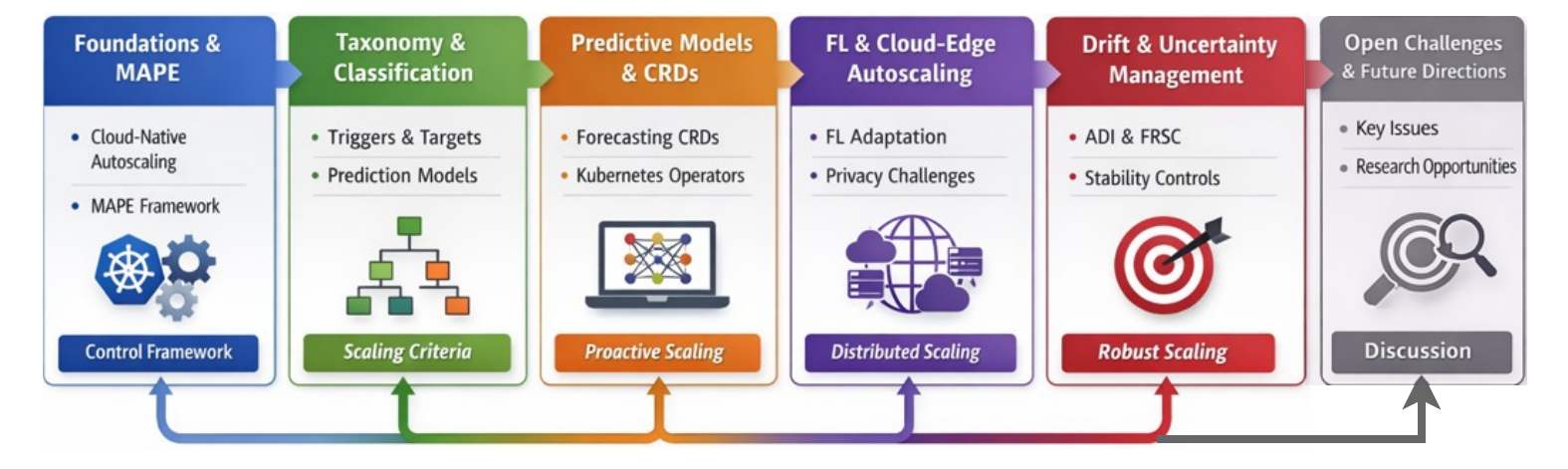}
    \caption{Proactive autoscaling roadmap from foundations to predictive, federated, and drift-aware control in cloud-edge systems.}
    \label{fig:workflow_concept}
\end{figure*}

Section~\ref{sec1} presents the introduction, research questions, systematic literature review, and workflow of the unified predictive autoscaling framework. Section~\ref{LiteratureSurvey} presents a comparative analysis of existing work and this work. This work is structured as a progressive pipeline that connects foundational concepts, modelling techniques, system integration, and advanced control mechanisms for intelligent Predictive autoscaling in cloud-native and federated cloud-edge environments, as shown in Figure \ref{fig:workflow_concept}. Specifically, Section~\ref{sec2} to Section~\ref{sec7} present the sequential stages of this framework. Section ~\ref{sec2} (Foundation \& MAPE) establishes the system-level basis by introducing cloud-native architectures and Kubernetes autoscaling mechanisms (HPA, VPA, Cluster Autoscaler), along with the MAPE control loop. This section defines how resource monitoring, decision-making, and execution are organised, forming the core control framework for all subsequent methods. Section ~\ref{sec3} (Taxonomy \& Classification) extends this work by systematically categorizing the various kinds of autoscaling along the lines of triggers, targets, prediction models and evaluation metrics. This offers a framework for analyzing the current approaches and their shortcomings, which will help determine what predictive and control strategies to use. Section ~\ref{sec4} (Predictive Models and CRDs) maps the taxonomy into implementation with the use of predictive autoscaling models (e.g., Informer, MV-Transformer) and the ability to use them in Kubernetes via CRDs and Operator-based reconciliation. This part implements the MAPE loop in the embedded forecasting, planning, and execution within the control plane to achieve proactive scaling. The framework is further extended in Section ~\ref{sec5} (FL and Cloud-Edge Systems) for federated learning environments, where the workloads are highly dynamic, considering their heterogeneity, communication variability and privacy constraints. It showcases the ability of predictive autoscaling to fit FL workload patterns and seamlessly integrate with distributed, multi-region infrastructures, covering cloud to edge deployment. In addition, the system is further improved in Section ~\ref{sec6} (Drift and Uncertainty Awareness) which deals with prediction errors and non-stationarity. It includes drift-aware and uncertainty-aware mechanisms such as ADI-based feedback and stability controllers (e.g., FRSC) to make better scaling decisions and maintain robustness in the presence of varying workloads. Section~\ref{sec7} outlines open challenges and research directions, identifying some of the limitations and opportunities for future developments in the field of intelligent predictive autoscaling. Section~\ref{sec8} provides a section-wise discussion of the key findings, implications, and research challenges identified throughout the review, and section~\ref{sec9} concludes the paper by summarizing the main contributions and key outcomes of the study.

\section{Related Surveys} \label{LiteratureSurvey}
Autoscaling has become an essential component of modern cloud-native computing environments due to the rapid growth of distributed applications, heterogeneous infrastructures, and continuously varying workloads~\cite{11417814, 10213996}. Traditional static resource provisioning mechanisms are insufficient for handling workload fluctuations efficiently, often resulting in over-provisioning, under-utilization, SLA violations, and increased operational cost. Consequently, autoscaling has evolved from simple threshold-based reactive mechanisms toward intelligent predictive frameworks integrating machine learning, deep learning, cloud-native orchestration, and federated cloud-edge computing paradigms. Recent advancements in Kubernetes orchestration, microservice architectures, edge intelligence, and predictive-driven resource management have further accelerated the development of adaptive autoscaling frameworks capable of supporting highly dynamic distributed systems.

Early research in autoscaling mainly focused on reactive and rule-based resource allocation strategies for cloud infrastructures. Qu et al.~\cite{qu2018auto} investigated autoscaling approaches for web applications using MAPE-loop architectures and categorized reactive and proactive scaling mechanisms for cloud systems. Similarly, Gar\'i et al.~\cite{gari2021reinforcement} explored reinforcement learning-based autoscaling frameworks and demonstrated the potential of intelligent adaptive policies for cloud elasticity management. Their study highlighted the capability of RL-based approaches to dynamically optimize resource allocation under uncertain cloud workloads. However, these studies mainly concentrated on centralized cloud infrastructures and did not comprehensively investigate Kubernetes-native orchestration, federated cloud-edge systems, or drift-aware autoscaling mechanisms.

The increasing adoption of cloud-native network functions has motivated extensive research on Kubernetes-based autoscaling mechanisms. Van Do et al.~\cite{van2025properties} investigated the properties and performance characteristics of Kubernetes Horizontal Pod Autoscaling (HPA) algorithms for scaling cloud-native network functions. Their work analyzed the mathematical behavior of HPA tolerance parameters and demonstrated that only a finite set of parameter intervals produces distinct scaling decisions, thereby simplifying parameter configuration for Kubernetes administrators. The study further examined session-based service workloads and established theoretical upper and lower performance bounds for HPA-based scaling. While the work provides valuable insights into the operational characteristics and optimization of reactive Kubernetes autoscaling, it primarily focuses on HPA parameter behavior and cloud-native network functions. In contrast, the present survey extends beyond reactive autoscaling by incorporating proactive forecasting models, deep learning and Transformer-based prediction techniques, CRD/Operator-driven orchestration, federated learning environments, drift-aware scaling mechanisms, and cloud-edge resource management frameworks.

In recent years, the growth in cloud-native systems has led to the incorporation of machine learning and predictive analytics in autoscaling systems. Kashyap et al.~\cite{kashyap2023prediction} summarized the prediction-based scheduling and workload forecasting methods for cloud systems with a focus on proactive resource management by leveraging time-series analysis. However, most of the previous ones were only focused on centralized cloud systems, and failed to consider federated cloud-edge orchestration, drift-aware scaling correction, and uncertainty-aware autoscaling mechanisms.

Autoscaling has been a major focus of study for cloud-native and microservice architectures, particularly regarding its implementation with AI. Jeong and Jeong~\cite{jeong2025autoscaling} gave a thorough survey of the cloud-native autoscaling techniques, classified the various scaling algorithms, and covered various topics on lifecycle-aware optimization, datasets, and cyber threats related to autoscaling. Similarly, Cai et al.~\cite{cai2025deep} introduced a deep learning and feedback control based autoscaling framework for cloud-native microservices that combines intelligent forecasting with adaptive control, enhancing QoS and resource utilization. Additionally, Meng et al.~\cite{meng2025catscaler} proposed CATScaler, a proactive autoscaling framework based on Transformer, which leverages convolution-augmented temporal modeling to predict workloads in a cloud-native application. While these studies showed high prediction accuracy, these studies were mainly for cloud-native microservices and only partially explored federated learning-aware orchestration and adaptive drift correction mechanisms.

\begin{table*}[!ht]
\centering
\caption{Feature-Based Comparison of Existing Surveys and Recent Autoscaling Studies}
\label{tab:feature_comparison}
\renewcommand{\arraystretch}{1.2}

\resizebox{\textwidth}{!}{
\begin{tabular}{|l|c|c|c|c|c|c|c|c|c|}
\hline
\textbf{Reference} &
\textbf{Reactive} &
\textbf{Proactive} &
\textbf{Hybrid} &
\textbf{K8s} &
\textbf{FL} &
\textbf{Predict-based Scaling} &
\textbf{CRD/Operator} &
\textbf{Drift-aware} &
\textbf{Cloud-Edge}
\\
\hline

Qu et al. (2018)~\cite{qu2018auto}
& \cmark & \cmark & \cmark & \LEFTcircle & \xmark & \LEFTcircle & \xmark & \xmark & \cmark
\\ \hline

Gar\'i et al. (2021)~\cite{gari2021reinforcement}
& \xmark & \xmark & \xmark & \LEFTcircle & \xmark & \LEFTcircle & \xmark & \xmark & \LEFTcircle
\\ \hline

Verma and Bala (2021)~\cite{verma2021auto}
& \cmark & \cmark & \LEFTcircle & \xmark & \xmark & \xmark & \xmark & \xmark & \cmark
\\ \hline

Dogani et al. (2022)~\cite{dogani2022k}
& \LEFTcircle & \cmark & \LEFTcircle & \LEFTcircle & \xmark & \xmark & \xmark & \xmark & \cmark
\\ \hline

Mampage et al. (2022)~\cite{mampage2022holistic}
& \LEFTcircle & \xmark & \xmark & \LEFTcircle & \xmark & \xmark & \xmark & \xmark & \cmark
\\ \hline

Kashyap et al. (2023)~\cite{kashyap2023prediction}
& \LEFTcircle & \cmark & \LEFTcircle & \LEFTcircle & \xmark & \cmark & \xmark & \xmark & \cmark
\\ \hline

Tari et al. (2024)~\cite{tari2024auto}
& \cmark & \cmark & \cmark & \LEFTcircle & \xmark & \LEFTcircle & \xmark & \xmark & \LEFTcircle
\\ \hline

Pham and Kim (2024)~\cite{pham2024elastic}
& \LEFTcircle & \cmark & \LEFTcircle & \cmark & \cmark & \LEFTcircle & \xmark & \xmark & \cmark
\\ \hline

Dogani and Khunjush (2024)~\cite{dogani2024proactive}
& \LEFTcircle & \cmark & \LEFTcircle & \LEFTcircle & \cmark & \LEFTcircle & \xmark & \xmark & \cmark
\\ \hline

Van Do et al. (2025)~\cite{van2025properties}
& \cmark & \xmark & \LEFTcircle & \cmark & \xmark & \xmark & \xmark & \xmark & \cmark \\ \hline

Jeong and Jeong (2025)~\cite{jeong2025autoscaling}
& \cmark & \cmark & \cmark & \cmark & \xmark & \LEFTcircle & \xmark & \xmark & \cmark
\\ \hline

Cai et al. (2025)~\cite{cai2025deep}
& \cmark & \cmark & \cmark & \cmark & \xmark & \cmark & \xmark & \xmark & \xmark
\\ \hline

Meng et al. (2025)~\cite{meng2025catscaler}
& \LEFTcircle & \cmark & \LEFTcircle & \cmark & \xmark & \cmark & \xmark & \xmark & \LEFTcircle
\\ \hline

Patni et al. (2025)~\cite{patni2025predictive}
& \LEFTcircle & \cmark & \LEFTcircle & \xmark & \cmark & \LEFTcircle & \xmark & \xmark & \cmark
\\ \hline

Kumar et al. (2026)~\cite{kumar2026critical}
& \xmark & \xmark & \xmark & \cmark & \xmark & \xmark & \cmark & \xmark & \cmark \\ \hline 

KhalilAbadi et al. (2026)~\cite{khalilabadi2026federated} 
& \xmark & \xmark & \xmark & \xmark & \cmark & \cmark & \xmark & \xmark & \cmark
\\ \hline

Chung et al. (2026)~\cite{chung2026decentralized} 
& \xmark & \xmark & \xmark & \xmark & \cmark & \LEFTcircle & \xmark & \xmark & \xmark
\\ \hline

Fu et al. (2026)~\cite{10948452} 
& \xmark & \xmark & \xmark & \xmark & \cmark & \LEFTcircle & \xmark & \xmark & \xmark
\\ \hline

Laidi et al. (2026)~\cite{10948452} 
& \xmark & \xmark & \xmark & \xmark & \cmark & \LEFTcircle & \xmark & \xmark & \LEFTcircle
\\ \hline 

Ren et al. (2026)~\cite{10930890} 
& \xmark & \xmark & \xmark & \xmark & \cmark & \LEFTcircle & \xmark & \xmark & \xmark
\\ \hline

Wang et al. (2026) ~\cite{10960683} 
& \xmark & \xmark & \xmark & \xmark & \cmark & \LEFTcircle & \xmark & \xmark & \cmark
\\ \hline

\textbf{This Work}
& \textbf{\cmark} & \textbf{\cmark} & \textbf{\cmark} & \textbf{\cmark} & \textbf{\cmark} & \textbf{\cmark} & \textbf{\cmark}
& \textbf{\cmark} & \textbf{\cmark} \\ \hline

\end{tabular}
}
\begin{flushleft}
\footnotesize
\textbf{Notation:} $\checkmark$ = Fully addressed; $\LEFTcircle$ = Partially addressed; $\times$ = Not addressed;
K8s = Kubernetes;
FL = Federated Learning;
DL = Deep Learning;
CRD = Custom Resource Definition.
\end{flushleft}
\end{table*}

Federated learning brings other challenges to cloud-edge systems, such as workload heterogeneity, communication overhead, privacy preservation, and adaptive resource management. To overcome these challenges, Pham and Kim~\cite{pham2024elastic} developed an elastic federated learning framework that works together with Kubernetes' Vertical Pod Autoscaler (VPA) for adaptive resource allocation in edge environments. Likewise, Dogani and Khunjush~\cite{dogani2024proactive} proposed a proactive autoscaling framework for containerized edge applications with workload prediction and adaptive scaling policies, which is based on federated learning. Patni et al.~\cite{patni2025predictive} further investigated predictive virtual machine scaling for federated learning over edge-cloud infrastructures and demonstrated the effectiveness of proactive scaling for improving latency and resource efficiency~\cite{8556457}. While these studies made significant strides in federated autoscaling research, they were mostly targeted at particular scaling goals and did not have a common framework that incorporates predictive deep learning, Kubernetes CRDs, Operator reconciliation workflows, drift-aware correction, and uncertainty-aware adaptive control.

Serverless and Function-as-a-Service (FaaS) environments also received considerable attention in autoscaling research. Tari et al.~\cite{tari2024auto} and Mampage et al.~\cite{mampage2022holistic} comprehensively reviewed autoscaling mechanisms in serverless computing and categorized scaling approaches into machine learning-based, framework-based, and model-based techniques. Their studies highlighted the importance of elastic scaling for serverless applications and identified key research challenges in adaptive resource management. However, these works mainly focused on serverless infrastructures and did not comprehensively investigate federated learning-aware autoscaling or cloud-edge orchestration mechanisms.

In addition, IoT-cloud resource management and QoS-aware scaling have remained important research directions. Verma and Bala~\cite{verma2021auto} reviewed autoscaling and VM migration techniques for IoT-cloud applications and emphasized future workload prediction for maintaining QoS continuity. Despite these contributions, the work did not address modern Transformer-based forecasting architectures, Kubernetes-native automation, or federated cloud-edge autoscaling frameworks. Kumar et al.~\cite{kumar2026critical} presented a comprehensive review of modern Kubernetes environments, covering runtime scheduling, container management, storage, networking, and event-driven processing. The study examined recent advancements such as JobSet, In-place Pod Resizing, enhanced autoscaling, and nftables-based kube-proxy, while highlighting challenges in device-aware scheduling, adaptive resource provisioning, scalable event handling, and orchestration resilience across cloud–edge infrastructures. However, the survey primarily focuses on Kubernetes infrastructure and orchestration rather than intelligent autoscaling. In contrast, this work specifically investigates predictive autoscaling, Kubernetes-native CRD/Operator frameworks, federated learning workloads, drift-aware scaling, and cloud–edge resource management. 

More Recent survey studies have extensively explored federated learning, privacy-preserving intelligence, and cloud–edge collaborative computing. KhalilAbadi et al.~\cite{khalilabadi2026federated} presented a comprehensive review of federated learning across cloud–edge–fog ecosystems, emphasizing privacy, scalability, aggregation strategies, and model drift challenges. Chung et al.~\cite{chung2026decentralized} investigated decentralized federated learning under Non-IID data distributions, highlighting architectural, optimization, security, and scalability issues without relying on centralized servers. Fu et al.~\cite{10948452} focused on differential privacy mechanisms in federated learning and proposed a taxonomy based on privacy guarantees and protection levels. Laidi et al.~\cite{10948452} examined the trade-offs among privacy preservation, model performance, and network efficiency in IoT-based federated learning environments. Ren et al.~\cite{10930890} reviewed Federated Foundation Models (FedFM), discussing scalability, trustworthiness, communication efficiency, and emerging technologies such as quantum computing. Wang et al.~\cite{10960683} surveyed federated analytics techniques and applications for privacy-preserving distributed data processing. Although these surveys offer valuable insights into federated learning, privacy, and distributed intelligence, they provide limited discussion on Kubernetes-native predictive autoscaling, CRD/Operator-driven orchestration, drift-aware resource scaling, and autonomous cloud–edge resource management, which are the primary focus of this survey.

Overall, existing literature demonstrates significant advances in reactive, proactive, and cloud-native autoscaling for distributed cloud and edge environments. Nevertheless, most studies investigate individual aspects of autoscaling, including workload forecasting, Kubernetes orchestration, federated learning, serverless elasticity, or cloud-edge resource management, in isolation. Limited attention has been given to unified frameworks that simultaneously integrate predictive forecasting, Kubernetes-native, CRD/Operator automation, federated learning-aware orchestration and scaling, drift-aware adaptation, and cloud-edge resource optimization. Consequently, there remains a substantial research gap in developing intelligent, autonomous, privacy concern, and self-adaptive autoscaling frameworks capable of operating effectively across heterogeneous federated cloud-edge computing environments.

Table~\ref{tab:feature_comparison} presents a feature-based comparison of representative surveys and recent autoscaling studies. The comparison highlights the evolution of autoscaling research from traditional reactive and proactive scaling approaches toward Kubernetes-native orchestration, predictive forecasting, federated learning, and cloud-edge resource management. It also demonstrates that advanced capabilities such as CRD/Operator-based automation, drift-aware adaptation, and unified federated autoscaling remain insufficiently explored in existing literature. In contrast, this work provides a unified taxonomy that integrates predictive autoscaling, Kubernetes-native orchestration, federated learning-aware resource management, deep learning and Transformer-based forecasting, CRD and Operator-driven automation, and drift-aware adaptive control within a single comprehensive framework for next-generation cloud-edge systems.

\section{Foundations of Cloud-Native and MAPE-Guided Autoscaling Frameworks} \label{sec2}
The fundamental principles of autoscaling in cloud native environments are based on container, microservices, and automated orchestration principles~\cite{deng2024cloud,gannon2017cloud}. It is important to understand these building blocks to be able to investigate the operation of scalability mechanisms and how to expand them to include intelligent and predictive functionality. This section explains the structure of cloud native systems, the built-in Kubernetes autoscalers, core concepts of autoscaling, and metrics and challenges that drive modern scaling strategies~\cite{qu2018auto}. It also defines how resource monitoring, decision-making, and execution are organized, forming the core control framework for all subsequent methods. 

\subsection{Cloud-Native Autoscaling and Kubernetes Architecture}
\begin{figure*}
    \centering
    \includegraphics[width=0.9\linewidth]{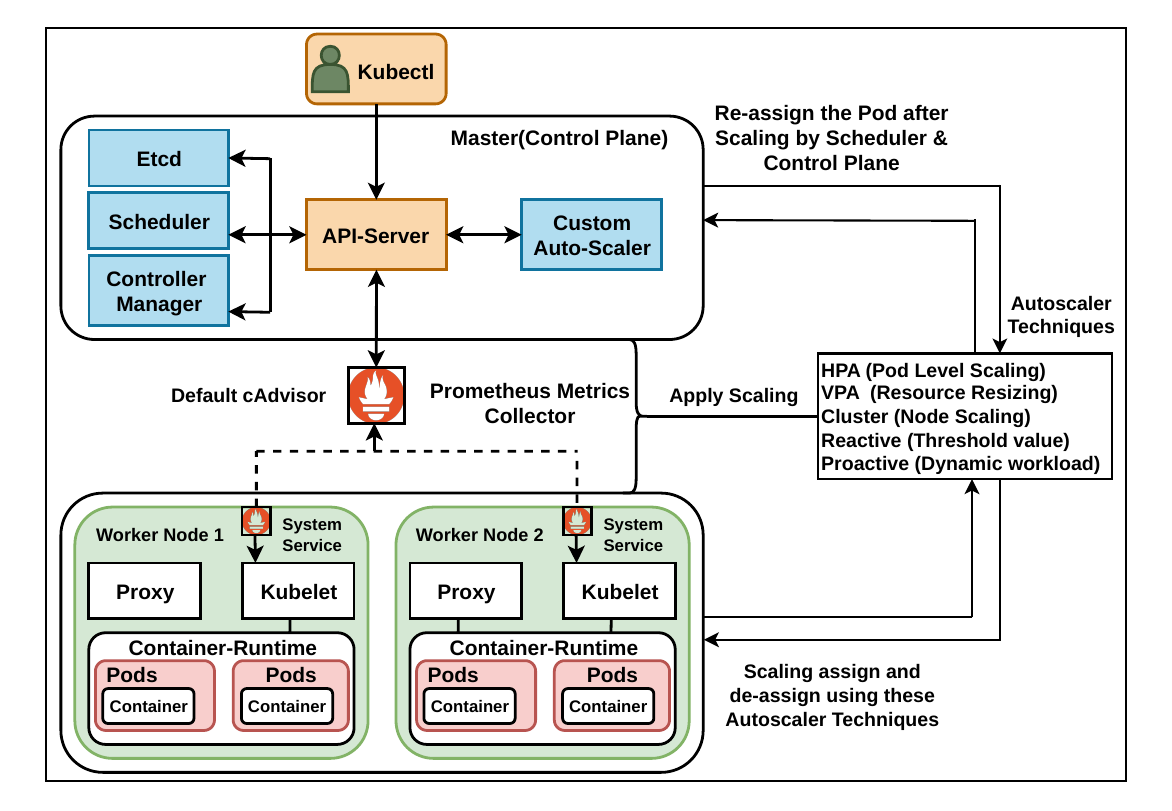}
    \caption{Kubernetes autoscaling workflow showing metric flow from worker nodes to autoscalers and scaling actions via the API Server, Scheduler, and nodes.}
    \label{fig1}
\end{figure*}

The subsection describes how cloud native autoscaling works by beginning with an explanation of the nature of microservices and container based workloads that enable fine grained scalability in a modern distributed system. It then looks at the Kubernetes execution framework shown in Figure~\ref{fig1} that describes how the components of the control plane interact with the worker nodes and the autoscaling components to facilitate automated resource provisioning \cite{chen2015self}. In addition, it explains the autoscaling techniques HPA, VPA and the Cluster Autoscaler, and how these techniques are coordinated on the Kubernetes control plane and worker nodes as illustrated in the Figure.

\begin{itemize}
    \item Cloud Native: Cloud-native systems adopt a modular architecture where applications are decomposed into independent microservices~\cite{gannon2017cloud,kumar2023optimal}. Each microservice operates within a container, ensuring isolation, portability, and consistent execution across heterogeneous environments~\cite{burns2016borg,bernstein2014containers}. Autoscaling decisions rely on key metrics such as CPU and memory utilization, latency, throughput, and quality of service indicators (e.g., error rate, tail latency, and SLA violations), enabling accurate and balanced resource scaling~\cite{zhai2025energy}. Applications are deployed as containers and orchestrated by Kubernetes, where containers are grouped into Pods, the smallest schedulable units. A Pod may contain one or more tightly coupled containers sharing network and storage resources. Microservices communicate through lightweight interfaces, allowing independent deployment, scaling, and fault recovery. This design enables fine-grained scaling of individual services rather than the entire application, improving overall efficiency and elasticity.  ng the entire application, thereby improving efficiency and elasticity.
    
    \item Kubernetes Framework: Kubernetes follows a cluster-based architecture that separates the control plane from workload execution~\cite{k8sArchGuide2024, wijesekera2025kubernetes}. The control plane manages system state, scheduling, and lifecycle operations. The API Server serves as the central interface for resource management~\cite{menouer2021kcss, turin2020formal}, while the Scheduler assigns Pods to nodes based on resource availability and policies. The Controller Manager continuously reconciles the system toward the desired state, and etcd stores cluster configuration as a fault-tolerant distributed key-value store~\cite{verma2015large}. Worker nodes execute application workloads. Each node runs a Kubelet agent to manage Pods, a container runtime (e.g., Docker or containerd) to run containers, and kube-proxy to handle networking and service communication. These components enable scalable and automated orchestration of containerized applications.

    Kubernetes provides core abstractions for application management. Pods are the basic execution units, Deployments manage stateless applications, and StatefulSets support stateful services. Services provide stable networking and load balancing, while ConfigMaps and Secrets handle configuration and sensitive data~\cite{hightower2017kubernetes,k8sObjects2023}. Together, these abstractions ensure scalable, resilient, and maintainable cloud-native systems.
\end{itemize}

\subsubsection{Horizontal Pod Autoscaler (HPA)}

The HPA automatically adjusts the number of pod replicas based on real time metrics such as CPU usage, memory usage, or custom application metrics. It is highly effective for stateless microservices with uniform workloads~\cite{hpaDocs2023}. However, the reliance on threshold driven decisions limits its responsiveness to sharp workload spikes, which may cause temporary degradation before scaling actions take effect~\cite{nguyen2020horizontal}.

\subsubsection{Vertical Pod Autoscaler (VPA)}

The VPA modifies CPU and memory resource requests for individual Pods based on historical usage patterns~\cite{pham2024elastic}. It improves resource efficiency for workloads that cannot be replicated horizontally. Despite its benefits, applying new resource values may require Pod restarts, which can disrupt applications that are sensitive to downtime~\cite{vpaDocs2023}.

\subsubsection{Cluster Autoscaler}

The Cluster Autoscaler operates at the worker node level, dynamically adjusting cluster size based on resource demand~\cite{wijesekera2025kubernetes}. It provisions new nodes when Pods cannot be scheduled due to insufficient resources and removes underutilized nodes to optimize cost. Compared to HPA and VPA, its response is slower due to the overhead of provisioning cloud instances, making it more suitable for workloads with gradual variability rather than sudden spikes~\cite{clusterAutoscaler2023}.

Figure~\ref{fig1} illustrates the Kubernetes autoscaling architecture and component interactions. Client requests are processed by the API Server, which coordinates with the Scheduler, Controller Manager, and etcd to maintain cluster state. The Scheduler assigns Pods to worker nodes, where execution is managed by the Kubelet and container runtime. Autoscaling decisions are driven by metrics collected from the Metrics Server and evaluated by mechanisms such as HPA~\cite{hpaDocs2023,nguyen2020horizontal}, VPA~\cite{vpaDocs2023}, Cluster Autoscaler~\cite{clusterAutoscaler2023}, and reactive–proactive techniques~\cite{kumar2024optimizing}. These decisions are fed back to the API Server, triggering scheduling updates and node scaling. This forms a continuous feedback loop for dynamic adjustment of pods, resources, and cluster capacity.

Finally, Kubernetes provides a flexible autoscaling framework through containerized microservices, declarative management, and control loops. While HPA, VPA, and Cluster Autoscaler address scaling at different levels, they are limited by reactive behavior, latency, and metric dependency. These challenges have led to the development of intelligent and predictive autoscaling approaches using custom operators, forecasting models, and adaptive control strategies, as illustrated in Figure~\ref{fig1}.

\subsection{Monitor-Analyse-Plan-Execute (MAPE) Guided Autoscaling Frameworks}

Autoscaling enables the dynamic adjustment of computing resources in response to changing workload demands. In cloud-native environments, this process involves continuously monitoring system behavior, analyzing conditions, planning scaling actions, and executing them through orchestration mechanisms~\cite{qu2018auto}. While traditional approaches rely on static threshold rules, the increasing variability and complexity of modern workloads have driven the adoption of adaptive, model-driven strategies. MAPE control loop provides a structured foundation for designing such intelligent predictive autoscaling frameworks~\cite{villegas2012maestro}.

\begin{figure}
    \centering
    \includegraphics[width=0.99\linewidth]{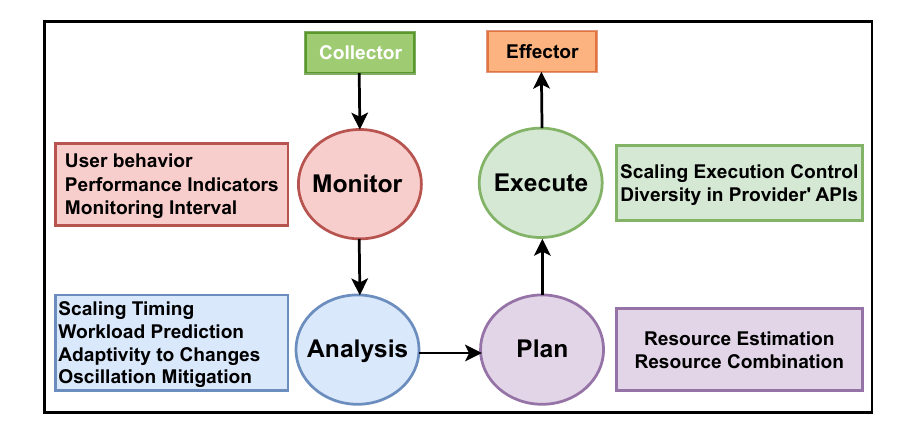}
    \caption{MAPE loop showing how metrics flow through monitor–analyze–plan–execute stages to guide resource adjustments via the effector.}
    \label{fig2}
\end{figure}

\begin{figure*}
    \centering
    \includegraphics[width=0.9\linewidth]{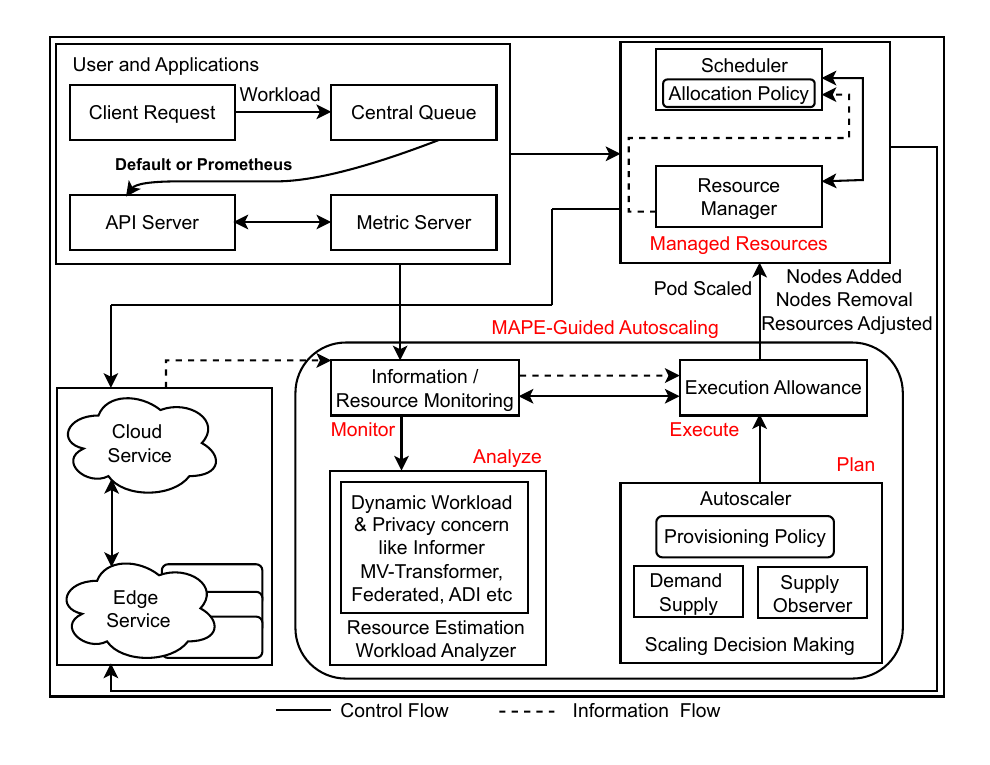}
    \caption{Enhanced MAPE-guided autoscaling architecture for cloud–edge and federated environments, showing workload flow and adaptive, privacy-aware scaling across distributed systems.}
    \label{fig3}
\end{figure*}

The MAPE loop is shown in Figure~\ref{fig2} in which system metrics are measured using sensors and the measured values are continuously fed back to update the resource(s) settings. You can gather metrics from a variety of sources, including Kubernetes Metrics Server, Prometheus, and custom instrumentation, and include applications metrics as well as CPU utilization, memory usage, latency, and throughput. The analysis of these observations involves rule-based, statistical or predictive approaches which are used to identify workload patterns, resource bottlenecks and potential overload situations. This analysis is used to inform the planning stage to take informed decisions about scaling (scaling up and down pod replicas, changing resources, or expanding nodes), with stability and policy compliance. The execution phase then uses Kubernetes controllers, operators, or custom resource definitions to implement the decisions, and the reconciliation process will ensure convergence towards the desired state of the system.

The enhanced MAPE-guided autoscaling architecture in cloud, edge, and federated environments is illustrated in Figure~\ref{fig3}. Dynamic workloads are generated by incoming user requests and application activities. Workload traces may be collected directly from Kubernetes environments using monitoring tools such as Prometheus and cAdvisor, which provide real-time metrics including CPU, memory, network, and throughput utilisation. Alternatively, publicly available datasets generated from large-scale production environments, such as Google Cluster Trace, Bitbrains, Microsoft Azure, Alibaba Cluster Trace, and NASA HTTP datasets, can be used for workload modelling, forecasting, and autoscaling evaluation. These metrics are continuously monitored through a metrics collection pipeline and forwarded to the analysis layer for further processing~\cite{kumar2025multivariate}. The \textbf{Monitoring} phase supports reactive autoscaling by providing real-time visibility into system behaviour and detecting sudden workload fluctuations. In the \textbf{Analysis phase}, advanced forecasting models such as Informer, MV-Transformer, LSTM, and Bi-LSTM analyse historical and real-time workload patterns to predict future resource demand, enabling proactive scaling decisions. For federated learning environments, this stage additionally considers client heterogeneity, varying participation rates, communication delays, and privacy constraints, which may be addressed through privacy-preserving techniques such as differential privacy~\cite{zhai2025energy}. The \textbf{Planning phase} combines current observations with forecasted demand to determine optimal scaling actions, thereby creating a hybrid autoscaling strategy that balances performance, resource utilisation, operational cost, and system stability. Finally, the \textbf{Execution phase} applies these decisions through Kubernetes-native mechanisms such as the Horizontal Pod Autoscaler (HPA), Vertical Pod Autoscaler (VPA), and Cluster Autoscaler. These components dynamically adjust pod replicas, resource allocations, and cluster capacity, ensuring efficient workload distribution and seamless resource management across cloud, edge, and cloudlet infrastructures.

In summary, MAPE is a modular framework that can be extended, and it decouples monitoring, decision-making and execution, enabling the autoscaling process to be flexible. The design ensures that machine learning models and sophisticated optimization methods can be easily integrated, so that resources can be managed adaptively, proactively, and with privacy concerns.

\subsubsection{Reactive Autoscaling}
Reactive autoscaling is based on actual time metric thresholds. At some point when the resource consumption is beyond a pre-defined threshold, the system initiates a scaling action \cite{kumar2025multivariate,kumar2025optimizing}. Typically, it is either a CPU or memory threshold that triggers the HPA. Reactive methods are simple, stable, and common to practitioners~\cite{lorido2014review}. But they only respond when there's an overload of work. Such delay can lead to under-provisioning, latency or degradation of service. If there is no smoothing or hysteresis mechanism, then there can be oscillations due to the changing workload.

\subsubsection{Proactive Autoscaling}
Proactive autoscaling anticipates what is likely going to happen in the future and prepares resources accordingly. The strategy is based on forecasting techniques like statistical models, machine learning models, or deep learning architectures~\cite{kumar2024optimizing, kumar2025multivariate, kumar2025optimizing}. Proactive autoscaling can also help to minimize latency and enhance user experience by provisioning resources in advance. But the effectiveness relies on the quality of monitoring data and predictions. Combination of proactive techniques and Kubernetes operators and custom controllers enables automatic scaling at the orchestration level~\cite{ghobaei2020machine}. This provides a basis for independent intelligent scaling pipelines.

\subsubsection{Hybrid and Event-Driven Scaling}
Hybrid autoscaling is Reactive and Proactive integrated together to obtain a both Reactive and Proactive response for immediate response and long term stability~\cite{kakkad2025intelligent}. In this method, proactive forecasting forecasts the scaling action corresponding to the plan, and reactive triggering is performed for sudden increases in numbers that cannot be predicted.In this approach, the planned scaling action is determined through the proactive forecasting, whereas the unpredictable increases of numbers are dealt with through the reactive triggering. Event driven scaling scales based on events in addition to resource metrics, like user events, application specific indicators, queue length, or message volume. For asynchronous or burst heavy workloads, frameworks such as KEDA can work with events from messaging systems, application telemetry, or application middleware to enable autoscaling~\cite{vu2022predictive}.

Cloud native environments have a number of challenges with autoscaling in the MAPE loop. Sudden changes in workload may make it hard for reactive methods to keep up with, and heterogeneous application behaviors are not conducive to devising universal scaling rules~\cite{qu2018auto}. Stateful services also make scaling more complex because of consistency constraints, and the delay in starting up containers, scheduling latency~\cite{verma2025optimal} and provisioning nodes can impact performance. Furthermore, fairness and isolation issues exist in multi-tenant clusters and energy, bandwidth, and intermittent connectivity are issues in edge environments~\cite{zhai2025energy}. Challenges such as these underscore the need for adaptive, predictive and powerful autoscaling solutions.

MAPE-guided autoscaling addresses these limitations by integrating reactive, proactive, and hybrid strategies within a structured control loop~\cite{kumar2025multivariate,kumar2025optimizing}. By separating monitoring, analysis, planning, and execution, the framework enables flexible integration of predictive models, anomaly detection, and intelligent decision policies. As illustrated in Figure~\ref{fig3}, this layered design connects the MAPE pipeline with Kubernetes autoscalers and operator-driven mechanisms, supporting scalable and responsive resource management across cloud and edge environments. This approach advances autoscaling beyond static thresholds toward learning-based, context-aware, and policy-driven systems.

\section{Taxonomy of Autoscaling Techniques} \label{sec3}
\begin{figure*}
    \centering
    \includegraphics[width=0.9\linewidth]{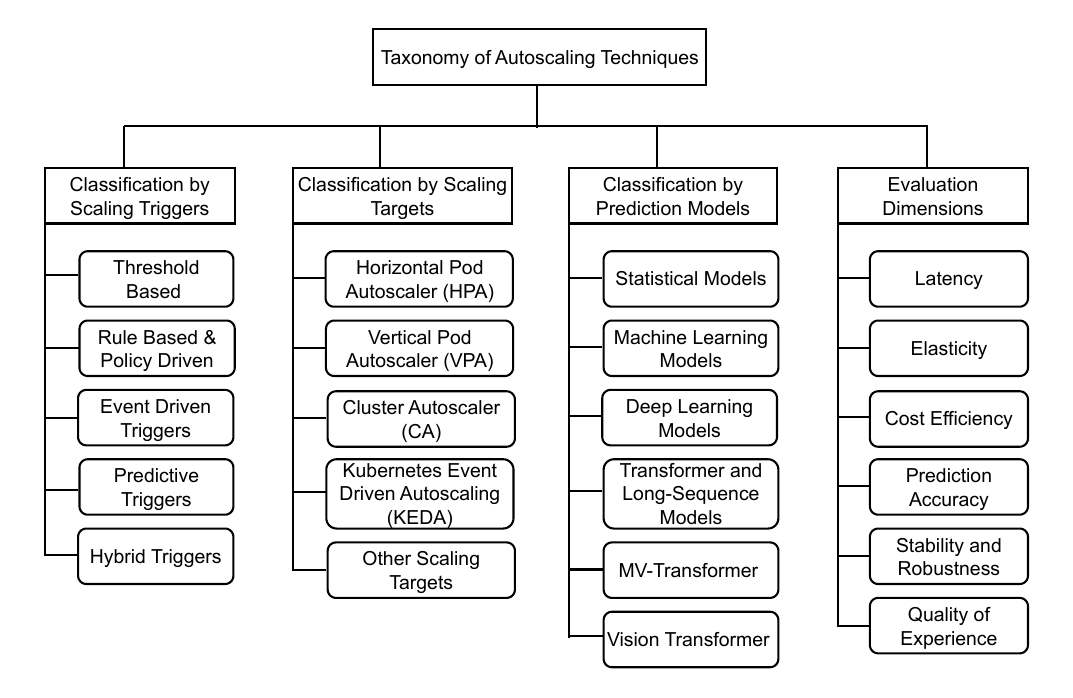}
    \caption{Overview of the autoscaling taxonomy, illustrating the relationships among scaling triggers, scaling targets, prediction models, and evaluation dimensions}
    \label{fig4}
\end{figure*}

The taxonomy of autoscaling techniques provides a structured way to understand how different autoscaling strategies operate in cloud and edge environments~\cite{liang2026autoscaling, alharthi2024auto}. Autoscaling approaches vary widely in how they detect workload patterns, select the scaling target, and determine the appropriate scaling action \cite{lorido2014review, peng2025global, molleti2022kubernetes}. This taxonomy can be organised into four major dimensions: (i) scaling triggers that determine when scaling is activated, (ii) scaling targets such as pods, containers, nodes, or event driven services, (iii) prediction models covering statistical forecasting, machine learning, and deep learning techniques, and (iv) evaluation dimensions that measure latency, elasticity, accuracy, stability, and cost efficiency. As illustrated in Figure~\ref{fig4}, the autoscaling taxonomy defines how workload signals transition through triggers, prediction models, and scaling targets to produce adaptive resource management decisions~\cite{cai2025deep}. 

By categorising autoscaling techniques across these dimensions, researchers can better compare existing solutions, understand their strengths and limitations, and identify opportunities for improvement. The taxonomy also clarifies how different strategies address challenges such as rapidly changing workloads, latency sensitive applications, resource variability at the edge, model complexity, and privacy requirements in federated environments. Overall, this classification forms the basis for analysing the full landscape of autoscaling mechanisms used across modern distributed computing platforms. The final comparative summary of autoscaling techniques across triggers, scaling targets, prediction models, and evaluation dimensions is shown in Table~\ref{tab1}. 

\subsection{Classification by Scaling Triggers}
Scaling triggers are the conditions that will cause autoscaling to be activated and what will happen when workloads change~\cite{waghmare2025comprehensive}. These triggers are critical as the accuracy and stability of the scaling decisions rely on the ability of these triggers to reflect the behaviour of the system.
\begin{itemize}
\item Threshold Based Triggers: Triggers based on threshold are the simplest and most commonly used mechanism. When one of the metrics, like CPU utilisation, memory usage, request rate, or queue depth, exceeds a specific threshold, they trigger scaling~\cite{archana2025threshold}. While simple to implement, threshold rules are ineffective for sharp spikes, non-stationary workloads and multi-variable dependencies. They need manual tuning, and can cause slow reactions or oscillation when workloads are changing rapidly.

\item  Rule Based and Policy Driven Triggers: These triggers are built from a combination of several metrics or conditions, and are expressed using logical expressions~\cite{li2025investigation}. For instance, scaling only happens under high load conditions on the CPUs and when the latency of the requests increases. Policies are more stable than thresholds, but still rely on a predetermined logic. They can be helpful in the context of known workloads, but have limited value when behaviour is outside of the expected range.

\item Event Driven Triggers: Event driven autoscaling responds to events from outside the system, instead of resource metrics. This can be anything from message arrival on Kafka queues, sensor data spikes, or shifts in request patterns~\cite{gattupalli2025serverless}. Event driven triggers are particularly valuable in microservices and serverless applications where loads occur in spikes. Kubernetes Event Driven Autoscaling (KEDA) can handle such triggers, and can handle a fast response in an asynchronous environment.

\item Predictive Triggers: They are triggers that start scaling according to the forecast workloads, not current workloads, ~\cite{kumar2026critical}. They rely on analytic techniques such as statistical forecasting, machine learning, or deep learning models. Predictive triggers can greatly minimize under-provisioning and latency as resources are prepared beforehand. They are very dependent on the accuracy of the forecasts, the generality of the model, and the timeliness of the monitoring data.

\item Hybrid Triggers: These are the hybrid ones which are a mix of reactive and predictive triggers. The predictive part is responsible for load behaviour prediction for long-term and the reactive part is responsible for unexpected load response.The reactive part deals with unexpected load response and the predictive part deals with load behaviour prediction for long term. Hybrid triggers are robust and stable, and can be used in environments where there are both periodic and unpredictable patterns. To reduce the drawbacks of one or the other, many state of the art autoscaling systems use this hybrid approach.
\end{itemize}
In summary, the different types of triggers highlight the importance of making the right choice to ensure responsiveness, stability, and accurate forecasting, which are fundamental for robust autoscaling systems.

\subsection{Classification by Scaling Targets}

Scaling targets define which component of the system is adjusted when an autoscaling decision is executed. Each autoscaler in Kubernetes operates at a different layer of the stack and therefore addresses different performance and workload characteristics. Understanding these targets is essential for selecting the appropriate mechanism for cloud, edge, and federated environments.

\begin{itemize}
    \item Horizontal Pod Autoscaler (HPA): The HPA manages scaling at the pod replica level by increasing or decreasing the number of running Pods based on metrics such as CPU utilisation, memory consumption, request rate, or custom application indicators. HPA is well suited for stateless microservices that can be replicated easily. Its lightweight decision process enables responsive scaling for moderately variable workloads. However, because HPA is reactive and relies heavily on predefined thresholds, it may lag behind sudden workload surges or highly non-linear traffic patterns~\cite{hpaDocs2023, nguyen2020horizontal}.
    
    \item Vertical Pod Autoscaler (VPA): The VPA adjusts CPU and memory requests for individual Pods. Instead of creating new replicas, VPA reallocates resources to ensure that a Pod receives adequate computational capacity. This makes VPA suitable for stateful services, monolithic applications, and long-running processes that cannot be horizontally replicated. VPA analyses historical usage data to determine optimal values, although applying these changes may require Pod restarts. For this reason, VPA is often used in recommendation mode or scheduled during controlled maintenance intervals~\cite{pham2024elastic, vpaDocs2023}.

    \item Cluster Autoscaler (CA): The Cluster Autoscaler manages scaling at the infrastructure layer by adding or removing worker nodes. It is activated when the scheduler cannot place Pods due to insufficient resources or when nodes remain underutilised for an extended period. CA interacts with cloud provider APIs to provision or delete compute instances. While node-level scaling is slower than Pod-level scaling, it ensures that the underlying infrastructure capacity aligns with application-level autoscaling actions. When combined with HPA and VPA, the Cluster Autoscaler supports a fully coordinated, multi-layer autoscaling strategy~\cite{wijesekera2025kubernetes, clusterAutoscaler2023}.

    \item Kubernetes Event Driven Autoscaling (KEDA): KEDA enables autoscaling based on external event sources instead of resource metrics~\cite{shrestha2025enhancing}. It reacts to signals such as message queue depth, event frequency, or IoT sensor activity from systems like Kafka, RabbitMQ, Azure Service Bus, or AWS SQS. When event load increases, KEDA rapidly scales worker Pods. This event-driven model is particularly effective for asynchronous, batch, and microservices workloads where traffic arrives in bursts. KEDA works in conjunction with the HPA, allowing event-driven logic to seamlessly integrate with Kubernetes autoscaling workflows.

    \item Other Scaling Targets: Beyond these core mechanisms, modern cloud-native platforms support additional scaling targets~\cite{molleti2022kubernetes, jeong2025autoscaling}. These include node rebalancing to eliminate resource hotspots, service-level scaling across microservice pipelines, adjusting container concurrency (common in serverless workloads), GPU or accelerator scaling for AI/ML inference pipelines, and federated scaling across multiple clusters in cloud–edge environments. These emerging mechanisms reflect the increasing diversity and complexity of distributed systems.
\end{itemize}

Overall, these scaling targets create a multi-layer autoscaling ecosystem that allows Kubernetes to adapt resources at the container, pod, node, and event levels, enabling robust and workload-aware resource management in modern distributed environments.

\subsection{Classification by Prediction Models}

Prediction models are essential for proactive autoscaling, as they allow systems to anticipate the need for scaling up or down and make decisions accordingly, before the performance degrades~\cite{dang2022efficient,verma2021auto,radhika2021review}. With the emergence of dynamic and heterogeneous cloud, edge, and federated environments, these models have surpassed basic statistical methods, incorporating advanced deep learning architectures and multivariate models~\cite{khaleq2021intelligent, kumar2025multivariate}. They are summarized in Table~\ref{tab1} below by their main attributes.

\begin{itemize}
    \item Statistical Models: The first models that have been developed for workload prediction are the statistical models like ARIMA and Holt-Winters which assume linear relationships between the workloads and also find the patterns in time-series data of the workloads in the past~\cite{vu2022predictive,fog2025comparing}. They are efficient, and useful for workloads that have stable trends or seasonality. Their assumptions of linearity and stationarity, however, restrict their usefulness in dynamic cloud-native environments where the system exhibits bursty and multivariate behaviour, and are better suited for lightweight baselines or in resource-limited edge systems~\cite{kumar2025multivariate}.

    \item Machine Learning Models: In addition to statistical methods, machine learning models have been developed to address the non-linear relationships and interactions between various resource metrics, like CPU, memory, and network usage, that can be used for enhancing performance of resource utilization systems~\cite{pintye2024enhancing,thota2022intelligent,singh2023machine}. The supervised methods (such as Random Forest, SVR, GBM) can be used to accurately predict workloads, while the unsupervised and semi-supervised approaches can be used for workload clustering and anomaly detection in dynamic environments~\cite{da2022online}. Such models can be used to estimate the resource requirement and inform scaling decisions for heterogeneous Kubernetes clusters, but they can suffer from a problem of concept drift, cold-start issues, and changing workloads~\cite{kumar2024optimizing}.
    
    \item Neural Sequence and Deep Learning Models: Deep learning models such as LSTM, GRU, Bi-LSTM and CNN-LSTM can outperform other models in terms of prediction accuracy, as deep learning models learn temporal dependencies and complex non-linear patterns directly from workload traces~\cite{etemadi2021cost,dang2021deep,taha2024proactive,agarwal2024deep}. The long-term trends, bursty behaviour and multivariate interactions are well captured by these models and serve as an ideal fit for proactive autoscaling within a cloud-native system~\cite{kumar2025multivariate, cai2025deep}. Their increased complexity and computational requirements, however, can make them less feasible to deploy in resource-constrained settings.

    \item Transformer-based models are able to model long and complex workload sequences without sequential processing constraints due to its global dependency capturing ability provided by self-attention mechanisms~\cite{ding2024dynamic,meng2025catscaler}. Informer, Autoformer, and Time-Series Transformers are variants of AutoEncoder, which offer enhanced scalability and long-horizon forecasting, making them particularly suitable for large-scale cloud or edge systems for proactive autoscaling of these systems~\cite{kumar2025optimizing}. Unlike recurrent models, these models offer higher flexibility in cases of non-stationary and bursty workloads.

    \item MV-Transformer and Multivariate Forecasting: Multivariate Transformer (MV-Transformer) models extend attention mechanisms to jointly capture temporal dynamics and cross-metric dependencies across multiple resource signals~\cite{kumar2025multivariate, shahzaad2025service}. By modelling interactions among CPU, memory, network, and application-level metrics, these models provide a comprehensive representation of workload behaviour. This enables accurate and fine-grained proactive autoscaling decisions, particularly in complex Kubernetes, cloud-edge, and federated environments with highly correlated and non-stationary workloads~\cite{kumar2025multivariate,kumar2025optimizing}.
\end{itemize}

\begin{table*}[t]
\centering
\caption{Comparative of autoscaling techniques across triggers, scaling targets, prediction models, and evaluation dimensions.}
\label{tab1}
\renewcommand{\arraystretch}{1.25}
\begin{tabularx}{\textwidth}{p{3cm} X X X X}
\hline
\textbf{Category} & \textbf{Definition / Purpose} & \textbf{Strengths} & \textbf{Limitations} & \textbf{Typical Use Cases} \\
\hline

\textbf{Scaling Triggers}~\cite{waghmare2025comprehensive,
archana2025threshold,
li2025investigation,
kakkad2025intelligent,
gattupalli2025serverless} &
Mechanisms that activate autoscaling decisions (threshold, rule-based, event-driven, predictive, hybrid). &
Simple configuration; event triggers react quickly; predictive triggers offer foresight; hybrid improves overall stability. &
Threshold oscillations; event triggers rely on external systems; predictive triggers depend on model accuracy. &
Microservices and web apps (threshold); serverless (event-driven); cloud–edge forecasting (predictive); large distributed platforms (hybrid). \\

\hline
\textbf{Scaling Targets}~\cite{hpaDocs2023, nguyen2020horizontal,pham2024elastic, vpaDocs2023,wijesekera2025kubernetes, clusterAutoscaler2023,shrestha2025enhancing,molleti2022kubernetes, jeong2025autoscaling} &
Components targeted for scaling (HPA, VPA, Cluster Autoscaler, KEDA). &
Pod-level elasticity (HPA); resource rightsizing (VPA); node-level adaptability (Cluster Autoscaler); efficient event-driven scaling (KEDA). &
HPA is reactive; VPA may restart pods; Cluster Autoscaler is slow; KEDA depends on event sources and external queues. &
Stateless workloads (HPA); stateful services (VPA); large clusters (CA); streaming/queue-driven systems (KEDA). \\

\hline
\textbf{Statistical Models}~\cite{kumar2024optimizing, vu2022predictive,fog2025comparing} &
ARIMA, Holt–Winters for linear or seasonal workloads. &
Low cost, interpretable, suitable for simple trends. &
Weak for non-linear, multivariate, or bursty patterns. &
Baseline forecasting, IoT workloads, predictable periodic services. \\

\hline
\textbf{Machine Learning Models}~\cite{pintye2024enhancing, thota2022intelligent, singh2023machine,da2022online} &
Random Forest, SVR, Gradient Boosting, kNN for non-linear pattern learning. &
Captures multivariate relations; good performance with moderate data; reduces noise. &
Requires feature engineering; sensitive to concept drift; limited long-term memory. &
Microservice autoscaling, latency prediction, moderate-scale cloud workloads. \\

\hline
\textbf{Deep Learning Models}\cite{etemadi2021cost,dang2021deep,taha2024proactive,agarwal2024deep,kumar2024optimizing,kumar2025multivariate,kumar2025optimizing,dogani2022k,vu2022predictive,ouhame2021efficient} &
LSTM, GRU, Bi-LSTM, CNN–LSTM for sequential and multivariate forecasting. &
Models long-range patterns; minimal need for handcrafted features; robust to noise. &
High training cost; requires large datasets; may be slow on edge devices. &
Proactive autoscaling, Kubernetes pod rightsizing, edge computing forecasts. \\

\hline
\textbf{Transformer Models}~\cite{ding2024dynamic,meng2025catscaler} &
Attention-based models (Transformer, Informer, Autoformer) for long-sequence prediction. &
Parallel sequence processing; captures global dependencies; handles irregular workloads. &
High memory usage; complex tuning; attention cost scales with sequence length (except Informer). &
Long-horizon autoscaling, cloud traffic forecasting, federated workload modelling. \\

\hline
\textbf{MV-Transformers}~\cite{kumar2025multivariate,kumar2024optimizing,shahzaad2025service} &
Multivariate Transformers with cross-metric attention. &
Captures temporal + inter-metric correlations; high accuracy; strong for proactive autoscaling. &
Complex architecture; requires large compute; training may need distributed nodes. &
CRD-based predictive scaling, Kubernetes operators, multi-cluster orchestration. \\

\hline
\textbf{Evaluation Dimensions}\cite{kumar2024optimizing,kumar2026critical,kumar2025multivariate,kumar2025optimizing,jeong2025autoscaling} &
Metrics for assessing autoscaling performance (latency, elasticity, cost, accuracy, stability, QoE). &
Multi-dimensional assessment enables holistic comparison of autoscalers. &
Trade-offs difficult to optimise simultaneously; cost–performance balance is workload-dependent. &
Benchmarking cloud, edge, and federated autoscaling frameworks; SLA-driven optimisation. \\

\hline
\end{tabularx}
\end{table*}

\subsection{Evaluation Dimensions}

Evaluating autoscaling techniques requires a multidimensional perspective because scaling decisions influence system behaviour at several layers, including performance, resource utilisation, cost, and user experience. The prediction models discussed in the previous subsections, ranging from statistical methods to deep learning and Transformer based architectures, directly affect these evaluation dimensions. Likewise, the choice of scaling triggers and scaling targets determines how effectively an autoscaling system adapts to changing workloads. This subsection describes the primary metrics used to assess autoscaling quality and explains how they relate to the taxonomy of techniques presented earlier~\cite{kumar2026critical,kumar2025multivariate,jeong2025autoscaling}.

\begin{itemize}

\item Latency: In cloud and edge systems, latency is a key measure of performance that is the time it takes for the system to process a request or task. Latency can be understood in two ways when it comes to autoscaling: the application response latency and scaling reaction latency. To prevent congestion, prediction models like LSTM, GRU, and MV Transformers predict future demand and allocate resources accordingly, lowering latency. On the other hand, if all triggers are reactive, then a latency spike may occur during a workload burst since scaling actions only start when there is a violation of the thresholds.

\item Elasticity: Elasticity is the ability of the resource allocation to closely conform to the actual load on a resource. An autoscaling system is elastically scalable, providing additional resources when demand rises and letting them go when demand falls. In this aspect, the statistical models tend to be less effective because of their inability of dealing with non linear and multivariate patterns. Deep learning and Transformer based models improve elasticity by making more accurate predictions, allowing for fine grained and proactive scaling actions at the pod, node, and cluster level. Each autoscaler–like HPA, VPA, and Cluster Autoscaler–has different capabilities in relation to elasticity, which can be used for different types of scaling.

\item Cost Efficiency: Too little or too much scaling can affect the cost efficiency of the operation: too much scaling can result in resource wastage, too little can result in performance degradation that may result in penalties or SLA violation. Pro-active models such as Informer and MV Transformer assist to reduce costs by forecasting demand trends and allocating resources accordingly. Additionally, event driven autoscalers such as KEDA can be used to further optimize cost by deploying resources on demand of a specific event. The analysis of trade-off between resource utilization and performance guarantees is needed for evaluation in this dimension.

\item Prediction Accuracy: Prediction accuracy is key to proactive autoscaling, and it will dictate the effectiveness of a model's predictive capabilities. Predictions ensure that the right resources are provisioned and minimize the risk of under provisioning and over provisioning, which in turn improves latency, elasticity, and cost-efficiency. Performance of the model can be quantified using various metrics such as MSE, RMSE, MAE and MAPE. Compared to classical models, deep learning and Transformer models typically outperform them in terms of accuracy because they better capture long-term dependencies and multivariate interactions. With FL based predictors, these benefits are extended to distributed environments where data privacy and heterogeneity should be preserved.

\item Stability and Robustness: Stability is the capacity of an autoscaling system to not oscillate between too large and too small. Short term fluctuations in values may cause oscillation if short term values are used as threshold based triggers. Robustness is the ability of the autoscaler to resist the effects of workload noise, sudden peaks, model prediction errors and node failures. There are techniques to improve this dimension like hybrid scaling strategies, uncertainty aware models, and drift aware mechanisms like ADI that use confidence levels or dynamic corrections to determine the scaling.

\item Quality of Experience (QoE): QoE is a measure that reflects the quality perceived by the user of a service, and is composed of a set of performance metrics such as response time, error rate, jitter and availability. The vision workloads handled by Vision Transformers have a particular need for high QoE due to their use in real-time applications like self-navigating and surveillance. One of the most important dimensions in the cloud edge and federated environments is autoscaling systems that can keep up QoE despite variation in workloads or limitations in resources.

\end{itemize}

These are the different dimensions on which auto-scaling mechanisms can be evaluated in general. They provide an overview of the impact that the selection of the prediction model, the scaling trigger and the scaling target have on the performance, cost and satisfaction of the system. These dimensions provide a clearer view into how cloud, edge, and federated platforms can be designed and compared with regards to intelligent predictive autoscaling capabilities. Table~\ref{tab1} gives the last summary of the predictive autoscaling models and Kubernetes-native automation techniques.

The answers to \textbf{RQ1} can be gained by the insights that these different types of autoscaling techniques provide, which is how accurate and responsive they are when using the different predictive models. For stable and seasonal workloads, interpretable baselines can be obtained using statistical models like ARIMA and Holt-Winters. Machine learning techniques are able to model non-linear relationships between CPU, memory, network and application metrics, making their system more adaptable under dynamic environments. To capture the temporal patterns and multivariate dependencies, neural models such as LSTM, GRU, Bi-LSTM and CNN-LSTM are implemented to promote proactive autoscaling. Based on transformer architectures, these are further extended by attention mechanisms, which allow to make accurate long horizon and multi-scale predictions, and multivariate Transformers can predict context-aware scalings by jointly analysing correlated metrics. In general, these models help boost prediction accuracy, minimize under-provisioning and latency, and facilitate stable, efficient and adaptive auto-scaling in a complex cloud-native environment.

\section{Predictive Autoscaling Models and Custom Resource Definitions} \label{sec4}
This section synthesises three key research directions that define intelligent Predictive autoscaling in cloud-native environments. First, proactive long-sequence forecasting using the InformerAutoScale framework~\cite{ding2024dynamic} leverages efficient attention mechanisms to predict future workload demand, enabling advance resource provisioning and reducing delays associated with reactive scaling. Second, multivariate prediction through the MV-Transformer within a MAPE control loop~\cite{kumar2025multivariate} jointly models CPU, memory, disk, network, and request rate metrics. By capturing cross-metric dependencies and temporal patterns, it improves scaling accuracy for complex distributed workloads~\cite{arbat2022wasserstein,wu2025elastic}. Third, predictive intelligence is integrated into the Kubernetes control plane using Custom Resource Definitions (CRDs) and Operator-based reconciliation~\cite{shahzaad2025service,ding2024dynamic,meng2025catscaler}. In this approach, forecasting outputs and scaling policies are encoded as custom resources, while operators continuously monitor system state and enforce scaling decisions. This enables low-latency actuation, consistent policy execution, and seamless integration with Kubernetes resource management~\cite{wijesekera2025kubernetes,kubernetesExtendKubernetes,miracapillo2025enabling}. 

\begin{figure}
    \centering
    \includegraphics[width=0.99\linewidth]{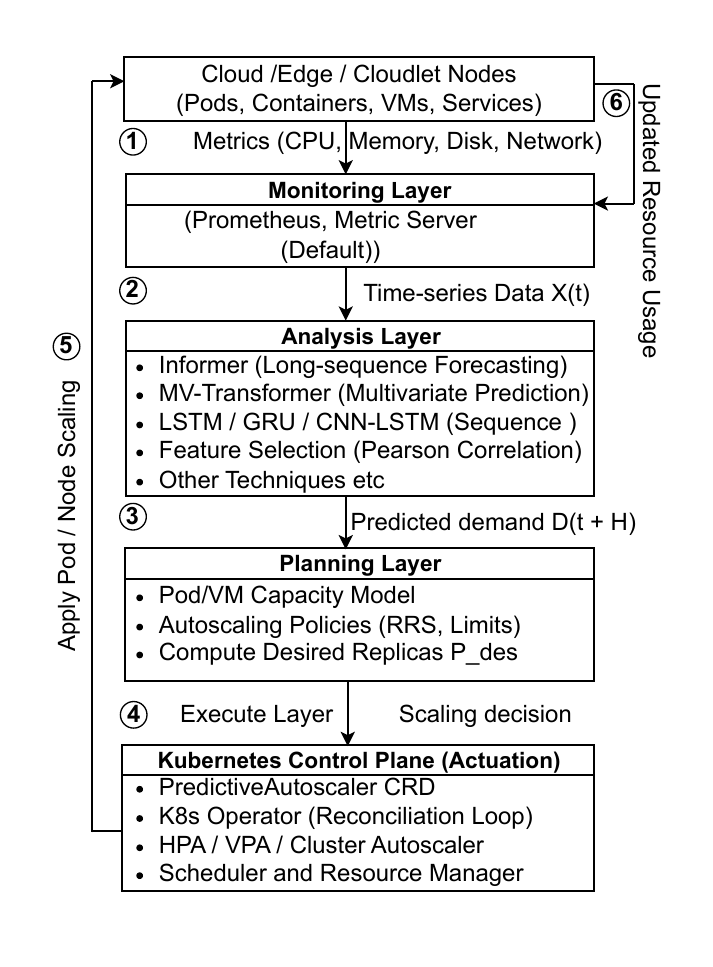}
    \caption{Predictive autoscaling pipeline integrating Informer and MV-Transformer with Kubernetes CRDs and Operators in a MAPE-guided workflow for proactive resource management across cloud–edge environments.}
    \label{fig5}
\end{figure}

\begin{figure*}
    \centering
    \includegraphics[width=0.99\linewidth]{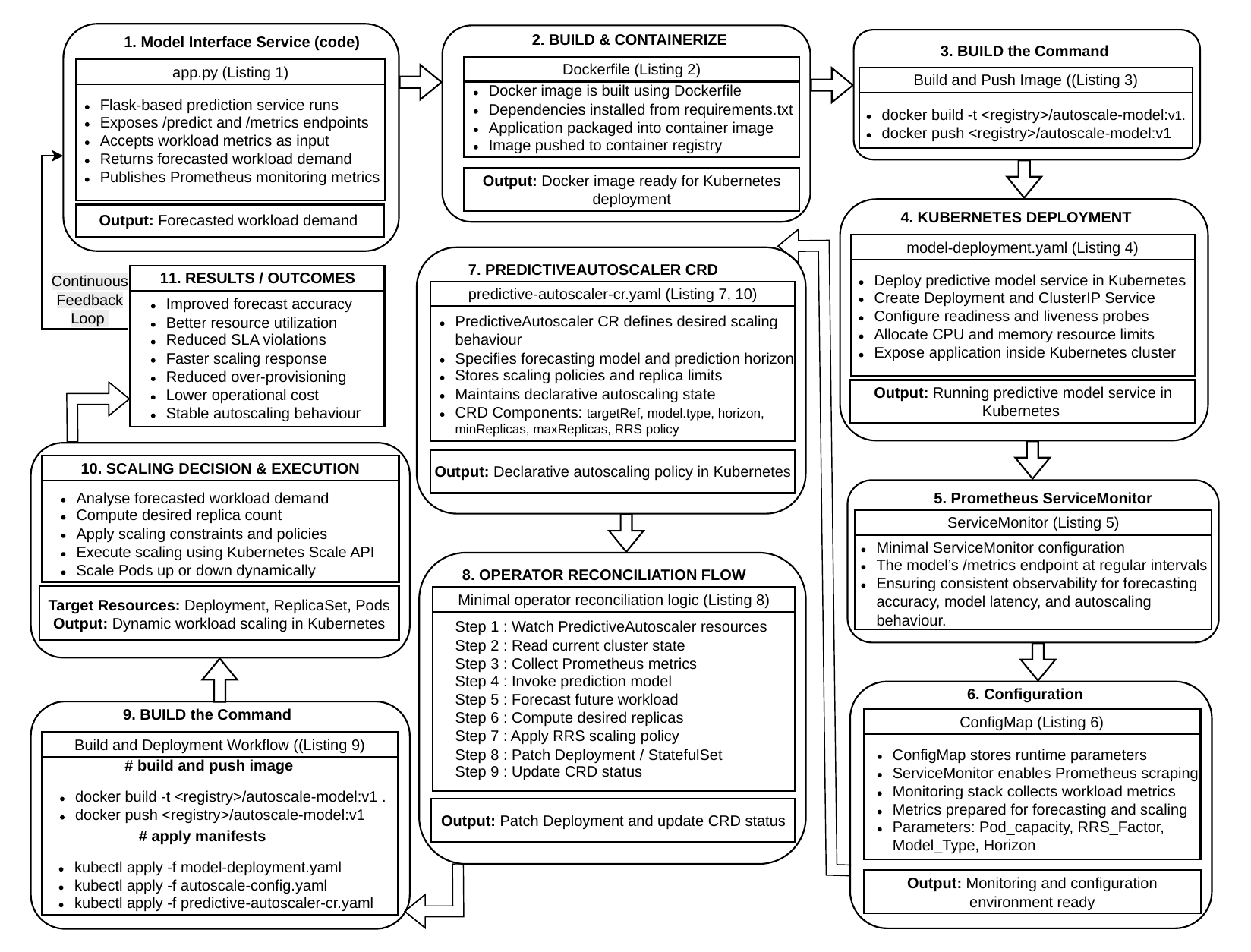}
    \caption{End-to-end predictive autoscaling workflow integrating forecasting services, Kubernetes deployment, monitoring, CRD-based policy management, Operator reconciliation, scaling execution, and continuous feedback for autonomous cloud-native resource management.}
    \label{fig:predictiveworkflow}
\end{figure*}

Figure~\ref{fig5} illustrates how monitoring data is transformed into predictive insights and translated into Kubernetes scaling actions using Informer forecasting, MV-Transformer analysis, and CRD–Operator reconciliation. A summary of predictive autoscaling models and Kubernetes-native automation techniques is presented in Table~\ref{tab:crdandoperator}. and Figure~\ref{fig:predictiveworkflow} presents the complete predictive autoscaling workflow for Kubernetes-based cloud-native environments. The process begins with workload monitoring and forecasting through the model inference service, followed by containerization and Kubernetes deployment. Configuration components such as ConfigMap and ServiceMonitor enable runtime parameter management and Prometheus-based observability. Predictive scaling policies are defined using the PredictiveAutoscaler CRD, while the Operator reconciliation loop continuously monitors metrics, invokes forecasting models, computes replica decisions, and applies scaling actions. Finally, workload scaling outcomes are fed back into the monitoring system, enabling continuous learning and adaptive autoscaling.

Together, these approaches form a unified predictive autoscaling pipeline: workload signals are collected through monitoring, analysed using advanced forecasting models, translated into scaling plans, and executed via Operator-driven reconciliation and Kubernetes-native actuation. This integration of long-sequence forecasting, multivariate modelling, and CRD-based orchestration provides a robust foundation for next-generation autoscaling in cloud-native and edge environments. 

\subsection{Proactive Long-Sequence Forecasting}

Reactive methods face difficulties in dealing with unpredictable workloads, Informer-based autoscaling works by forecasting future workload demands at long time horizons and proactively allocating resources, overcoming these challenges~\cite{kumar2025optimizing,ding2024dynamic}. To enable scalable forecasting in cloud environments, Informer leverages efficient attention mechanisms that capture long-range temporal dependencies, while limiting the computational complexity. It is embedded into the MAPE loop, which makes scaling decisions based on the metrics measured from the work loaded on the system, leading to better latency, avoiding under-provisioning and steady resource allocation.

Multivariate models are used to accurately predict correlated resource metrics, such as CPU, memory, disk, and network usage, by modeling them together with the MV-Transformer, as detailed in~\cite{arbat2022wasserstein,wu2025elastic}. Moreover, neural sequence models like LSTM, GRU, Bi-LSTM, and CNN-LSTM can capture non-linear temporal patterns and workload variability, and make scalable decisions in dynamic and bursty workloads~\cite{taha2024proactive}. These models are capable of learning temporal dependencies and cross-metric relationships to enhance the accuracy of provisioning, minimize SLA breaches, and promote system stability~\cite{shahzaad2025service,meng2025catscaler}.

Various metrics such as provisioning accuracy, temporal stability, prediction error (e.g., mean square error, root mean square error, mean absolute error), and elastic speedup are used to assess efficiency and responsiveness, as detailed by Kumar et al.~\cite{kumar2024optimizing} and Qu et al.~\cite{qu2018auto}. Other scale-in strategies, like the Resource Removal Strategy, enhance stability even more by avoiding oscillations when presented with dynamic workloads~\cite{kumar2025multivariate}. The following section uses Docker, Kubernetes, and cloud-native stack integration to explore the use of these predictive autoscaling techniques in real world scenarios.

 \subsection{Deployments: Docker, Kubernetes, \& Cloud-Native Stacks} \label{sec4.5}
The Deployment segment involves using Docker, Kubernetes, and Cloud-native stacks, Docker, Kubernetes, and the Cloud-native Stacks are the tools included in the Deployments segment. If you want to use predictive autoscaling in a cloud-native environment, you need to be able to integrate workload monitoring, workload forecasting models, decision making, and Kubernetes orchestration into a single operational context. While there may be some differences between platforms, most predictive autoscaling architectures follow a similar approach of taking some observables from the workload and making decisions about scaling resources automatically \cite{miracapillo2025enabling,wijesekera2025kubernetes}.

The deployment involves the monitoring stage first. Active containers and Kubernetes pods are constantly monitored for resource utilization metrics, including CPU usage, memory consumption, network throughput, disk activity, requests received, and app-specific performance metrics. These measurements are collected via monitoring tools like Prometheus and Kubernetes Metrics Server and saved as time-series data. This monitoring infrastructure enables the past and current observations to be used for workload analysis and prediction \cite{hpaDocs2023,vpaDocs2023,clusterAutoscaler2023}.

After collecting workload data, the forecasting stage is initiated. A prediction service is used to send historical resource traces to which workload forecasting models are installed. Forecasting can be done through statistical models like ARIMA, deep learning models like LSTM and CNN-LSTM, and attention-based models like Informer and MV-Transformer \cite{yuan2024time,kumar2024optimizing}. The forecasting service uses the last observed behavior of the workloads and predicts the resources that will be needed in the future within a given forecasting horizon. The autoscaling system, therefore, does not respond to the actual utilization of resources, but instead predicts future workload requirements \cite{kumar2024optimizing,yuan2024time}.

The expected workload demand is then forwarded to the adaptation layer where scaling decisions are created. This step transforms the outputs of forecasting to actionable resource management policies. We need a model to estimate capacity to meet the anticipated demands, which is a capacity estimation model. The number of replicas, pods, or nodes to be deployed equals the workload forecast, service level requirements, and constraints on resource allocation. The system can be more stable in case of workload fluctuations due to number of additional mechanisms, including the Resource Removal Strategy (RRS) \cite{kumar2024optimizing,miracapillo2025enabling} that mitigates oscillatory scaling behavior.

After the scaling decision is made, the execution part applies the necessary changes to the resources. These decisions are implemented in the cluster by Kubernetes-native components like Horizontal Pod Autoscalers (HPA), Vertical Pod Autoscalers (VPA), Cluster Autoscalers (CA), Custom Resource Definitions (CRDs), and Operators \cite{kubernetesExtendKubernetes, clusterAutoscaler2023, vpaDocs2023}. The Kubernetes control plane will continuously compare the desired system state with the current cluster state, and execute the scaling operations. They can be created, deleted, scaled up, or scaled down based on the scaling target selected \cite{miracapillo2025enabling,wijesekera2025kubernetes}.

Forecasting services are generally packaged as Docker containers for easy deployment. The characteristics of portability, isolation, and reproducibility of containerization, and software dependency management are easier than ever. Prediction models can be packaged into Docker images and deployed consistently to cloud, edge and hybrid infrastructures. Kubernetes then orchestrates these containers, managing scheduling, service discovery, load balancing, fault recovery and lifecycle management. This container-driven architecture allows forecasting services to scale without the application workloads affecting the container's size or count \cite{kubernetesExtendKubernetes, hpaDocs2023}.

Observability is not just about workload forecasting and resource scaling; it is also a key aspect of ensuring reliable operation. All of the telemetry data related to forecasting latency, forecasting accuracy, frequency of scaling, resource utilization, and application performance are gathered constantly by monitoring systems. These observations can be used for administrators and controllers to assess the effectiveness of autoscaling, identify anomalies, detect model drift, and ensure optimal scaling in the future \cite{hpaDocs2023,clusterAutoscaler2023,kumar2024optimizing}.

Today, predictive autoscaling systems also utilize declarative configuration management. Forecasting periods, model selection policies, scaling thresholds, pod capacities, reconciliation periods and scaling constraints are usually configured in Kubernetes-native resources. This enables operators to dynamically change the autoscaling behavior without container image rebuilds or changes to the application's source code \cite{kubernetesExtendKubernetes,miracapillo2025enabling}.

Predictive autoscaling is a closed loop process in terms of implementation. Workload metrics are gathered using monitoring services, first. Second, the forecast models rely on past observations. Thirdly, scaling policies translate these forecasts into resource allocation decisions. Fourth, the scaling actions are executed by Kubernetes controllers and operators. Finally, newly created workload metrics are returned to the monitoring system, establishing a feedback loop that facilitates adaptive and autonomous resource management \cite{kumar2024optimizing,yuan2024time,wijesekera2025kubernetes}.

This deployment workflow serves as the operational infrastructure for adding advanced forecasting models like Informer, MV-Transformer, LSTM and Federated learning aware predictors to Kubernetes environments. Predictive autoscaling brings together all three concepts in a single architecture, turning resource management from a reactive process into a proactive and intelligent control model that can be used to manage resources for dynamic cloud-native and cloud-edge applications. In \textbf{Appendix~A}, there are examples of how to apply it to the detailed implementation, the Docker configuration, the Kubernetes manifests, the CRD specification, the monitoring configuration, and the workflows of an operator.

\subsection{Custom Resource Definitions for Autoscaling}
\label{sec4.6}
Custom Resource Definitions (CRDs) provide one of the most important mechanisms for extending Kubernetes beyond its built-in resource management capabilities \cite{kubernetesExtendKubernetes,miracapillo2025enabling}. In predictive autoscaling systems, CRDs allow forecasting models, scaling policies, workload requirements, and control parameters to be represented as native Kubernetes resources \cite{kumar2024optimizing,wijesekera2025kubernetes}. This enables predictive intelligence to be integrated directly into the Kubernetes control plane rather than operating as an external component. As a result, predictive autoscaling becomes fully declarative, observable, and manageable through standard Kubernetes workflows \cite{yuan2024time,miracapillo2025enabling}.

A custom autoscaling resource is normally created in the implementation process. The predictive autoscaling CRD presents new fields to specify forecasting-based resource management, which are not found in traditional autoscaling CRDs like the Horizontal Pod Autoscaler (HPA) or Vertical Pod Autoscaler (VPA) \cite{hpaDocs2023,vpaDocs2023}. Some of these fields can be the type of prediction model, forecasting horizon, pod capacity, workload characteristics, scaling constraints, replica boundaries, confidence thresholds and stability-control parameters. After being registered to the Kubernetes API server, the custom resource becomes a first-class Kubernetes object, and can be managed using standard Kubernetes commands and APIs \cite{kubernetesExtendKubernetes,wijesekera2025kubernetes}.

Once the CRD is deployed, users can define instances of the custom resource to define the behavior they want to enable for specific workloads. This declarative specification is a configuration interface for forecasting services with Kubernetes resource controllers. The desired state determines the desired behavior of predictive autoscaling, and information on runtime, such as forecasts, scaling recommendations, confidence intervals, and usage statistics of the resource, can be stored in the resource status field. This separation between desired state and observed state is based on the principles of Kubernetes design, and it also offers a structured way of managing the policies for predictive autoscaling \cite{miracapillo2025enabling,kubernetesExtendKubernetes}.

The next step is the interaction between the CRD and the forecasting system. Continuous monitoring components gather workload metrics about running infrastructure resources and applications. The measurements are then sent to a predictive model that can be either Informer, LSTM, Transformer, CNN-LSTM or MV-Transformer \cite{yuan2024time,kumar2024optimizing}. The forecasting service is able to analyse workload pattern histories and produce future resource demand estimates. The forecasting results are not directly applied to the cluster resources; they are instead stored in the CRD framework and Kubernetes scaling decisions are part of the Kubernetes control plane state \cite{kumar2024optimizing,wijesekera2025kubernetes}.

After the forecasting information is available, the Operator for Kubernetes takes care of decisions. Operators constantly check for CRD instances via Kubernetes watch objects. The Operator pulls the most up-to-date CRD configuration, and determines what scaling actions are necessary based on workloads, resource conditions, and scaling policies, as they change.The Operator pulls the latest CRD configuration, and checks for necessary scaling action based on workload, resource conditions, and scaling policies as they change, \cite{miracapillo2025enabling,kubernetesExtendKubernetes}. The Operator uses predicted demand to decide how many replicas are needed, then to calculate resource allocation needs and to apply stability-control policies like Resource Removal Strategy (RRS) and to create an appropriate scaling plan \cite{kumar2024optimizing}.

Predictive autoscaling is based on the reconciliation process. The Operator reconciles the cluster with the desired state defined in the CRD, during each cycle of reconciliation. Any discrepancies are automatically corrected if they are detected. The actions can include scaling pods horizontally, adjusting resource requests vertically, provisioning more worker nodes, or scaling several autoscaling components at the same time. The cluster slowly moves towards the desired predictive scaling state through continuous reconciliation \cite{miracapillo2025enabling,wijesekera2025kubernetes}.

Extensibility is one of the main benefits of CRD based architectures. Extension of the CRD schema allows embedding new prediction models, optimization algorithms, uncertainty-aware controllers, drift-detection mechanisms, federated learning policies, and application-specific constraints without changing any of the core components of Kubernetes \cite{kubernetesExtendKubernetes,kumar2024optimizing}. This agility allows researchers and practitioners to try out new, high-level autoscaling methods without disrupting the old Kubernetes infrastructure.

CRDs further enhance transparency and operational governance. The outputs that can be used to forecast the scale up or scale down of a deployment, the scaling recommendations to be considered, and what decisions have been made about scaling, can all be viewed by the administrator using the normal Kubernetes tools, since they are part of the Kubernetes resources. Historical decisions can be audited, monitored, and analysed to assess forecasting performance, map out operational bottlenecks, and enhance future scaling strategies \cite{miracapillo2025enabling,wijesekera2025kubernetes}. This transparency can be especially useful in production settings where explainability and accountability of autonomous decisions are gaining in significance.

In a large-scale cloud-native scenario, CRDs also enable multi-cluster and cloud-edge deployments. Multiple Operators can work together and take advantage of predictive autoscaling across geographically distributed clusters without changing the declarative control model. Such functionality is particularly crucial in federated learning systems, edge computing platforms, and distributed AI applications, where workloads are often highly dynamic and may span multiple resource domains or locations \cite{kumar2024optimizing,yuan2024time}.

Overall, CRDs enable the conversion of predictive autoscaling from a set of standalone forecasting services into a Kubernetes-native control system. Declarative specifications, continuous reconciliation, Operator-based automation and predictive workload intelligence build a scalable and extensible base for autonomous management of resources in cloud-native, cloud-edge and federated computing environments. \textbf{Appendix~B} contains detailed CRD schemas, resource definitions, reconciliation workflows and implementation examples.

\begin{table*}[!ht]
\centering
\caption{Comparative analysis of predictive autoscaling models, Kubernetes-native automation, and deployment mechanisms.}
\label{tab:crdandoperator}
\renewcommand{\arraystretch}{1.25}
\begin{tabularx}{\textwidth}{p{3cm} X X X X}
\hline
\textbf{Category} &
\textbf{Definition / Purpose} &
\textbf{Strengths} &
\textbf{Limitations} &
\textbf{Typical Use Cases} \\
\hline

\textbf{Informer-Based Forecasting}
~\cite{ding2024dynamic,kumar2025optimizing}
&
Attention-based long-sequence forecasting model for proactive workload prediction and advance resource provisioning.
&
Captures long-range temporal dependencies; efficient attention mechanism; supports long-horizon forecasting; reduces scaling delay.
&
Requires large training datasets; complex hyperparameter tuning; prediction quality degrades under sudden workload shifts.
&
Cloud workload forecasting, proactive Kubernetes autoscaling, long-horizon resource planning.
\\

\hline

\textbf{MV-Transformer}
~\cite{kumar2025multivariate,arbat2022wasserstein,wu2025elastic}
&
Multivariate Transformer model that jointly analyses CPU, memory, disk, network, and request-rate metrics.
&
Captures cross-metric dependencies; improves forecasting accuracy; supports complex distributed workloads.
&
High computational cost; increased model complexity; requires multiple correlated metrics.
&
Cloud-native applications, microservices, multi-resource autoscaling, edge-cloud environments.
\\

\hline

\textbf{Deep Learning Forecasting Models}
~\cite{taha2024proactive,shahzaad2025service,meng2025catscaler}
&
Neural sequence models such as LSTM, GRU, Bi-LSTM, and CNN-LSTM for workload prediction.
&
Learns non-linear patterns; handles bursty workloads; improves provisioning accuracy.
&
Training overhead; sensitive to data quality; may require retraining under workload drift.
&
Short- and medium-term workload prediction, SLA-aware autoscaling, resource forecasting.
\\

\hline

\textbf{PredictiveAutoscaler CRD}
~\cite{kumar2024optimizing,yuan2024time,wijesekera2025kubernetes}
&
Custom Resource Definition that stores forecasting models, scaling policies, capacities, and replica constraints as Kubernetes-native resources.
&
Declarative configuration; versionable and auditable; integrates predictive intelligence into Kubernetes APIs.
&
Requires CRD schema management; introduces additional operational complexity.
&
Predictive autoscaling platforms, cloud-native automation, Kubernetes-based orchestration.
\\

\hline

\textbf{Operator Reconciliation}
~\cite{kubernetesOperatorPattern,dame2022kubernetes,yu2025arsc}
&
Continuous observe-predict-plan-act control loop that enforces scaling decisions based on CRD specifications.
&
Automated execution; low-latency control; supports asynchronous processing and model selection.
&
Additional controller overhead; reconciliation delays may affect responsiveness under extreme workloads.
&
Autonomous autoscaling, cloud-edge orchestration, multi-cluster resource management.
\\

\hline

\textbf{Model Control System (MCS)}
~\cite{kumar2024optimizing,kumar2025multivariate}
&
Framework that manages multiple forecasting models and dynamically selects the best-performing model for inference.
&
Adaptive model selection; improved forecasting robustness; supports heterogeneous workloads.
&
Requires model monitoring and validation; increased resource consumption.
&
Hybrid forecasting environments, adaptive autoscaling systems, intelligent resource management.
\\

\hline

\textbf{Monitoring and Observability}
~\cite{hpaDocs2023,vpaDocs2023}
&
Prometheus-based collection of workload, latency, and system metrics used by predictive autoscaling frameworks.
&
Real-time visibility; supports forecasting, debugging, and performance analysis.
&
Storage overhead; monitoring infrastructure complexity.
&
Cloud-native monitoring, autoscaling analytics, performance management.
\\ \hline

\textbf{Containerised Model Services}
~\cite{hpaDocs2023,kumar2026critical}
&
Dockerised forecasting services exposing REST/gRPC APIs for predictive inference.
&
Portable; scalable; easy integration with Kubernetes deployments.
&
Network overhead for remote inference; requires image maintenance and registry management.
&
Model serving, edge deployment, predictive autoscaling pipelines.
\\

\hline

\textbf{Configuration and Policy Management}~\cite{kumar2024optimizing}
&
ConfigMaps and policy resources used to manage forecasting parameters, pod capacities, and scaling strategies.
&
Centralised configuration; dynamic updates; reusable across components.
&
Configuration inconsistency can affect scaling decisions; requires governance.
&
Production Kubernetes environments, predictive scaling platforms.
\\

\hline

\textbf{Multi-Cluster Predictive Scaling}
~\cite{yu2025arsc,miracapillo2025enabling}
&
Extension of predictive autoscaling across federated and geographically distributed Kubernetes clusters.
&
Global resource optimisation; coordinated scaling; improved fault tolerance.
&
Cross-cluster communication overhead; increased operational complexity.
&
Edge-cloud systems, geo-distributed applications, federated infrastructure.
\\

\hline
\end{tabularx}
\end{table*}

\subsection{Kubernetes Operator Reconciliation for Predictive Scaling} \label{sec4.7}
The operator reconciliation loop acts as the execution backbone of predictive autoscaling, connecting forecasting models with Kubernetes-based scaling actions \cite{kubernetesOperatorPattern,kubernetesOverview,kubernetesAutoscalingWorkloads}. It follows a continuous observe-analyze-act cycle to ensure that the cluster state aligns with the desired state defined in Custom Resource (CR) objects~\cite{kumar2024optimizing,dame2022kubernetes,yu2025arsc}.

\begin{itemize}
    \item Watch: Continuously monitors changes in \texttt{PredictiveAutoscaler} CR instances and workload metrics.
    
    \item Read: Retrieves CR specifications, current pod status, historical workload metrics, and scaling policies.
    
    \item Predict: Invokes the Model Control System to estimate future workload demand using the most suitable predictive model.
    
    \item Plan: Computes the desired number of replicas based on predicted demand, applying strategies such as Resource Removal Strategy (RRS) to ensure stable scale-in decisions.
    
    \item Act: Executes scaling actions via the Kubernetes Scale API or by updating workload resources (e.g., Deployment, StatefulSet).
    
    \item Update: Stores predictions, selected models, and scaling decisions in the CRD status for monitoring and traceability.
    
    \item Parallel Execution: Utilizes asynchronous or coroutine-based processing to enable concurrent metric collection, model inference, and scaling operations.
    
    \item Model Selection: Maintains multiple candidate models and dynamically selects the best-performing one based on validation accuracy, allowing adaptation to varying workload patterns.
    
    \item Multi-cluster Support: Extends autoscaling decisions across federated or geo-distributed clusters for coordinated resource management in edge-cloud environments.
    
    \item Outcome: Enables an autonomous, adaptive, and scalable autoscaling system by integrating Kubernetes-native control with predictive intelligence.
\end{itemize}
To maintain responsiveness at scale, operators use asynchronous routines or coroutine-based pipelines that allow metric collection, model inference, and scaling logic to execute concurrently without blocking. The Model Control System usually keeps several candidate models and may keep a min heap ordered by the validation error of each model (this way, the best one is always used for the inference). This enables the autoscaler to adaptively use different models; e.g., for long-term forecasting, Transformers can be used, while for short-horizon or patterns with a high volatility, LSTM variants. In addition to single cluster deployments, federation-aware operators are able to propagate predictive-scaling decisions to multiple clusters, allowing for coordinated scaling for geo-distributed applications. This brings the autoscaling capabilities of predictive scaling to multi-region, edge-cloud, and hybrid deployments where resource governance is crucial.

In summary, the reconciliation loop brings together Kubernetes-native automation, advanced forecasting models, making it an autonomous, adaptive, and extensible scaling ecosystem.

Taking a closer look at the Informer-based long-sequence forecaster, MV-Transformer multivariate predictor, and CRD Operator reconciliation pattern, the three examples show how predictive modelling and Kubernetes-native automation can help achieve proactive autoscaling. Each component directly contributes to answering \textbf{RQ2} and demonstrates the interplay of forecasting, multivariate learning and integration of the control plane to create an intelligent predictive autoscaling pipeline. Under-provisioning and reactive delays are prevented with long-horizon forecasting by provisioning resources in advance. By supporting cross-metric dependencies for CPU, memory, disk and network workloads, multivariate prediction results in more accurate predictions. In the meantime, Operator integrates prediction and scaling policies, executed with low latency, into Kubernetes with the help of CRD. These mechanisms contribute to stability and minimize resource waste and service quality, giving us a clear and satisfying answer to \textbf{RQ2}.

\section{Autoscaling in FL and Cloud-Edge Systems} \label{sec5}
\begin{figure*}
    \centering
    \includegraphics[width=0.99\linewidth]{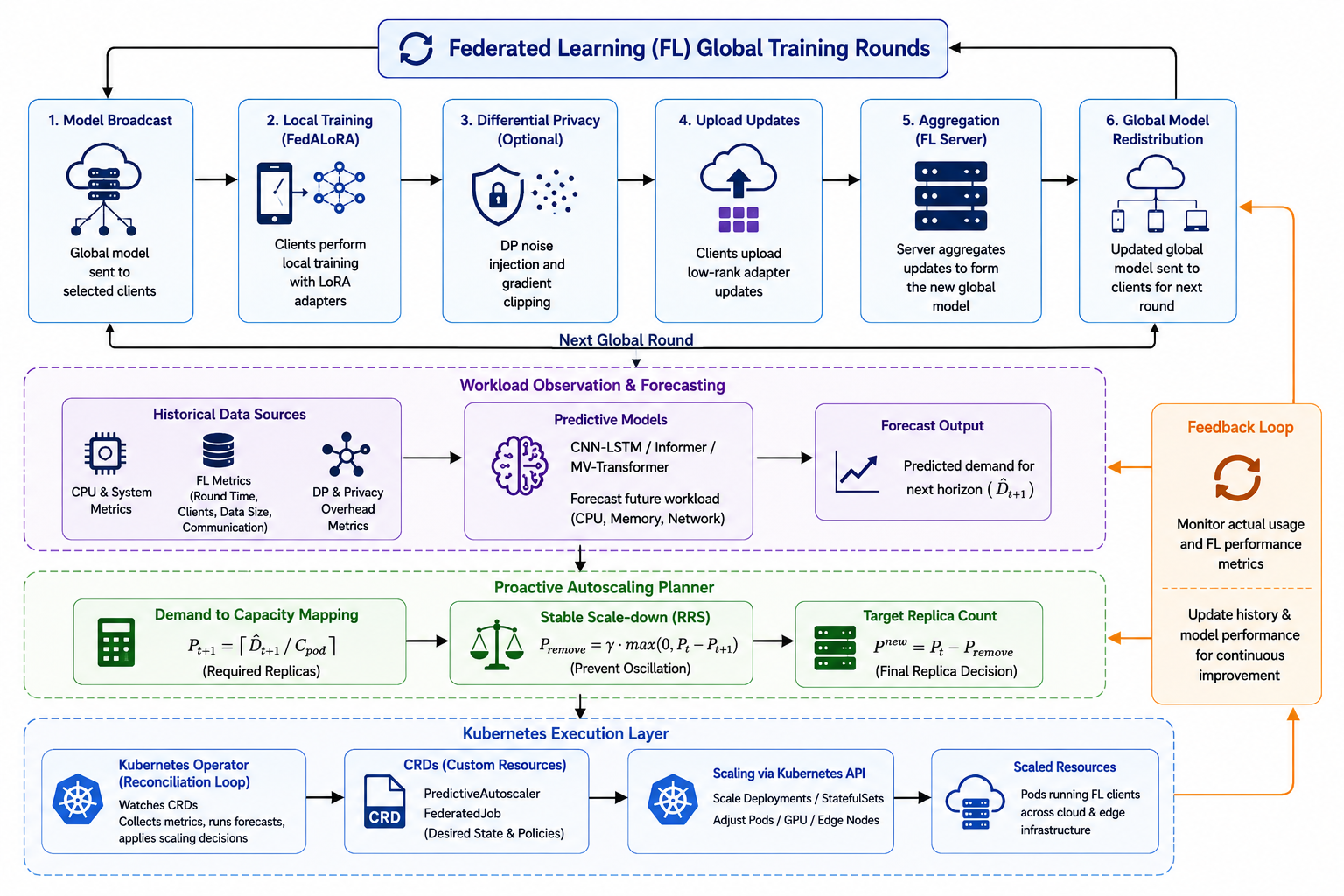}
    \caption{Interactive workflow of predictive autoscaling for federated learning (FL) in cloud–edge environments using forecasting models, Kubernetes Operators, and proactive resource management.}
    \label{flinteractautoscaling}
\end{figure*}

Unlike traditional cloud-native applications \cite{dogani2024proactive,pham2024elastic}, FL has a fundamentally different workload pattern. Unlike the conventional method of request driven operations, FL performs training rounds in a discrete and highly variable manner, with the participation of clients, data diversity, and communication overheads \cite{albaseer2023data,liu2025fedinv}. With resource-limited edge environments and privacy-preserving features like DP, FL workloads can be more dynamic and irregular, which poses significant challenges for autoscaling. This section combines the predictive autoscaling techniques from recent advances in the field of proactive forecasting \cite{kumar2025optimizing} and multivariate workload modelling \cite{kumar2025multivariate} and Kubernetes-native predictive control for distributed learning systems \cite{kumar2024optimizing}. The Figure \ref{flinteractautoscaling} shows an end-to-end example of interaction between federated learning workflows and predictive autoscaling mechanisms in cloud–edge systems. It starts with FL global training rounds with model broadcast, local client training, differential privacy processing, uploading of updates to the global server, aggregation, and global model redistribution. Forecasting models like CNN–LSTM, Informer, and MV-Transformer are used to constantly monitor and analyse the generated workload patterns to predict the future CPU, memory and network requirements. Using capacity mapping and scale-down control strategies, the proactive autoscaling planner uses these predictions to determine stable replica allocation. Kubernetes Operators, CRDs then make decisions on scaling of pods, GPUs and edge nodes, and feedback loops continuously improve forecasting accuracy and scaling stability, ensuring efficient, privacy-aware and resource-efficient FL orchestration. 

\begin{figure}[ht]
    \centering
    \includegraphics[width=0.95\linewidth]{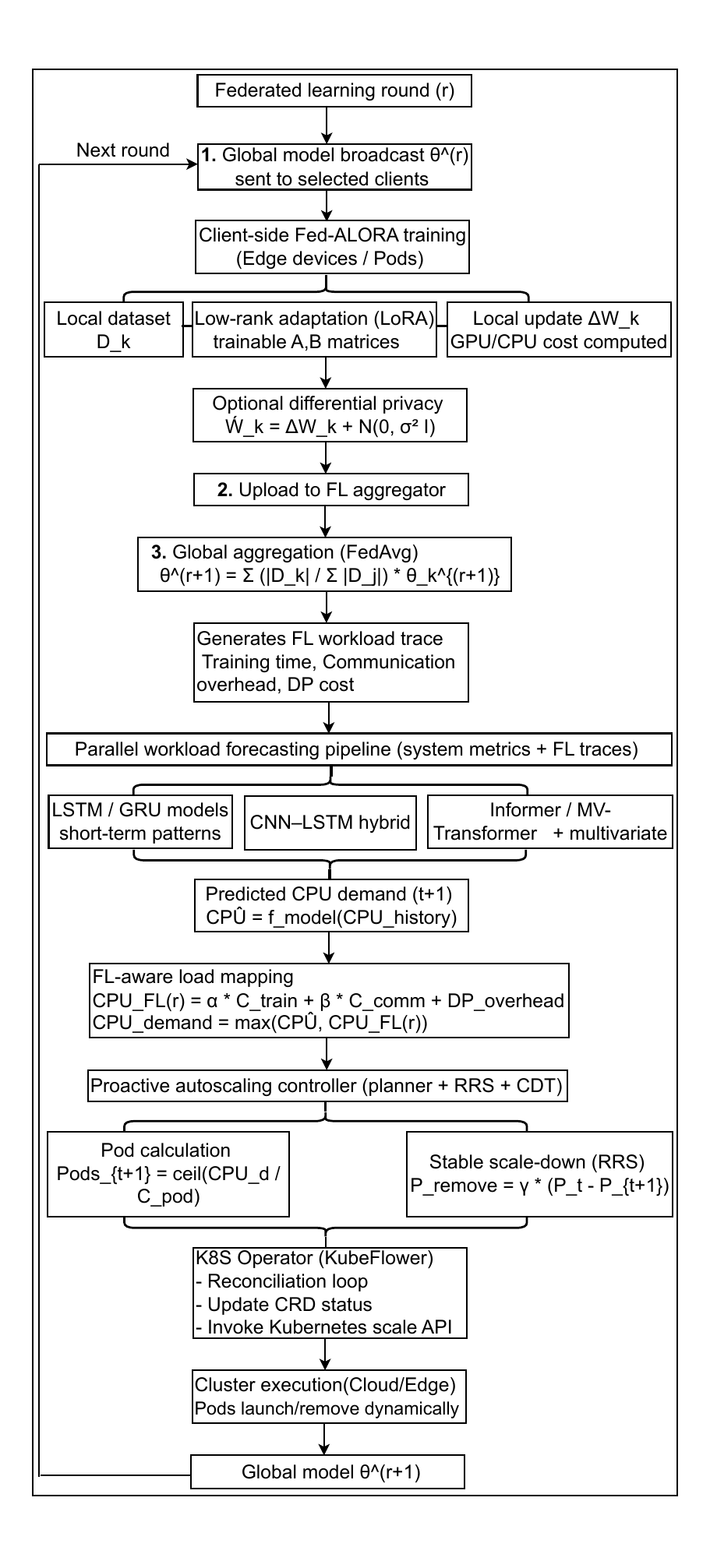}
    \caption{End-to-end FL autoscaling pipeline: forecasting-driven scaling via Kubernetes Operators and CRDs.}
    \label{fig6}
\end{figure}

The architecture shown in Figure~\ref{fig6} depicts a single approach to combining FL dynamics with proactive autoscaling in cloud–edge systems. It illustrates the iterative nature of FL, starting with broadcasting the global model to selected clients, then local FedALoRA training \cite{yi2025fedalora,luo2025dynamicfedpeft} and (optionally) DP processing, and finally uploading the low-rank adapter updates to aggregate on the server. In parallel to this, workload forecasting models analyze the CPU traces and FL induced demands from the history to predict the needs of resource in the future. The predictions are used by a proactive autoscaling planner to use capacity formulas and stable scale-down logic, which is then used to determine the optimum replica count.

The Kubernetes scheduler then implements these decisions, expanding the number of pods in the cluster and the global model is sent back to clients to trigger the next iteration. This end-to-end pipeline is designed to highlight the strong relationship between learning dynamics and resource management, enabling efficient and privacy-preserving FL operations in diverse edge settings. The architecture shown in Figure~\ref{fig6} is synthesized from recent advances in proactive autoscaling system \cite{kumar2024optimizing, kumar2025multivariate, kumar2025optimizing} and federated training frameworks \cite{ye2024openfedllm, yi2025fedalora}. Table~\ref{tab:fl} provides a summary of the key concepts on FL workloads and predictive autoscaling across cloud–edge systems. 

\subsection{Overview of Federated Learning}

Federated Learning is a distributed training paradigm where multiple clients collaboratively train a global model without sharing raw data, exchanging only model updates to preserve privacy~\cite{ye2024openfedllm,yi2025fedalora,dogani2024proactive}. Training proceeds in iterative global rounds consisting of model broadcast, local training, update upload, aggregation, and redistribution. Each round introduces cyclic yet bursty workloads combining computation, communication, and aggregation costs, resulting in highly variable resource demand. These characteristics make FL particularly challenging for traditional autoscaling approaches and motivate predictive, workload-aware scaling strategies~\cite{pham2024elastic,cheng2023proscale}.

\subsection{FL Workload Issues: Heterogeneity, Uncertainty, Privacy}

FL workloads are inherently non-stationary due to device heterogeneity, dynamic client participation, and privacy-preserving mechanisms such as differential privacy (DP)~\cite{yi2025fedalora,ali2025recent,radhika2021review,albaseer2023data}. Variations in compute capability, dataset size, and network conditions lead to unpredictable training times and communication overhead, while DP introduces additional computational cost and variability. These factors produce bursty CPU, memory, and network usage patterns that reactive autoscalers cannot handle effectively, often resulting in delayed scaling, resource inefficiency, and unstable system behaviour.

\subsection{Proactive Autoscaling for FL Workloads}
To handle FL workload variation, the proactive autoscaling is proposed, which relies on predicting the future resource demand through a model, such as CNN-LSTM~\cite{yi2025fedalora} and Informer~\cite{luo2025dynamicfedpeft,pham2024elastic} or MV-Transformer. Workload metrics are projected in advance and resources are allocated according to the number of required replicas, allowing resources to be provisioned prior to the start of FL. Combining system-level predictions with FL-specific demand, proactive autoscaling mitigates resource under-provisioning, enhances resource usage and levels training performance. Techniques like Resource Removal Strategy (RRS) also help in improving stability by avoiding oscillations during scale-in operations~\cite{patni2025predictive,mukherjee2025amas}.

\subsection{Differential Privacy and its Impact on Scaling}

Nevertheless, Differential Privacy (DP) adds noise to the model updates for FL, thus resulting in higher computational overheads and resource requirements~\cite{liu2025fedinv, yi2025fedalora, ye2024openfedllm}. Adding DP operations like clipping gradients, adding noise, adds to training time and CPU usage, resulting in a higher and more varied load pattern. Therefore, the overhead caused by DP has to be taken into account when designing an autoscaling system to prevent scaler from under-provisioning during active FL rounds and to guarantee stable convergence~\cite{baek2021enhancing, xu2023federated, wu2022adaptive}.

\subsection{KubeFlower Operator and Container Isolation Mechanisms} \label{sec5.5}
KubeFlower expands the Kubernetes Operator-CRD paradigm to orchestrate FL, fine-tuning and predictive autoscaling in a single control plane. Drawing on previous CRD-based predictive autoscaling frameworks \cite{kumar2024optimizing}, KubeFlower introduces domain-specific CRDs, such as the \texttt{PredictiveAutoscaler} and \texttt{FederatedJob} CRD, which represent both training behavior and resource policies~\cite{parra2024kubeflower,Ahmadpanah_Mirabi_Sahafi_Erfani_2025,almosti2025analysis,wu2025containerized}. The operator continuously compares the desired system state to cluster state and allows for fully autonomous FL orchestration.

\subsubsection{Operator Architecture and Reconciliation Logic}
An Operator in Kubernetes acts as an event-driven controller that observes Custom Resource (CR) instances and executes a reconciliation loop~\cite{kubernetesOperatorPattern,dame2022kubernetes}. For KubeFlower, each FL round triggers a new reconciliation cycle:

\begin{itemize}
  \item \textbf{Metrics Collection:} System metrics (CPU, memory, throughput) and FL metrics (training cost $C^{\text{train}}_k$, communication cost $C^{\text{comm}}_k$, DP overhead) are collected from participating pods. For a client $k$, observed demand at time $t$ may be denoted as:
  \begin{equation}
    \omega_{k,t} = \{CPU_{k,t}, Mem_{k,t}, IO_{k,t}, \Delta W_k\}.
  \end{equation}

  \item \textbf{Prediction via Model Control System:}  
  The operator invokes a Model Control System (MCS) which maintains multiple forecasting models (CNN-LSTM, Informer, MV-Transformer). Each model $f_m$ is ranked by validation error using a min-heap:
  \begin{equation}
    \text{heap} = \min\{ \text{MSE}(f_m) \},
  \end{equation}
  ensuring that the current best model is used to compute
  \begin{equation}
    \hat{D}_{t+H} = f_{\text{best}}(X_{t-w:t}),
  \end{equation}
  where $X_{t-w:t}$ is the sliding window of multivariate workload traces.

  \item \textbf{Planning and Stable Scaling:}  
  The operator converts the predicted demand into desired replicas using the proactive capacity model:
  \begin{equation}
    P_{t+1} = 
    \left\lceil 
      \frac{\hat{D}_{t+H}}{C_{\text{pod}}}
    \right\rceil.
  \end{equation}
  To prevent oscillation and ensure safe scale-in, the Resource Removal Strategy (RRS) is applied:
  \begin{equation}
    P_{\text{remove}} 
    = \gamma \cdot \max(0, P_t - P_{t+1}),
    \label{eq:rrs_kubeflower}
  \end{equation}
  where $0 < \gamma \le 1$ modulates the scale-in rate. The new replica count becomes
  \begin{equation}
    P^{\text{new}} = P_t - P_{\text{remove}}.
  \end{equation}

  \item \textbf{Execution via Kubernetes APIs:}  
  Scaling decisions are applied using the Kubernetes Scale API or via patches to the \texttt{Deployment} or \texttt{StatefulSet} object:
  \begin{equation}
    \texttt{kubectl\_scale}(P^{\text{new}}).
  \end{equation}

  \item \textbf{CRD Status Update:} The CRD status update records the forecasted demand, the selected prediction model, latency statistics, and the final replica count applied by the autoscaler. By maintaining this information within the CRD status field, the system establishes a consistent and authoritative source of truth for operational behaviour. This structured record supports governance and auditability by allowing operators to trace how scaling decisions were made, and it aids in diagnosing issues such as model drift or unexpected prediction deviations. Additionally, the shared status facilitates effective coordination across multiple controllers, ensuring that different components of the Kubernetes control plane operate with aligned and up-to-date information.

\end{itemize} 

\begin{table*}[!ht]
\centering
\caption{Comparative analysis of federated learning workloads, predictive autoscaling mechanisms, and Kubernetes-native orchestration across cloud-edge environments.}
\label{tab:fl}
\renewcommand{\arraystretch}{1.25}
\begin{tabularx}{\textwidth}{p{3cm} X X X X}
\hline
\textbf{Category} & \textbf{Definition / Purpose} & \textbf{Strengths} & \textbf{Limitations} & \textbf{Typical Use Cases} \\
\hline

\textbf{Federated Learning Overview} ~\cite{ye2024openfedllm,yi2025fedalora,dogani2024proactive,zhao2022edge,pham2024elastic,albaseer2023data,liu2025fedinv}
&
Distributed training paradigm where clients collaboratively train a global model through model broadcast, local training, update upload, aggregation, and redistribution without sharing raw data.
&
Preserves privacy; reduces data movement; supports distributed intelligence across edge devices.
&
Communication intensive; synchronization overhead; sensitive to client participation variability.
&
Cross-device FL, healthcare analytics, smart cities, edge AI, privacy-preserving learning.
\\ \hline

\textbf{FL Workload Challenges}~\cite{yi2025fedalora,ali2025recent,radhika2021review,albaseer2023data}
&
FL workloads exhibit heterogeneity, non-IID data distributions, fluctuating communication delays, and dynamic client participation.
&
Captures realistic distributed learning environments; improves model generalization across diverse data sources.
&
Highly non-stationary resource demand; unpredictable CPU, memory, network, and GPU utilization.
&
Large-scale federated systems, mobile edge networks, IoT ecosystems, distributed LLM fine-tuning.
\\
\hline

\textbf{Proactive Autoscaling for FL}~\cite{yi2025fedalora,luo2025dynamicfedpeft,pham2024elastic,patni2025predictive,mukherjee2025amas,samfass2021predictive,zheng2024lion,hu2023federated,zhang2022federated,tun2021federated}
&
Forecasts future workload demand using CNN-LSTM, Informer, MV-Transformer, and related predictive models to provision resources before workload surges occur.
&
Reduces under-provisioning and over-provisioning; improves training throughput; supports stable convergence.
&
Prediction errors may cause resource inefficiency; model retraining may be required under workload drift.
&
Federated training clusters, cloud-edge orchestration, proactive Kubernetes autoscaling.
\\
\hline

\textbf{Differential Privacy (DP) Impact}~\cite{liu2025fedinv,yi2025fedalora,ye2024openfedllm,baek2021enhancing,xu2023federated,wu2022adaptive,el2022differential,liberti2024federated}
&
DP introduces gradient clipping and noise injection to protect client privacy during model updates.
&
Provides strong privacy guarantees; improves regulatory compliance and data protection.
&
Additional computation and communication overhead; increased training latency and resource consumption.
&
Privacy-preserving healthcare, finance, government, and sensitive FL deployments.
\\
\hline

\textbf{KubeFlower Operator and Kubernetes-Native Control}~\cite{kumar2024optimizing,parra2024kubeflower,Ahmadpanah_Mirabi_Sahafi_Erfani_2025,almosti2025analysis,wu2025containerized,kubernetesOperatorPattern,dame2022kubernetes,nakata2021concentrated,mavridis2023orchestrated,morcillo2025privacy,sultana2023high}
&
Extends the Operator-CRD paradigm to integrate FL orchestration, workload prediction, and autoscaling into a unified Kubernetes control plane.
&
Automated reconciliation; policy-driven scaling; improved observability and governance; low-latency control.
&
Increased operational complexity; CRD and Operator maintenance overhead.
&
Federated Kubernetes platforms, predictive autoscaling systems, autonomous cloud-native operations.
\\
\hline

\textbf{Container Isolation and Sandboxing}~\cite{nakata2021concentrated,mavridis2023orchestrated,morcillo2025privacy,zhai2025energy}
&
Uses Docker, CRI-O, and containerd to isolate FL clients, datasets, DP mechanisms, and optimizer states.
&
Strong security and fault isolation; resource fairness; scalable parallel execution.
&
Additional runtime overhead; resource fragmentation may occur under heavy workloads.
&
Multi-tenant FL platforms, privacy-preserving edge computing, distributed AI services.
\\
\hline

\textbf{Multi-Region and Cross-Cluster Autoscaling}~\cite{sultana2023high}
&
Coordinates predictive autoscaling decisions across geographically distributed Kubernetes clusters using shared forecasting signals.
&
Improves global resource utilization; supports large-scale federated deployments; enhances load balancing.
&
Cross-region communication latency; increased orchestration complexity.
&
Global FL deployments, multi-cloud systems, geo-distributed edge infrastructures.
\\
\hline

\textbf{Evaluation Dimensions}~\cite{kumar2024optimizing,kumar2025multivariate,kumar2025optimizing}
&
Measures FL autoscaling effectiveness using latency, convergence time, resource utilization, prediction accuracy, cost, privacy overhead, and stability.
&
Provides comprehensive assessment of both learning and infrastructure performance.
&
Trade-offs among privacy, accuracy, cost, and scalability are difficult to optimize simultaneously.
&
Benchmarking FL autoscalers, Kubernetes-based orchestration frameworks, cloud-edge performance studies. \\
\hline
\end{tabularx}
\end{table*}

\subsubsection{Container Isolation and Sandboxing for FL}

KubeFlower relies on container runtime isolation mechanisms such as Docker, CRI-O, and containerd to execute FL clients inside independent sandboxes~\cite{nakata2021concentrated,mavridis2023orchestrated,morcillo2025privacy}. Each container encapsulates the client's private dataset $D_k$, its local LoRA parameters $(A_k, B_k)$, the mechanisms responsible for DP noise generation, and the local optimizer state. This encapsulation strengthens security by ensuring that no client can access another client's data or gradients, supports fault isolation so that slow or dropped clients do not affect the global workflow, and enforces strict CPU, memory, and GPU isolation to maintain fairness across participants. The approach also enables large-scale parallelism, allowing hundreds of FL clients to run simultaneously through horizontal scaling.

For container $k$, a resource quota $Q_k$ is enforced as:
\begin{equation}
Q_k = \{ CPU_k^{\max}, Mem_k^{\max}, GPU_k^{\max} \},
\end{equation}
preventing noisy neighbours from interfering with federated updates and ensuring consistent performance across heterogeneous client workloads~\cite{zhai2025energy}.

\subsubsection{Multi-Region and Cross-Cluster FL Autoscaling}

KubeFlower supports deployment across multiple federated Kubernetes clusters using frameworks such as KubeFed or multi-cluster service meshes, enabling FL workloads to be orchestrated across geographically distributed regions~\cite{sultana2023high}. In this setting, predictive autoscaling signals can be exchanged between clusters so that resource decisions are made with a global perspective rather than in isolation. The aggregated scaling signal, computed as
\begin{equation}
\Pi_{c}^{\text{global}} = \text{Aggregate}\big( \Pi_{c_1}, \Pi_{c_2}, \ldots, \Pi_{c_n} \big),
\end{equation}
combines the predicted replica demands $\Pi_{c_i}$ from individual clusters into a unified control value. This global prediction mechanism allows to provision it in a coordinated way between regions, to support wide-area FL workloads, and to proactively consider cross-cluster traffic and model aggregation loads. This ensures better load balancing, efficient utilization of distributed compute resources, and faster convergence in large-scale cross-region training scenarios.

KubeFlower combines predictive modelling, autoscaling logic, FL orchestration and container-level isolation, enabling Kubernetes to become a fully self-contained platform for privacy-preserving and resource-efficient federated large model training \cite{11075838}. Operator-CRD integration allows for fast reconciliation, model-aware scaling and isolation of different types of edge clients.

\textbf{RQ3 Answer:} FL workloads are dynamic and heterogeneous, unable to be effectively managed by reactive autoscaling. A set of different client resources, along with the variations of network, non-IID data and differential privacy (DP) overhead, produces non stationary workload patterns, causing delays in scaling up, under-provisioning during rounds of high intensity, and over-provisioning during idle periods. To overcome these challenges, there is a new paradigm called predictive autoscaling, which can predict the system-level and FL-specific demand. Temporal dynamics, cross-metric dependency and workload spikes are captured by models like Informer, MV-Transformer, CNN-LSTM, which allow for proactive resource provisioning. This decreases latency and stabilises FL convergence and resource efficiency. This is extended even more by Kubernetes-native mechanisms. CRDs integrate predictive policies into the control plane, while operator reconciliation provides for consistent and low-latency scale actions. The techniques include Resource Removal Strategy (RRS), which decreases the oscillation and enhance the stability. In conclusion, combining predictive models with the Kubernetes-native orchestration provides an efficient, stable, and private way of achieving federated workload autoscaling in cloud–edge environments, which answers \textbf{RQ3}.

\section{Drift-Aware and Uncertainty-Aware Autoscaling} \label{sec6}

Autoscaling in cloud–edge FL systems is challenging due to highly dynamic and heterogeneous workloads, influenced by varying client capabilities, communication delays, and differential privacy (DP) overhead~\cite{baek2021enhancing,xu2023federated,wu2022adaptive}. These factors produce non-stationary resource utilisation, making static or threshold-based scaling ineffective \cite{el2022differential,liberti2024federated}.

Although predictive models such as Informer, MV-Transformer, and CNN-LSTM (Section~\ref{sec4}) improve forecasting accuracy, prediction errors accumulate across FL rounds~\cite{meng2025catscaler,arbat2022wasserstein}. Underestimation leads to stragglers and delayed convergence, while overestimation causes resource waste and increased cost~\cite{zhai2025energy, wu2025elastic}. This persistent mismatch between predicted and actual demand is referred to as \emph{autoscaling drift}, which destabilises system performance over time.

To address this, drift-aware and uncertainty-aware frameworks extend predictive autoscaling with error-aware feedback mechanisms. These systems monitor prediction deviations, quantify drift, and apply real-time corrections to scaling decisions. Round stability controllers further mitigate stragglers by enforcing minimum resource guarantees and targeted adjustments. 

\begin{figure*}
    \centering
    \includegraphics[width=0.99\linewidth]{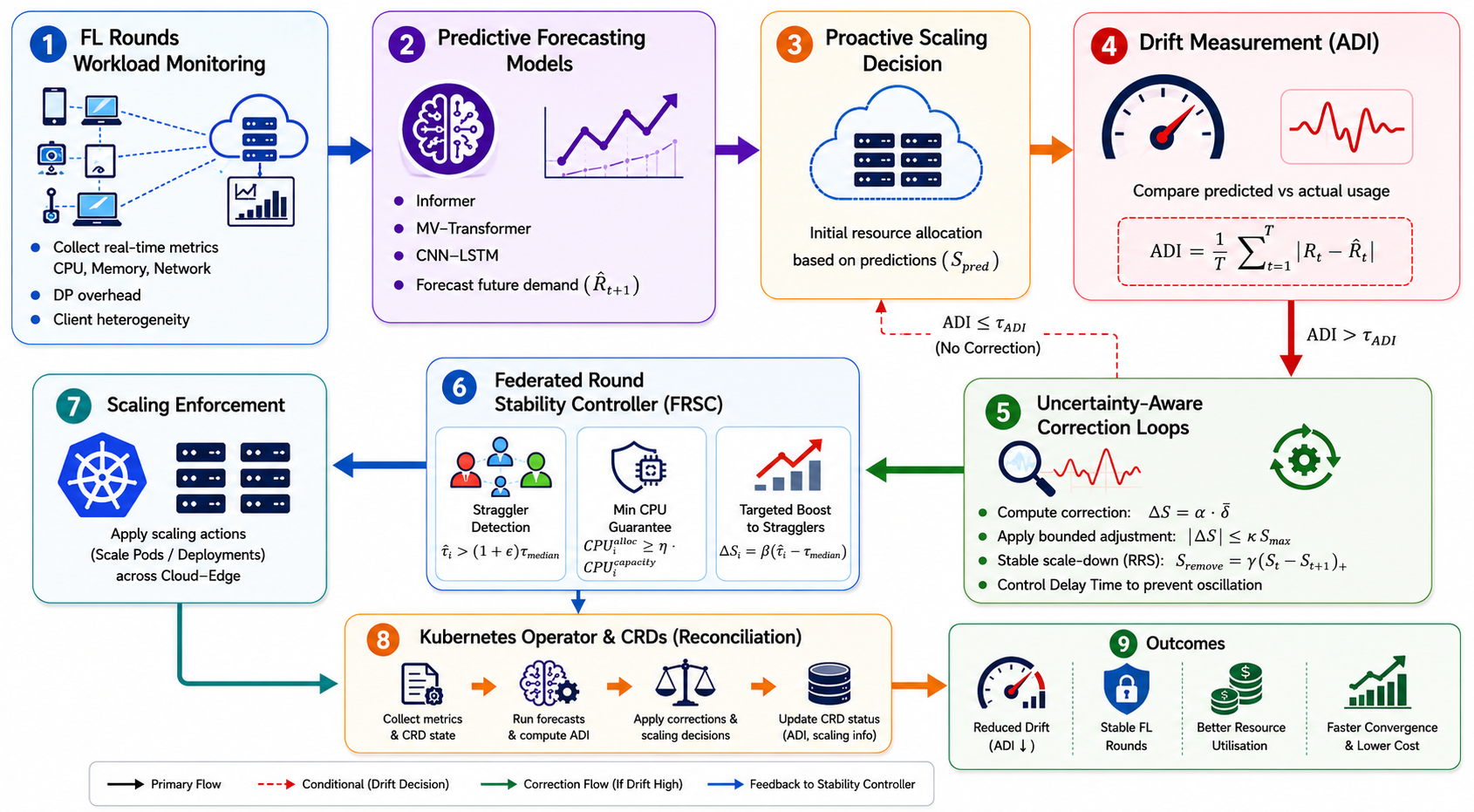}
    \caption{Drift-aware and uncertainty-aware autoscaling workflow for federated learning in cloud–edge environments.}
    \label{fig:Drift-aware and uncertainty}
\end{figure*}

The Figure \ref{fig:Drift-aware and uncertainty} illustrates a closed-loop drift-aware autoscaling framework designed for federated learning (FL) systems operating across heterogeneous cloud–edge environments. The workflow begins with FL workload monitoring, where real-time CPU, memory, communication, and differential privacy (DP) overhead metrics are collected from distributed clients during each training round. These workload traces are analysed by predictive forecasting models such as Informer, MV-Transformer, and CNN–LSTM to estimate future resource demand and generate proactive scaling decisions. The Autoscaling Drift Index (ADI) then measures the mismatch between predicted and actual resource utilisation to detect accumulated forecasting drift. When the drift exceeds a predefined threshold, uncertainty-aware correction loops apply bounded feedback adjustments, stable scale-down mechanisms, and oscillation control strategies to refine scaling actions. In parallel, the Federated Round Stability Controller (FRSC) identifies potential stragglers and enforces minimum CPU guarantees together with targeted resource boosts to maintain synchronised FL rounds. The corrected scaling decisions are subsequently enforced through Kubernetes Operators and CRDs, which reconcile cluster state and update runtime scaling information. Overall, the framework forms a continuous predict–measure–correct–stabilise control cycle that improves robustness, resource efficiency, and convergence stability for privacy-aware federated learning workloads.

\begin{figure}[ht]
    \centering
    \includegraphics[width=0.95\linewidth]{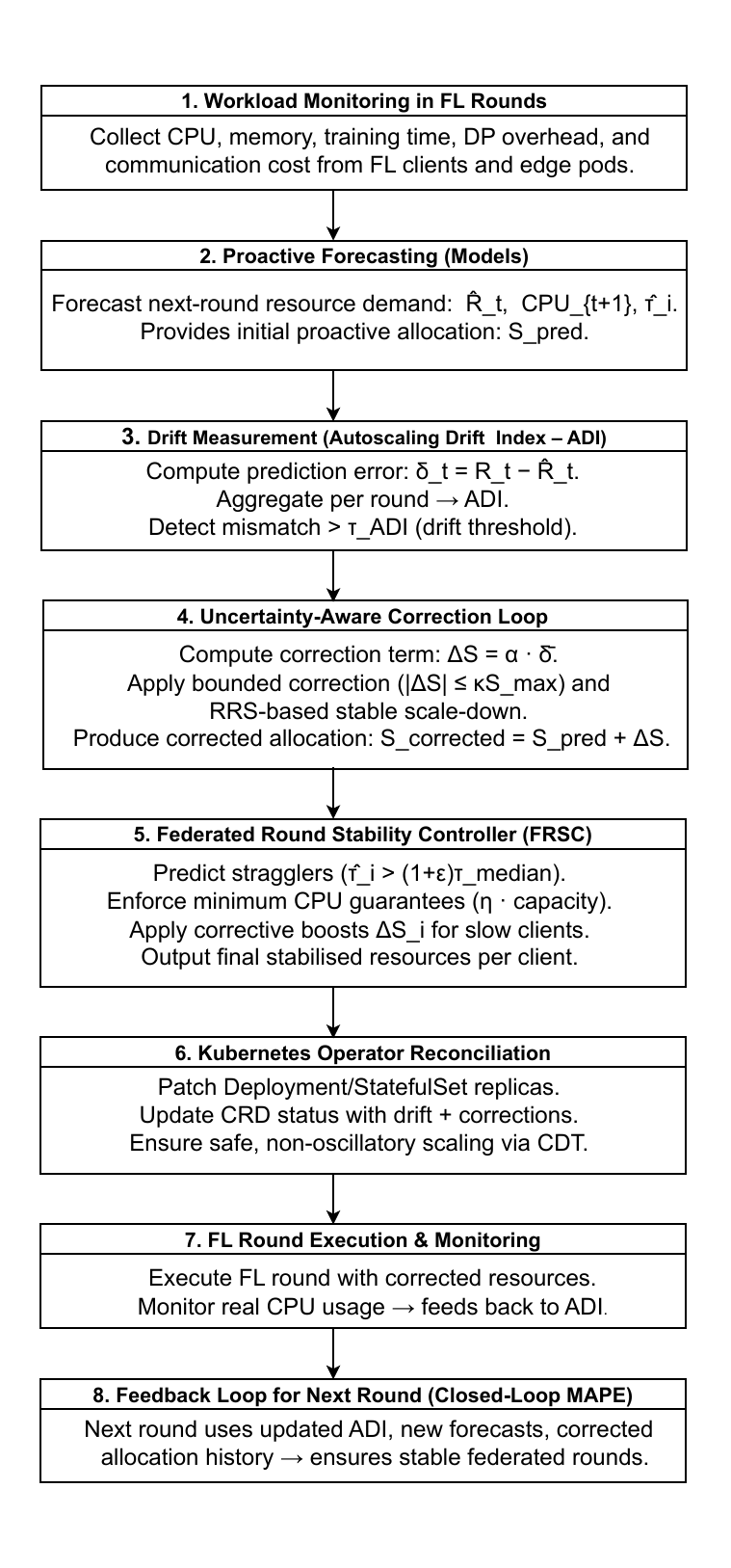}
    \caption{Drift-aware and uncertainty-aware autoscaling pipeline for federated learning.}
    \label{fig7}
\end{figure}

Figure~\ref{fig7} illustrates the complete drift-aware autoscaling workflow synthesised from recent advances in predictive autoscaling \cite{kumar2025optimizing, kumar2025multivariate}, federated optimisation with heterogeneous clients \cite{ye2024openfedllm,yi2025fedalora}, and uncertainty-aware control mechanisms such as ADI and FRSC. Starting from real-time workload monitoring in federated rounds, proactive forecasting models produce initial resource allocations, which are then refined using drift measurement and bounded correction loops that compensate for accumulated prediction errors \cite{dogani2024proactive,zhao2022edge}. The Federated Round Stability Controller further mitigates straggler slowdowns by enforcing minimum CPU guarantees and corrective boosts. Kubernetes Operator reconciliation applies these decisions at runtime and updates CRD state for continuity \cite{pham2024elastic,albaseer2023data,liu2025fedinv}. Together, these components form a closed-loop predict–measure–correct–stabilise cycle that ensures stable and efficient autoscaling under heterogeneous, non-IID, and DP-enhanced FL workloads. The Summary of drift-aware and uncertainty-aware autoscaling is shown in table~\ref{tab6}.

\subsection{Definition and Sources of Autoscaling Drift} \label{sec6.1}
Autoscaling drift refers to the cumulative mismatch that develops between predicted resource demand and the actual utilisation observed during execution~\cite{zhong2025data,song2023follow,berral2021theta,hua2024humas}. If $\hat{R}_t$ denotes the forecasted CPU requirement at time $t$ and $R_t$ represents the realised usage, then the instantaneous prediction error is
\begin{equation}
\delta_t = R_t - \hat{R}_t.
\end{equation}
Within a single FL round, these errors accumulate into a drift term that captures the overall deviation between expected and observed behaviour:
\begin{equation}
\label{eq:cumdrift}
\Delta_{\text{drift}}^{(r)} = \sum_{t \in \text{round}\ r} |\delta_t|.
\end{equation}
A larger value of $\Delta_{\text{drift}}^{(r)}$ indicates a stronger misalignment between prediction and resource consumption, which directly affects the efficiency and stability of autoscaling decisions.

The origin of drift can be traced to several fundamental characteristics of FL systems. A first source is client heterogeneity. Devices participating in a training round differ greatly in their processing capabilities, meaning the time required for local computation depends on both data volume and available compute capacity:
\begin{equation}
T_{k}^{\text{train}} = \frac{E_k \cdot |\mathcal{D}_k|}{C_{k}^{\text{CPU}}},
\end{equation}
where $E_k$ is the number of local epochs and $C_{k}^{\text{CPU}}$ the computational throughput of client $k$. These disparities create unpredictable variations in local training duration, which directly shape round-level resource demand.

A second source of drift comes from non-IID and time-varying workloads. Differences in local data distributions and fluctuating client participation lead to irregular cycles of computation and communication. The total round cost can be expressed as
\begin{equation}
\tau^{(r)} = T_{\text{train}}^{(r)} + T_{\text{comm}}^{(r)},
\end{equation}
highlighting how both training and communication contribute to variability. Shifts in either component influence the actual CPU usage pattern, making predictions inherently uncertain.

A third and increasingly significant source of drift is privacy-induced variability. When DP is applied, local gradients are modified by adding Gaussian noise
\begin{equation}
\tilde{g}_k = g_k + \mathcal{N}(0,\sigma^2 I),
\end{equation}
which expands computation time and increases the dispersion of CPU utilisation across clients. These effects magnify the unpredictability of resource consumption within each round.

Taken together, these factors make drift an unavoidable property of FL workloads. Without explicit correction mechanisms, the deviation accumulates over successive rounds, causing autoscaling decisions to diverge progressively from actual requirements and leading to instability, inefficiency, and degraded convergence behaviour.

\subsection{Autoscaling Drift Index (ADI)} \label{sec6.2}
The ADI presents a formal and interpretable metric of the variance between the actual resource demand and the value of autoscaling in a FL round~\cite{chen2023edgefd,li2025unsupervised}. The ADI is the measures taken within a round of $T$ measurement intervals, given by
\begin{equation}
\label{eq:adi}
\text{ADI} = \frac{1}{T} \sum_{t=1}^{T} |R_t - \hat{R}_t|,
\end{equation}
where $R_t$ is the measured CPU usage and $\hat{R}_t$ is the forecasted required CPU usage at time~$t$. The absolute prediction error is therefore added up over the course of a round and divided by the length of the round, resulting in a short and compact scalar that indicates the size of the prediction-resource mismatch.

ADI has a number of features that make it an ideal choice for drift-aware autoscaling. If the predicted demand equals realised consumption during the whole round – that is, if there is no difference between the two – then the value is zero, and the forecast is perfectly aligned. An index that sums absolute deviations over time will have a higher penalty for long-term underestimation or overestimation, but will not penalize short-term shifts as much, and is thus sensitive to the overall drift that is common in synchronous FL situations. Furthermore, the formulation is valid for Kubernetes Operators' periodic reconciliation cycles, and ADI can be computed on-the-fly and utilized instantly inside the control plane to evaluate if the scaling actions are still aligned with the actual workload behavior.

For operation, a drift correction threshold $\tau_{\text{ADI}}$ is set, and if it is exceeded, the correction has to be made. The decision rule is given by
\begin{equation}
\mathbf{1}_{\text{correct}} =
\begin{cases}
1, & \text{if } \text{ADI} > \tau_{\text{ADI}},\\[3pt]
0, & \text{otherwise},
\end{cases}
\end{equation}
Shows whether prediction error is out of acceptable limits or not. The correction mechanism adjusts the initial scaling decision to cancel out the measured drift when it is activated, thus avoiding any accumulation of errors that could lead to destabilising round lengths or stragglers. A benefit of using feedback based on ADI is that it mitigates drift across different clients, as demonstrated in empirical studies, rather than being considered as an inherent consequence of forecasting uncertainty. 

\begin{table*}[!ht]
\centering
\caption{Comparative analysis of drift-aware and uncertainty-aware autoscaling mechanisms for federated learning and cloud-edge environments.}
\label{tab6}
\renewcommand{\arraystretch}{1.25}
\begin{tabularx}{\textwidth}{p{3cm} X X X X}
\hline
\textbf{Category} & \textbf{Definition / Purpose} & \textbf{Strengths} & \textbf{Limitations} & \textbf{Typical Use Cases} \\
\hline

\textbf{Autoscaling Drift} ~\cite{zhong2025data,song2023follow,berral2021theta,hua2024humas}
&
Represents the cumulative mismatch between predicted resource demand and actual resource utilization during execution. Drift accumulates across FL rounds due to forecasting inaccuracies and workload variability.
&
Provides a measurable indicator of prediction quality; reveals long-term deviation between planning and execution.
&
Difficult to eliminate completely in highly dynamic environments; accumulates over time if left uncorrected.
&
Predictive autoscaling, cloud-native resource management, FL workload monitoring, cloud-edge orchestration.
\\
\hline

\textbf{Primary Drift Sources} ~\cite{zhong2025data,song2023follow,schmahl2023cyclic,patchipala2023tackling,berral2021theta,hua2024humas}
&
Drift originates from client heterogeneity, non-IID and time-varying workloads, communication variability, and DP-induced computational overhead.
&
Provides insight into root causes of scaling instability and prediction degradation.
&
Sources often interact simultaneously, making precise modelling and mitigation challenging.
&
Federated learning systems, heterogeneous edge environments, privacy-preserving distributed training.
\\
\hline

\textbf{Autoscaling Drift Index (ADI)} ~\cite{chen2023edgefd,li2025unsupervised}
&
A quantitative metric that measures average deviation between predicted and observed resource demand over a monitoring interval or FL round.
&
Simple to compute; interpretable; supports real-time drift monitoring and threshold-based control.
&
Requires careful threshold tuning; may not capture all workload characteristics when used alone.
&
Kubernetes Operator feedback loops, online drift detection, predictive autoscaling evaluation.
\\
\hline

\textbf{Uncertainty-Aware Correction Loops}~\cite{schmahl2023cyclic,patchipala2023tackling,bayram2024adaptive}
&
Feedback-driven mechanisms that adjust predicted scaling actions using observed prediction errors, bounded corrections, and stabilisation controls.
&
Improves robustness; reduces prediction-induced instability; adapts to changing workload conditions.
&
Additional controller complexity; improper tuning may introduce delayed responses or oscillations.
&
Closed-loop autoscaling systems, adaptive cloud services, dynamic FL environments.
\\
\hline

\textbf{Resource Removal Strategy (RRS)}~\cite{patni2025predictive,mukherjee2025amas}
&
Controlled scale-in mechanism that gradually removes excess resources rather than applying aggressive downscaling.
&
Reduces oscillations; prevents sudden under-provisioning; improves system stability.
&
May temporarily retain excess resources and increase short-term operational cost.
&
Predictive Kubernetes autoscaling, proactive resource management, FL round stabilisation.
\\
\hline

\textbf{Control Delay Time (CDT)}~\cite{patchipala2023tackling,schmahl2023cyclic}
&
Minimum waiting interval between consecutive scaling corrections to avoid rapid control actions.
&
Prevents excessive scaling fluctuations; improves controller stability.
&
Can delay reaction to genuine workload spikes if configured conservatively.
&
Real-time autoscaling controllers, Kubernetes reconciliation loops, edge resource management.
\\
\hline

\textbf{Federated Round Stability Controller (FRSC)}~\cite{foley2022openfl,park2021few,wang2024one,ma2023flamingo,wang2024fedcda,zou2024optscaler,ullah2018control,weber2024combining}
&
Stability controller that predicts stragglers, enforces minimum resource guarantees, and applies targeted corrective boosts to synchronise FL rounds.
&
Reduces round-time variance; mitigates stragglers; improves convergence stability and resource fairness.
&
Requires accurate training-time prediction and additional monitoring overhead.
&
Synchronous FL systems, distributed edge learning, large-scale federated training.
\\
\hline

\textbf{Drift-Aware FL Autoscaling}~\cite{zhong2025data,li2025unsupervised,wang2024fedcda}
&
Combines predictive forecasting, ADI monitoring, uncertainty-aware correction, and FRSC mechanisms to maintain alignment between resource allocation and workload demand.
&
Improves convergence, resource utilisation, stability, and scalability under heterogeneous conditions.
&
More complex architecture requiring forecasting, monitoring, and feedback integration.
&
Privacy-preserving FL, cloud-edge orchestration, autonomous Kubernetes-based learning platforms.
\\
\hline

\textbf{Evaluation Dimensions}~\cite{kumar2026critical,zhong2025data,li2025unsupervised}
&
Measures effectiveness using drift magnitude, ADI reduction, convergence time, resource utilisation, stability, oscillation frequency, and scaling accuracy.
&
Provides holistic assessment of drift-aware autoscaling performance.
&
Multiple metrics may have conflicting optimisation objectives.
&
Benchmarking adaptive autoscalers, cloud-native control systems, FL resource management studies.
\\
\hline
\end{tabularx}
\end{table*}

\subsection{Uncertainty-Aware Correction Loops} \label{sec6.3}
While the prediction-based autoscaling approach is a forward-looking approach to estimating the demand for resources, prediction errors are inevitable in FL networks, which introduce significant uncertainty to the demand for resources due to client behaviour, network conditions and privacy mechanisms~\cite{schmahl2023cyclic,patchipala2023tackling}. The prediction-based autoscaling approach is forward-looking and estimates the demand for resources, but prediction errors are not excluded in FL networks because of substantial uncertainty in the demand for resources due to client behaviour, network conditions and privacy mechanisms. Let $S_{\text{pred}}$ represent the proactive scaling decision derived from a forecasting model, e.g., LSTM, Informer or MV-Transformer \cite{bayram2024adaptive}. Drift-aware autoscaling enhances this prediction with a correction feature that explicitly incorporates recent differences between the predicted and actual demands. The correction term is defined as
\begin{equation}
\label{eq:correction}
\Delta S = \alpha \cdot \bar{\delta},
\end{equation}
In it, the moving average of the past prediction errors is $\bar{\delta}$, and the feedback gain is $\alpha$ which determines the sensitivity of the controller to the accumulation of prediction error. The new resource allocation is
\begin{equation}
\label{eq:corrected}
S_{\text{corrected}} = S_{\text{pred}} + \Delta S,
\end{equation}
Allowing the system to react not only to the forecasting signal but also to the actual behaviour in the previous round.

Due to the synchronous nature of FL, small errors can build up and cause instabilities in the system if corrections are made too strongly. A bounded correction rule is used by the controller to limit the size of the changes:

\begin{equation}
|\Delta S| \le \kappa S_{\max},
\end{equation}
where $\kappa$ represents the maximum fraction of the resource that may be used. This is to avoid indicating overshoots that may destabilize the system, as well as ensuring the smoothness of drift compensation between consecutive rounds.

The system has a Resource Removal Strategy (RRS) that is used when reducing resources:
\begin{equation}
S_{\text{remove}} = \gamma \cdot (S_t - S_{t+1})_+, \quad 0<\gamma<1,
\end{equation}
so that only a small amount of the excess resources is taken out each cycle. This scale-in behaviour will be amortised which helps reduce the risk of repetitive under-provisioning especially for synchronous FL where slow devices are stragglers. The controller also imposes a minimum time between two consecutive corrective actions: Control Delay Time (CDT). This time buffer helps to avoid quick oscillations when predictions may fluctuate from one reading to the next, or when the devices themselves vary.

Together, these mechanisms transform the autoscaling process from a simple predict-and-execute sequence into a closed-loop predict–measure–correct–stabilise pipeline. This design significantly enhances the resilience under uncertainty, and also guarantees the reliability of proactive scaling, even when FL workloads are very variable in time.

\subsection{Federated Round Stability Controller (FRSC)} \label{sec6.4}

The FRSC is designed to stabilise synchronous FL rounds by preventing the emergence of stragglers and by ensuring consistent training times across heterogeneous clients~\cite{foley2022openfl,park2021few}. In synchronous FL, the global round cannot complete until every participating device has finished its local update \cite{wang2024one,ma2023flamingo}. For this reason, even a single under-provisioned or overloaded client can delay the entire round~\cite{ wang2024fedcda,zou2024optscaler}. The FRSC operates as a lightweight control layer that anticipates potential delays, enforces minimum resource guarantees, and applies targeted corrective boosts to clients predicted to become stragglers~\cite{ullah2018control,weber2024combining}.

The first component of FRSC involves predicting which clients are likely to fall behind. Let $\hat{\tau}_i$ denote the forecasted local training time for client~$i$, computed from its expected CPU availability and workload cost. The controller compares this quantity against the median predicted training time across all clients. A device is marked as a potential straggler if
\begin{equation}
\hat{\tau}_i > (1+\epsilon)\tau_{\text{median}},
\end{equation}
where $\epsilon$ defines the tolerance window. This simple yet effective rule identifies clients that are likely to prolong the round before the slowdown materialises.

To ensure that predicted stragglers and weaker devices receive sufficient resources, the FRSC applies minimum CPU guarantees. Each client is allocated at least a fraction $\eta$ of its hardware capacity:
\begin{equation}
\text{CPU}_{i}^{\text{alloc}} \ge \eta \cdot \text{CPU}_{i}^{\text{capacity}}.
\end{equation}
This baseline prevents starvation, reduces the risk of sudden slowdowns, and ensures that clients with low throughput still maintain predictable execution behaviour.

For devices identified as likely stragglers, the controller applies a corrective boost that increases their allocated resources in proportion to the severity of the predicted delay. The boost term is defined as
\begin{equation}
\Delta S_i = \beta (\hat{\tau}_i - \tau_{\text{median}}),
\end{equation}
where $\beta$ is a gain parameter which determines the level of correction. Essentially, this is a targeted adjustment that will keep the system balanced when it comes to round completion times, whilst keeping the overhead to a minimum, as boosts are only applied when needed. Empirical results indicate that FRSC minimizes the variance in round-time under irregular workloads due to DP noise and device heterogeneity. It reduces stragglers and aligns the executions over the rounds, which enhances synchronisation in synchronous FL, and works in tandem with the drift-aware correction mechanisms in Section~\ref{sec6.3}.

The overall performance of drift-aware autoscaling is compared to the FL performance under device heterogeneity and differential privacy constraints. The more this DP noise exists, the more overhead it requires and also the more variation in training time it can cause, increasing the chance of under-provisioning. A drift-aware controller dynamically adjusts the way resources are allocated to these privacy induced costs, and also helps to reduce all stragglers introduced by the diverse devices. It contributes to the improvements of convergence, resource utilization and robustness in dynamic cloud-edge FL environments by smoothing round durations and maintaining the synchronisation.

\textbf{Answer to RQ4}: Autoscaling drift is the difference between expected and actual resource usage which happens over time in FL and other dynamic workloads. It is caused by non-stationary usage, different privacy and communication between clients, non-IID and time-varying data, and DP overhead, resulting in a non-stationary utilisation and long-standing prediction errors. This drift can be corrected by uncertainty-aware correction loops, which track prediction error and provide adaptive feedback to adapt their resource allocation, thereby making autoscaling a closed-loop control loop. Moreover, round stability controllers ensure minimum resource guarantees and make targeted adjustments when executing, thus minimizing the straggler effects. These mechanisms are combined to provide robustness, stability of resource provisioning and efficient convergence in dynamic, privacy-preserving cloud–edge environments, thereby addressing \textbf{RQ4}.

\section{Open Challenges and Research Directions} \label{sec7}
\begin{figure*}
    \centering
    \includegraphics[width=0.99\linewidth]{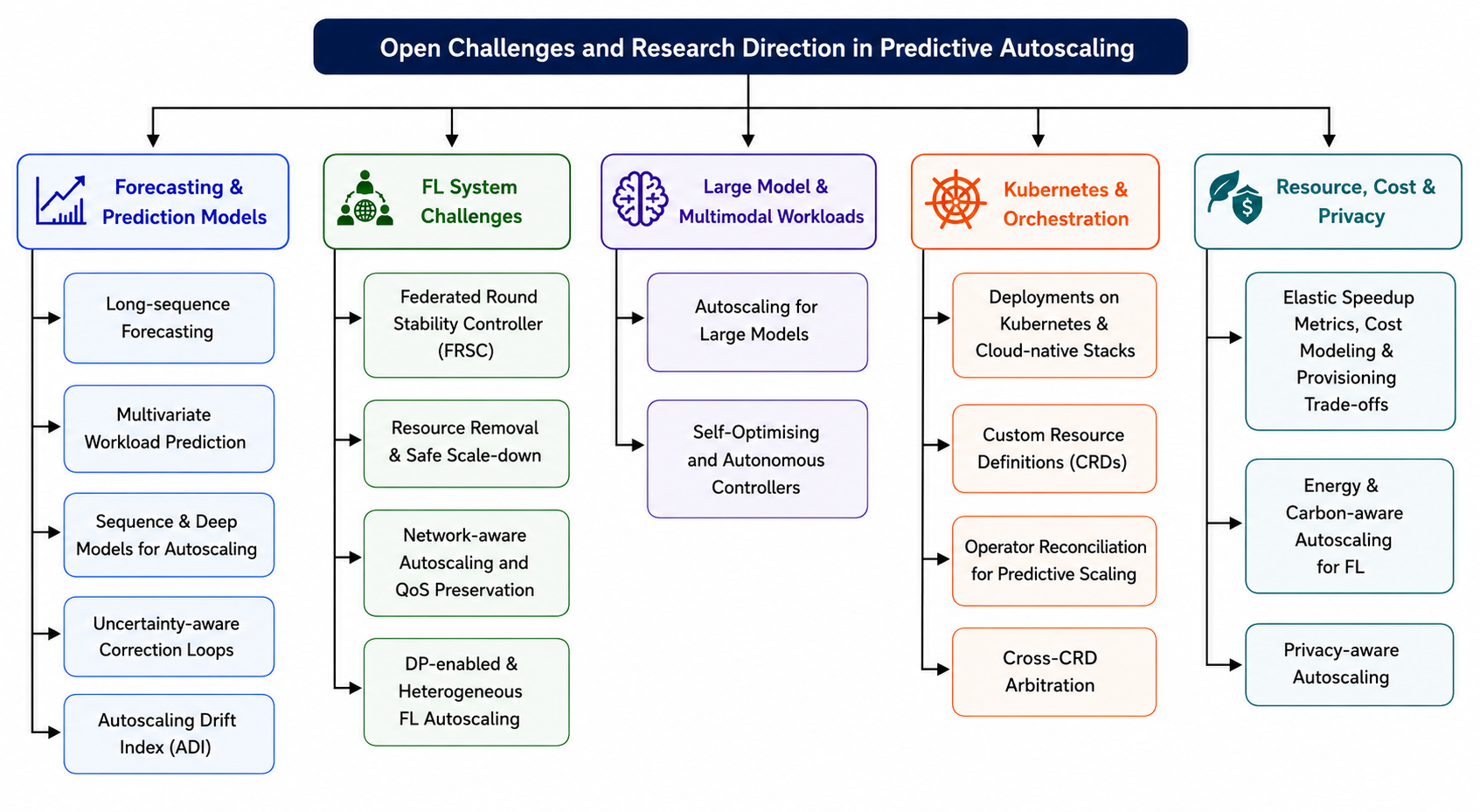}
    \caption{Hierarchical taxonomy of challenges and future research directions in predictive autoscaling for cloud-native and federated cloud–edge environments.}
    \label{research_and_directions}
\end{figure*}

While predictive models, kubernetes-native automation and federated orchestration have come a long way, they are still constrained in their ability to scale up and down with dynamic workloads, uncertainty, heterogeneity, and the constraints of the Kubernetes system itself. To overcome this, future auto-scaling frameworks need to be adaptive and uncertainty-aware, and move towards self-optimising systems that encompass prediction, control and orchestration. The most important open challenges and research directions are summarized in a shortened manner in Table~\ref{tab7}. Figure~\ref{research_and_directions} shows a hierarchical structure of the most significant research areas, such as forecasting models, federated learning issues, large model workloads\cite{11075838}, Kubernetes orchestration and resource-aware autoscaling. It also depicts the interactions between new research trends like uncertainty-aware scaling, autonomous controllers, privacy-aware autoscaling and predictive orchestration mechanisms.

\subsection{Forecasting and Prediction Models}

\begin{enumerate}
    \item  Proactive Long-Sequence Forecasting in Dynami Environments: This subsection discusses the key challenges and promising future research directions related to proactive long-sequence forecasting in dynamic and continuously shifting environments~\cite{kumar2025optimizing,ding2024dynamic}.

    \begin{itemize}

    \item Challenge: Informer and similar long-sequence forecasting models perform well on stationary or semi-stationary traces, but real cloud and federated workloads exhibit abrupt shifts, seasonal variations, client churn, mobility-driven spikes, and device-level failures. The accuracy of forecasts gets much worse when there is concept drift or structural change in workload distributions. Most current models aren't adaptive and can cause long horizon predictions to go stale, or even be misleading, resulting in ongoing resource misallocation over several cycles of auto-scaling.

    \item Research Directions:
    \begin{itemize}
        \item Online and continual learning for long-sequence forecasters: Model learning at runtime without requiring a complete retraining to counteract catastrophic forgetting and make the models more robust to changes in workload pattern.
        \item Hybrid local-global forecasting architectures: Use Informer for capturing long-term trends, short-term forecasting using CNN–LSTM modules, and adaptive forecast fusion mechanisms.
        \item Forecasting under uncertainty: Synthesize a probabilistic attention mechanism, quantile forecasting and the Bayesian Transformer variants to formally express prediction confidence for uncertainty-aware autoscaling.
        \item Resource-efficient inference: Study how to distill, prune and apply sparse attention for real-time forecasting on resource-limited edge clusters.
    \end{itemize}

    \end{itemize}

    Overall, the solution of such challenges is crucial for developing resilient, accurate and adaptive long-sequence forecasting systems that can facilitate next generation proactive autoscaling within dynamic cloud and federated environments.
\item Multivariate Workload Prediction at Scale: This problem is defined for multivariate workload prediction at scale, where autoscalers must reason over large, noisy, and highly correlated metric streams~\cite{shahzaad2025service,meng2025catscaler, kumar2025multivariate}. 
    \begin{itemize}

    \item Challenge: Cloud-native telemetry pipelines generate hundreds of interdependent and noisy metrics. MV-Transformers can capture these cross-metric relationships, yet high-dimensional attention incurs substantial computational overhead, increases sensitivity to noise, and heightens the risk of overfitting. Irregular sampling and missing metrics further reduce prediction stability and degrade end-to-end autoscaling performance.

    \item Research Directions:
    \begin{itemize}
        \item Causal feature selection: Move beyond correlation-based filtering toward causal discovery techniques to identify metrics with genuine predictive influence, thereby reducing noise and computation.
        \item Hierarchical multivariate models: Enable multi-level forecasting (node, service, and cluster levels) in a unified MV-Transformer with shared encoders and task-specific output heads.
        \item Robustness to irregular and missing streams: Design attention mechanisms that operate reliably under asynchronous sampling, imputation-free inference, and metric dropout conditions.
        \item Explainable multivariate attention: Provide transparent, interpretable cross-metric attributions using saliency analysis, temporal–causal reasoning, and counterfactual perturbation techniques.
    \end{itemize}
    \end{itemize}

     Overall, addressing these research gaps is essential for building scalable, reliable, and interpretable multivariate forecasting systems capable of supporting next-generation proactive autoscaling in large cloud clusters.
\item Sequence and Deep Models for Autoscaling: This problem is defined for sequence-based autoscaling in heterogeneous cloud, edge, and federated environments where lightweight models must balance predictive accuracy with strict resource constraints.

    \begin{itemize}

    \item Challenge: Neural sequence models such as LSTM, GRU, CNN-LSTM, and Bi-LSTM remain widely deployed in resource-limited clusters because of their low inference overhead \cite{etemadi2021cost,dang2021deep,taha2024proactive,agarwal2024deep}. Yet their narrow receptive fields, sequential computation bottlenecks, and limited ability to capture long-range dependencies restrict the effectiveness of proactive autoscaling, especially in federated, multimodal, and highly dynamic workload patterns~\cite{dogani2022k,vu2022predictive,ouhame2021efficient,cai2025deep}.

    \item Research Directions:
    \begin{itemize}
         \item Dynamic model governance: Incorporate performance monitors and drift detectors that automatically switch between forecasters (LSTM, MV-Transformer, Informer) based on workload regime and prediction confidence.
        \item Federated forecasting models: Train autoscaling forecasters themselves using FL so that scaling policies adapt to the behaviour of each cluster without centralising telemetry data.
        \item Reinforcement learning for policy optimisation: Replace static threshold-based rules with RL-driven autoscaling controllers optimised jointly for latency, cost, elasticity, and system stability.
    \end{itemize}

    \end{itemize}

    Overall, advancing these capabilities will enable sequence models to play a more adaptive and scalable role in predictive autoscaling across distributed cloud, edge, and federated infrastructures.

\item Uncertainty-Aware Correction Loops: This problem is defined for uncertainty-aware correction loops that must stabilise prediction-driven autoscaling under fluctuating workloads and evolving error patterns~\cite{schmahl2023cyclic,patchipala2023tackling,zhong2025data,li2025unsupervised,bayram2024adaptive}.

    \begin{itemize}
    \item Challenge: Prediction-based scaling without correction gradually accumulates deviation across rounds, while overly aggressive correction risks oscillatory behaviour or potential violations of service-level requirements.

    \item Research Directions:
    \begin{itemize}
        \item Bayesian control loops: Incorporate posterior predictive distributions from forecasting models to determine stable and optimal correction magnitudes under uncertainty.
        \item Multi-controller coordination: Coordinate correction strategies across multiple clusters or edge regions to prevent conflicting actions and ensure globally consistent behaviour.
        \item Joint optimisation of prediction and correction: Train forecasting models using control-aware objectives that penalise unstable drift and favour smoother, more reliable autoscaling responses.
    \end{itemize}
    \end{itemize}

    A unified uncertainty-aware correction mechanism can ultimately enhance the resilience and stability of prediction-driven autoscaling systems.

\item Autoscaling Drift Index (ADI): This problem concerns the need for a unified metric that quantifies how far autoscaling behaviour deviates from ideal forecasts, especially in dynamic cloud-edge-federated environments~\cite{hua2024humas, zhong2025data, song2023follow, chen2023edgefd, li2025unsupervised}.

    \begin{itemize}
    \item Challenge: The ADI offers a promising formal measure of forecast-resource mismatch, but its practical use requires adaptive thresholding, noise-aware interpretation, and seamless integration across multi-layer controllers. Without careful tuning, ADI may misidentify transient deviations as drift or overlook structural changes in workload behaviour.

    \item Research Directions:
    \begin{itemize}
        \item Adaptive ADI thresholds: Dynamically adjust $\tau_{\text{ADI}}$ based on statistical confidence intervals, workload volatility patterns, and model uncertainty estimates.
        \item Semantic decomposition of drift: Break down measured drift into meaningful components, such as privacy noise, client heterogeneity, network congestion, or telemetry jitter, to support targeted corrective actions.
        \item ADI-guided model selection: Use ADI as a feedback mechanism to trigger model switching, selective retraining, or hybrid forecast fusion when drift crosses acceptable limits.
    \end{itemize}

    \end{itemize}

    Overall, advancing ADI into a robust and actionable control metric will significantly strengthen predictive autoscaling stability across cloud, edge, and federated infrastructures.
\end{enumerate}

\subsection{FL System Challenges}
\begin{enumerate}
    \item Federated Round Stability Controller: This problem is defined for maintaining stable round durations in large-scale FL environments affected by client heterogeneity and communication variability~\cite{wang2024fedcda,zou2024optscaler,weber2024combining}.
    \begin{itemize}
    \item Challenge: Stragglers, heterogeneous devices, and fluctuating communication conditions lead to highly unstable round durations in synchronous FL, reducing throughput and slowing global convergence.
    \item Research Directions:
    \begin{itemize}
        \item Adaptive straggler prediction: Develop mobility-aware predictors using dynamic percentile thresholds, temporal smoothing, and client-side performance profiling.
        \item Cross-layer stability mechanisms: Integrate FRSC with network-layer congestion control, transport-level adaptation, and intelligent client scheduling policies.
        \item Extension to asynchronous FL: Adapt FRSC principles for partial participation, stale gradient handling, and event-driven aggregation rounds to improve robustness under real-world variability.
    \end{itemize}
    \end{itemize}
    Overall, strengthening round stability will be essential for enabling scalable, communication-efficient, and high-throughput FL deployments across heterogeneous edge and mobile environments.

\item Resource Removal Strategy and Safe Scale Down: The problem of designing an effective resource removal strategy and ensuring safe scale down arises because predictive autoscaling must reduce resources cautiously without disrupting ongoing FL operations.

    \begin{itemize}
    \item Challenges: Predictive autoscaling should be conservative when scaling up, as removing pods too early may interrupt federated rounds, end running rounds for the local metric and slow down the aggregation process \cite{hubalek2022towards,baek2021enhancing,xu2023federated}. The classic reactive autoscaler may deprovision too quickly and too drastically when CPU utilization dips in a short period of time, and even the proactive model could mis-predict the short-time drop-off where the FLs are not the same or the noise is due to DP. The challenge lies in the design of scale-in mechanisms that are minimised disruption and maintain the temporal stability over the synchronous FL rounds~\cite{wu2022adaptive,el2022differential,liberti2024federated,parra2024kubeflower}.

    \item Research Directions:
    \begin{itemize}
        \item Round-aware safe scale-in: Design policies that only do scale-in at safe time points (e.g., aggregate or communication phase).
        \item Provisioning that waits for local checkpoints or update uploads for provisioning to terminate replicas: Add pod draining logic which waits for local checkpoints or wait for uploads to update before terminating replicas.
        \item Adaptive RRS tuning: Dynamically tune the RRS coefficient $\gamma$ based on drifts trends and workload variations using reinforcement learning or online feedback.
        \item Lightweight state migration: Consider micro checkpointing or partial state migration to prevent losing computation during pod removal.
    \end{itemize}
    \end{itemize}

    Overall, a robust resource removal strategy should ensure that scale-down decisions preserve training progress, maintain system stability, and minimise the operational risks associated with dynamic FL workloads.

\item Network-Aware Autoscaling and QoS Preservation for FL: This problem is defined for FL systems where network conditions play a critical role in round completion time, yet most existing autoscaling approaches focus primarily on compute-based signals. 

    \begin{itemize}
        \item Challenges: FL introduces significant network variability due to irregular upload sizes, heterogeneous bandwidth, and unpredictable client participation \cite{koch2025intelligent,sofia2023dynamic,mr2025pod}. Although CPU-oriented forecasting can guide compute scaling, network congestion, packet delay, or latency spikes may still create stragglers and prolong aggregation \cite{ye2024openfedllm,yi2025fedalora,dogani2024proactive}. The challenge is to integrate network forecasting and Quality-of-Service (QoS) constraints into autoscaling decisions~\cite{cheng2023proscale}.

        \item Research Directions:
        \begin{itemize}
        \item Multivariate network-compute forecasting: Extend MV-Transformer models to jointly predict CPU utilisation, bandwidth variation, RTT, packet loss, and throughput dynamics.
        \item Network-aware pod placement: Use CRDs or custom schedulers to support topology-aware placement, such as co-locating aggregators with high-bandwidth or low-latency clients.
        \item QoS-driven scaling constraints: Incorporate Kubernetes QoS classes and CNI-level network metrics into autoscaling formulations to prevent QoS violations.
        \item Prediction of Large model update sizes: Model the variance of LoRA or adapter update sizes to proactively allocate network buffers and reduce I/O bottlenecks.
        \end{itemize}
    \end{itemize}
    Overall, integrating network intelligence into autoscaling improves round stability, reduces straggler effects, and enables more reliable FL across distributed cloud-edge environments.

\item DP-Enabled and Heterogeneous FL: This problem is defined for managing autoscaling in FL systems where DP and device heterogeneity jointly create unpredictable and highly variable computational demands~\cite{baek2021enhancing,xu2023federated,wu2022adaptive,el2022differential,liberti2024federated,yi2025fedalora,ye2024openfedllm,liu2025fedinv}.

    \begin{itemize}
        \item Challenge: DP adds substantial computation and memory cost that is dependent on noise. All of those, together with the varied hardware of the clients, varying network circumstances and inconsistent training speeds locally, create irregular workload surges, making accurate autoscaling choices difficult.
    
        \item Research Directions:
        \begin{itemize}
        \item Adaptive DP-cost modelling: Create running estimators that estimate the computational and communication cost due to the DP mechanisms, which will help in allocating replicas more prcisely.
        \item Privacy-aware scaling policies: Design autoscaling policies that take into account privacy budgets ($\epsilon, \delta$) along with compute requirements, balancing privacy guarantees with system performance.
        \item Balanced DP + FL optimisation: Investigate joint optimisation of client selection, DP noise levels, and autoscaling thresholds to achieve efficient scaling without compromising learning quality.
        \end{itemize}
    \end{itemize}
    
    Overall, advancing autoscaling for DP-enabled and heterogeneous FL systems will be essential for maintaining efficiency, fairness, and privacy guarantees in large-scale, real-world deployments.  
    
\end{enumerate}

\subsection{Large Model and Multimodal Workloads}
\begin{enumerate}
    \item Autoscaling for Large Model and Multimodal Workloads: This problem is defined for autoscaling systems that must efficiently manage the heavy, bursty, and highly variable resource demands introduced by LLMs and multimodal architectures\cite{yi2025fedalora,ali2025recent,radhika2021review,albaseer2023data,zhao2022edge,pham2024elastic,albaseer2023data,liu2025fedinv}.
        \begin{itemize}
        \item Challenge: LLM fine-tuning, multimodal pipelines, and vision-transformer inference create substantial GPU and memory requirements, exhibit irregular temporal behaviour, and impose coordination challenges across compute, storage, and network resources. These characteristics make traditional threshold-based scaling insufficient for maintaining performance and cost efficiency.

        \item Research Directions:
        \begin{itemize}
            \item Hierarchical autoscaling: Develop multi-level autoscaling strategies that coordinate CPU, GPU, specialised accelerators, and interconnect bandwidth for end-to-end efficiency.
            \item Model-sharding-aware schedulers: Design scaling policies that incorporate tensor parallelism, pipeline parallelism, LoRA adapter placement, and the memory footprint of KV-cache and attention layers.
            \item Workload characterisation: Construct large-scale, realistic traces of LLM and multimodal workloads to accurately benchmark next-generation autoscalers.
        \end{itemize}
        \end{itemize}
        Overall, advancing autoscaling mechanisms for LLM and multimodal workloads is critical for sustaining performance, reliability, and cost control in modern AI-driven cloud and edge infrastructures.
\item Self-Optimising and Autonomous Controllers: This problem is defined for emerging autoscaling systems that aim to operate with minimal manual intervention while continuously adapting to dynamic workloads and diverse deployment environments\cite{pham2024elastic,patni2025predictive,mukherjee2025amas,kumar2025optimizing,ding2024dynamic,kumar2025multivariate}.
        \begin{itemize}
        \item Challenge: Most existing autoscalers still depend on fixed thresholds, hand-crafted rules, and offline-tuned hyperparameters, limiting their ability to respond intelligently to non-stationary behaviour, evolving application characteristics, and cross-cluster heterogeneity.

        \item Research Directions:
        \begin{itemize}
            \item Safe reinforcement learning: Develop RL-based autoscaling policies that incorporate formal safety guarantees to prevent QoS violations, budget overruns, and unstable oscillations.
            \item Meta-learning for cross-cluster adaptation: Enable autoscalers to generalise across clusters, applications, and workload regimes by rapidly adapting their internal models with minimal retraining.
            \item Human-in-the-loop governance: Integrate interpretable controls, policy explanations, and operator override mechanisms to ensure transparency and operational trust.
        \end{itemize}
        \end{itemize}

    Overall, advancing self-optimising and autonomous controllers is essential for building reliable, adaptive, and fully intelligent predictive autoscaling frameworks that can operate robustly across cloud, edge, and federated environments.

\end{enumerate}

\subsection{Kubernetes and Orchestration}
\begin{enumerate}
    \item Deployments on Kubernetes, Docker, and Cloud-Native Stacks: This problem is formulated for predictive autoscaling systems under real operational conditions with significant noise, heterogeneity, and system-level constraints, which greatly impact performance~\cite{kumar2024optimizing,kumar2025multivariate,kumar2025optimizing}.

    \begin{itemize}

    \item Challenge: Most studies are conducted in minikube testbeds or synthetic traces, but in production, clusters have imperfect and noisy telemetry, restricted number of API calls, cluster instability due to eviction, and varying node configurations. If not properly engineered, and implemented, proactive autoscalers can bring scheduling overhead, extra control-plane pressure and unpredictable interactions with built-in Kubernetes controllers.

    \item Research Directions:
    \begin{itemize}
        \item Large-scale production validation: Test predictive autoscalers across managed Kubernetes deployments (EKS, AKS, GKE) to gain insight into the robustness of these autoscalers when presented with real workload fluctuations and infrastructure noise.
        \item Low-overhead monitoring architectures: Use adaptive sampling, sketch-based telemetry, and compression-aware pipelines to minimize Prometheus and Kubernetes metric collection overhead.
        \item Multi-node heterogeneity modelling: Create forecasters that take account of the differences between GPUs, CPU, TPU, node affinity rules, and resource fragmentation phenomena.
    \end{itemize}

    \end{itemize}

    To the end, progress in research toward deployment is critical to keep predictive autoscalers reliable, efficient, and production-ready in a variety of cloud-native situations.

\item Custom Resource Definitions for Autoscaling: This problem is defined for predictive autoscaling systems where CRDs are used to encode model logic, scale policies and telemetry pipelines within the Kubernetes control plane~\cite{kumar2024optimizing,yuan2024time,wijesekera2025kubernetes,kubernetesExtendKubernetes,miracapillo2025enabling}.

    \begin{itemize}
    \item Challenge: While the CRDs are a great tool for extending Kubernetes, the lack of consistent schema design, schema drift over time, and bad coordination between operators can result in conflicting scaling operations, policy violations, or unstable control loops in production clusters.

    \item Research Directions:
    \begin{itemize}
        \item Standardised CRD schemas: Community-developed, autoscaling CRD schemas, with a common structure that establishes predictive model metadata, uncertainty scores, differential-privacy parameters and scaling constraints.
        \item Admission controller and policy engine security and validation of CRD: Ensure that the admission controllers and policy engines validate the CRD fields (e.g., model signatures, replica bounds, or rate-stability settings) before the CRD is passed to the control plane.
        \item Cross-CRD arbitration: Design arbiration methods that interoperate between horizontal, vertical, and custom autoscalers that are attempting to change the same deployment.
    \end{itemize}

    \end{itemize}

    In summary, improving the design, validation and coordination of CRD are crucial to create more secure, robust and scalable predictive autoscaling solutions within Kubernetes.
\item Predictive Scaling: Operator Reconciliation is a problem in predictive autoscaling systems that require Kubernetes operators to coordinate multiple inferences, CRD updates, and scaling decisions in a timely manner while ensuring correctness under asynchronicity in cluster behaviour~\cite{kubernetesOperatorPattern,kubernetesOverview,kubernetesAutoscalingWorkloads,kumar2024optimizing,dame2022kubernetes,yu2025arsc}.

    \begin{itemize}

    \item Challenge: Kubernetes operators run under eventual consistency, event-driven triggers and race conditions. These issues are exacerbated with predictive autoscaling, as it consists of frequent inference, potentially conflicting or overlapping reconciliation loops, and interaction among multiple controllers.

    \item Research Directions:
    \begin{itemize}
        \item Reduce race conditions with concurrency-safe reconciliation: Implement operators that manage transactions, have deterministic ordering of loops, and have conflict-free replicated data structures.
        \item Non-blocking inference pipelines: Implement coroutine-based or async inference pipelines that execute concurrently to reconciliation loops, thus preventing an inference delay from impacting cluster-wide decisions.
        \item Operator verification: Use formal verification, model checking or policy invariants to ensure safety properties: no unsafe scale-down, respect cooldown periods, maintain QoS guarantees.
    \end{itemize}
    \end{itemize}

    In summary, enhancing operator reliability, concurrency handling and safety guarantees is essential to support stable and reliable predictive autoscaling in Kubernetes environments.

\item Cross-Layer Autoscaling: This problem is defined for coordinating autoscaling across compute, network, and edge-cloud layers~\cite{nakata2021concentrated,mavridis2023orchestrated,morcillo2025privacy,hubalek2022towards,baek2021enhancing,xu2023federated}.  
    \begin{itemize}
    \item Challenge: Autoscaling often assumes compute is the bottleneck. Real systems face synchronization bottlenecks, network congestion, and edge-cloud latency inconsistencies.

    \item Research Directions:
        \begin{itemize}
            \item Joint optimisation of compute and network: Predict both CPU/GPU and network throughput for coordinated scaling.
            \item Edge-cloud mobility-aware scaling: Adapt resources based on client mobility, handovers, and intermittent connectivity.
            \item Global autoscaling orchestration: Unified controllers that coordinate edge micro-clusters with cloud backends.
        \end{itemize}
    \end{itemize}

    Overall, cross-layer coordination ensures holistic resource management, reduces latency bottlenecks, and improves service reliability across distributed cloud-native systems. 

\end{enumerate}

\begin{table*}[t]
\centering
\caption{Challenges and research directions for autoscaling in cloud, edge, and federated environments.}
\label{tab7}
\begin{tabular}{p{0.26\textwidth} p{0.32\textwidth} p{0.34\textwidth}}
\hline
\textbf{Topic} & \textbf{Challenge} & \textbf{Research Directions} \\
\hline

Forecasting Models~\cite{kumar2025optimizing,ding2024dynamic,kumar2025multivariate} &
Degradation under drift, non-stationarity, and workload variability. &
Online learning, hybrid models, uncertainty-aware forecasting, lightweight inference. \\ \hline

Multivariate Prediction~\cite{kumar2025multivariate,arbat2022wasserstein} &
High-dimensional noisy metrics reduce stability and increase cost. &
Causal feature selection, hierarchical models, robust attention, explainability. \\ \hline

Sequence Models~\cite{dang2021deep,taha2024proactive,kumar2024optimizing} &
Limited long-range dependency modelling in dynamic environments. &
Dynamic model switching, federated learning, RL-based scaling policies. \\ \hline

Evaluation and Cost~\cite{qu2018auto,kumar2024optimizing} &
Traditional metrics ignore latency, energy, and cost trade-offs. &
Multi-objective metrics, tail-latency awareness, cost-aware scaling. \\ \hline

Cloud-Native Deployment~\cite{kumar2024optimizing,kumar2025multivariate} &
Real-world systems introduce noise, heterogeneity, and instability. &
Production validation, low-overhead monitoring, heterogeneity-aware models. \\ \hline

CRDs and Operators~\cite{wijesekera2025kubernetes,kubernetesExtendKubernetes} &
Conflicts and unstable control loops. &
Standardised CRDs, validation, concurrency-safe reconciliation. \\ \hline
 
Scale-Down (RRS)~\cite{hubalek2022towards,baek2021enhancing} &
Aggressive scale-in disrupts workloads and FL processes. &
Safe scale-in policies, checkpoint-aware removal, adaptive tuning. \\ \hline

Network-Aware Scaling~\cite{koch2025intelligent,pham2024elastic} &
Ignoring network leads to QoS degradation. &
Joint compute-network prediction, topology-aware scheduling, QoS constraints. \\ \hline

Energy and Sustainability~\cite{zhai2025energy,savazzi2022energy} &
Autoscaling ignores power and carbon constraints. &
Energy-aware forecasting, green scheduling, power-aware policies. \\ \hline

Drift and Uncertainty~\cite{zhong2025data,schmahl2023cyclic} &
Prediction errors cause instability or oscillations. &
Adaptive drift metrics, Bayesian correction, uncertainty-aware control. \\ \hline

FL and Heterogeneity~\cite{xu2023federated,wu2022adaptive} &
Device variability and DP overhead cause irregular demand. &
Privacy-aware scaling, adaptive cost modelling, fairness-aware policies. \\ \hline

Large Model and Multimodal Workloads~\cite{yi2025fedalora,ali2025recent} &
High GPU demand and bursty behavior challenge scaling. &
Hierarchical scaling, model-aware scheduling, workload characterisation. \\ \hline

Autonomous Controllers~\cite{pham2024elastic,patni2025predictive} &
Static rules fail under evolving workloads. &
Safe RL, meta-learning, human-in-the-loop control. \\ \hline

Cross-Layer Scaling~\cite{nakata2021concentrated,mavridis2023orchestrated} &
Compute-only focus ignores network and edge constraints. &
Joint optimisation across compute, network, and edge layers. \\ \hline

Security and Privacy~\cite{asadi2025wavesurfer,yenugula2023monitoring} &
Telemetry and models introduce attack surfaces. &
Robust forecasting, secure pipelines, auditable scaling. \\ \hline

\hline
\end{tabular}
\end{table*}

\subsection{Resource, Cost and Privacy}
\begin{enumerate}
    \item Elastic Speedup Metrics, Cost Modeling, and Provisioning Trade-offs: This problem is defined for evaluating predictive autoscaling systems where performance, energy, and cost must be analysed jointly rather than through isolated statistical metrics~\cite{kumar2024optimizing,kumar2025multivariate,kumar2025optimizing,qu2018auto,zhai2025energy}.

    \begin{itemize}

    \item Challenge: Conventional evaluation metrics such as MSE, MAE, $\Theta_O$, $\Theta_U$, and RMSE fail to capture user-perceived latency, energy expenditure, tail-latency behaviour, or cost-performance efficiency. Autoscaling decisions affect application response times in complex, non-linear ways, making it difficult to optimise latency, cost, and elasticity simultaneously.

    \item Research Directions:
    \begin{itemize}
        \item Unified evaluation metrics: Design multi-objective elastic speedup formulations that jointly account for utility, energy consumption, carbon impact, and QoS guarantees.
        \item Tail-aware provisioning: Integrate percentile-based latency constraints (e.g., 95th and 99th percentiles) into predictive autoscaling strategies to ensure reliability under bursty workloads.
        \item Cost-aware predictive autoscaling: Incorporate dynamic pricing models, such as spot instances, reserved capacity, and burstable VMs, into forecasting-driven scaling decisions to optimise monetary efficiency.
    \end{itemize}

    \end{itemize}

    Overall, advancing these evaluation frameworks will enable predictive autoscaling systems to balance performance, stability, and operational cost more effectively in real-world cloud deployments. 

\item Energy and Carbon-Aware Autoscaling for FL Training: This problem is defined for FL and Large model fine-tuning systems deployed across cloud and edge environments, where energy consumption and carbon emissions are becoming critical operational concerns. 
    \begin{itemize}
        \item Challenges: FL workloads, synchronous rounds, and DP mechanisms impose considerable energy overhead, especially on resource-limited edge devices \cite{bostandoost2024lacs,savazzi2022energy,seo2025pioneering}. However, most existing autoscalers optimise only latency or throughput and overlook energy budgets, carbon intensity, and the availability of renewable power. As sustainable computing gains prominence, autoscaling strategies must incorporate environmental objectives into their decision-making logic~\cite{yi2025fedalora,ali2025recent,radhika2021review,albaseer2023data,zhai2025energy}.

    \item Research Directions:
    \begin{itemize}
        \item Energy-aware forecasting: Learn energy-demand and carbon-intensity time-series jointly with traditional system metrics to guide sustainable scaling decisions.
        \item Green-aware scheduling: Shift non-urgent FL rounds or Large adaptation tasks to low-carbon time windows or to zones powered by renewable energy.
        \item Dynamic power caps: Adjust mini-batch sizes, LoRA ranks, local epochs, or training frequency to meet node-level or cluster-level power constraints.
        \item Energy-augmented CRDs: Extend PredictiveAutoscaler CRDs to encode sustainability goals, carbon limits, and power-aware scaling parameters.
    \end{itemize}
    \end{itemize}
Overall, these directions highlight the need for autoscalers that balance performance with sustainability, enabling environmentally responsible FL and Large model training across distributed cloud–edge infrastructures.

\item Privacy-Aware Autoscaling: This problem is defined for autoscaling frameworks that must remain secure and privacy-aware in multi-tenant cloud and federated environments\cite{deng2025secure,yi2025fedalora,dogani2024proactive,zhao2022edge,pham2024elastic,liu2025fedinv,cheng2023proscale}.  

\begin{itemize}
\item Challenge:  Attack surfaces such as adversarial poisoning, malicious client update and tampered metrics from forecasting models, telemetry streams and scaling decisions.

\item Research Directions:
    \begin{itemize}
        \item Adversarially robust forecasting: Model construction that is robust to poisoned metrics, adversarial time-series and malicious clients.
        \item Protect autoscaler pipelines: Secure telemetry, aggregation, attested model inference.
        \item Regulated, auditable autoscaling: Compliant design awareness autoscaling with auditable control loops and signed CRD updates.
    \end{itemize}
\end{itemize}

Overall, these methods guarantee that autoscaling keeps to be effective, reliable, and compliant even in the event of potential security threats and privacy limitations.

\end{enumerate}

What is clear across all of these forecasting models, control loops, privacy mechanisms, and Kubernetes-native orchestration is that it's time to get rid of "reactionary" and "predictive" autoscaling, and embrace models that are "self-correcting," "uncertainty-aware," "cross-layered," and "autonomously optimising. Going forward, it is crucial to integrate intelligent forecasting, adaptive control, resource-efficient orchestration and privacy-preserving mechanisms to enable highly dynamic cloud-native and federated cloud–edge environments. The challenges identified in this section will be key to building infrastructure that can be resilient and scalable, energy efficient, and performance-aware for a wide array of workloads, large models, and distributed FL applications. Next generation autoscaling architectures require adaptive and self-optimising architectures to tightly couple predictive, control and orchestration with system-awareness, particularly within federated, edge and cloud computing environments.


\section{Discussion} \label{sec8}
This review provides a thorough examination of predictive autoscaling for cloud-native and federated cloud-edge applications, and bridges the gaps between the following advancements in workload forecasting, Kubernetes orchestration, federated learning and drift-aware resource management. The analysis shows that there is a definite shift from these traditional threshold-based autoscaling towards more intelligent and adaptive scaling systems that can handle very dynamic, distributed applications.  

The findings show that cloud-native technologies—including containers, microservices, and Kubernetes—have revolutionized resource management, making it possible to scale resources finely and automatically. Although the traditional reactive autoscaling approach is still widely used and deployed because it is easy to implement, it can be effective for only short-term periods and is prone to resource oscillations and SLA violations when workload changes quickly. Predictive autoscaling has thus become a viable solution because it can make predictions based on the workload trends of the past to proactively trigger scaling operations. 

The review also illustrates that the techniques of workload forecasting have undergone considerable change over the years. The early statistical methods, such as moving averages and ARIMA models, were able to offer interpretable forecasting capabilities but could not address complex workload dynamics. Prediction accuracy was enhanced by using machine learning and deep learning models like LSTM, Bi-LSTM and CNN-LSTM that could learn nonlinear workload behaviours. Recently, Transformer-based architectures, such as Informer and MV-Transformer, have demonstrated outstand performance in capturing long-range temporal dependencies and relationships between the various workloads, making them well suited to cloud-native autoscaling applications. 

Another key takeaway is the increasing importance of intelligent resource management utilizing Kubernetes-native automation. By combining predictive models with Kubernetes, Horizontal Pod Autoscalers, Vertical Pod Autoscalers, Custom Resource Definitions (CRDs), Operators, and reconciliation loops allow for predictive scaling decisions directly in the orchestration layer. With this integration, Kubernetes is now an intelligent control framework that enables applications to be self-adaptive across cloud. 

The review also emphasizes on the growing relevance of federated learning in cloud-edge computing infrastructures. Federated learning adds novel features to cloud applications, such as heterogeneous workloads, variable participation of clients, communication delays, mechanisms for privacy protection, and non-IID data distributions. These attributes help make autoscaling and resource allocation more difficult. Predictive autoscaling mechanisms can have a substantial impact on the performance of federated learning by offering adaptive resource provisioning, lower latency in training, and more stable training rounds in distributed environments. 

Moreover, the analysis shows that it is not enough to accurately predict to guarantee a stable autoscaling under the very dynamic operating conditions. Errors in forecasts, shifts in workloads, client variations, and environmental uncertainty can add up over time, causing autoscaling drift and poor performance. Drift aware and uncertainty aware techniques like Autoscaling Drift Index (ADI), feedback-driven correction loops, Round Response Stability (RRS), and Federated Round Stability Control (FRSC) are increasingly becoming an integral part of future autoscaling frameworks. These mechanisms are constantly checking the accuracy of predictions, making decisions about scaling and assuring the stability of operations.

Overall, the reviewed studies demonstrate a gradual convergence of predictive analytics, Kubernetes-native orchestration, federated learning, and adaptive control mechanisms. This convergence forms the foundation of intelligent predictive autoscaling systems capable of supporting future cloud-edge infrastructures. The proposed unified framework presented in this review integrates these emerging directions into a single taxonomy-driven perspective, providing a roadmap for the development of scalable, adaptive, and autonomous resource management solutions for next-generation cloud-native and federated computing environments.

\section{Conclusions} \label{sec9}
This paper presented a comprehensive review of predictive autoscaling techniques for cloud-native and federated cloud-edge environments. The study analyzed the evolution of autoscaling from traditional threshold-based and reactive approaches toward predictive, Kubernetes-native, and drift-aware resource management frameworks. A four-dimensional taxonomy encompassing scaling triggers, scaling targets, prediction models, and evaluation metrics was proposed to systematically classify and compare existing autoscaling approaches. Furthermore, the review examined the integration of statistical, machine learning, deep learning, and Transformer-based forecasting models with Kubernetes orchestration mechanisms, including Horizontal Pod Autoscalers, Vertical Pod Autoscalers, Custom Resource Definitions (CRDs), Operators, and reconciliation loops. The analysis demonstrated that predictive autoscaling significantly improves resource utilization, QoS compliance, cost efficiency, and system responsiveness compared with traditional reactive methods. Moreover, emerging federated learning and cloud-edge applications introduce new challenges related to workload heterogeneity, privacy preservation, communication variability, uncertainty, and autoscaling drift. In response to these challenges, recent research has increasingly taken a drift-aware and uncertainty-aware approach to improve scaling stability and reliability, including with feedback-driven correction, Autoscaling Drift Index (ADI), and Federated Round Stability Control (FRSC). In summary, this review gives a consolidated view of the predictive autoscaling and lays the groundwork for developing intelligent, scalable, and autonomous resource management system in future cloud-native and federated cloud-edge computing environments.

\bibliographystyle{IEEEtran}
\bibliography{cas-refs}

\appendix[] \label{secappendics:Deployments}
\subsection{Deployments: Docker, Kubernetes, \& Cloud Native Stacks}
This subsection provides the practical guidance and code examples for deployment of the following predictive autoscaling components in Docker and Kubernetes \cite{hpaDocs2023}. Examples of the artifacts include an autoscaler model inference service that is lightweight, build instructions for the container, Kubernetes manifests for deployment and exposing the service, Prometheus friendly annotation for the metrics scraping, and an example of PredictiveAutoscaler custom resource \cite{kumar2026critical}. Examples are kept as simple as possible with the aim that they can be expanded for production use. A typical deployment is a modular design with responsibilities spread among a number of services. Monitoring is done by using a metric exporter and Prometheus for collecting time-series data \cite{hpaDocs2023, vpaDocs2023}. A dedicated model service is used for inference and exposes REST endpoints and/or gRPC endpoints. The output of forecasts are fed into an adaptation manager which then translates them into decisions regarding scaling, either by modifying CRDs or by sending direct scaling instructions. Finally, an operator or controller reconciles the PredictiveAutoscaler resource and takes action to scale in the Kubernetes cluster \cite{clusterAutoscaler2023}.

\subsubsection{Example model server and Dockerfile} Below is a small Python Flask based model server that returns a JSON forecast. In practice you will replace the toy predict function with the real model. The Listing \ref{lst1} illustrates a minimal Flask-based model server that exposes a \texttt{/predict} endpoint to return JSON forecasts consumed by the autoscaler or operator during reconciliation. Prometheus counters and histograms are integrated to monitor request frequency and latency, enabling in-cluster observability. Although the prediction logic is a lightweight placeholder, this is the point where real forecasting models such as Informer, MV-Transformer, or CNN-LSTM would be invoked for production autoscaling. The accompanying \texttt{Dockerfile} packages the microservice with Gunicorn for concurrent request handling, making the image suitable for deployment across Kubernetes clusters or edge nodes (Listing \ref{lst2}).

\begin{lstlisting}[caption={Flask-based model inference service with Prometheus metrics}, label={lst1}]

# app.py
from flask import Flask, request, jsonify
import time, random, prometheus_client

app = Flask(__name__)

# Prometheus metrics
REQUEST_COUNT = prometheus_client.Counter('model_requests_total', 'Total model requests')
REQUEST_LATENCY = prometheus_client.Histogram('model_request_latency_seconds', 'Request latency seconds')

@app.route('/metrics')
def metrics():
    return prometheus_client.generate_latest(), 200, {'Content-Type': 'text/plain; version=0.0.4; charset=utf-8'}

@app.route('/predict', methods=['POST'])
def predict():
    start = time.time()
    REQUEST_COUNT.inc()
    payload = request.get_json(force=True)
    # toy predictor: in practice call your informer / transformer / cnn-lstm model here
    horizon = payload.get('horizon', 1)
    last_val = payload.get('last', 100)
    forecast = int(last_val * (1 + 0.05 * random.uniform(-1,1)))  # replace with model inference
    duration = time.time() - start
    REQUEST_LATENCY.observe(duration)
    return jsonify({'forecast': forecast, 'horizon': horizon})
\end{lstlisting}

\begin{lstlisting}[caption={Dockerfile for model inference service}, label={lst2}]
# Dockerfile
FROM python:3.10-slim

WORKDIR /app
COPY requirements.txt .
RUN pip install --no-cache-dir -r requirements.txt

COPY app.py .

EXPOSE 5000
CMD ["gunicorn", "--bind", "0.0.0.0:5000", "app:app", "--workers", "2", "--threads", "4"]
\end{verbatim}

\begin{verbatim}
# requirements.txt
flask
gunicorn
prometheus_client
\end{lstlisting}

\paragraph{Build and push the image} Build the Docker image and push it to a container registry so that Kubernetes can pull and deploy it (Listing \ref{lst3}).
\begin{lstlisting}[caption={Docker build and push commands for autoscaling model service}, label={lst3}]
docker build -t <registry>/autoscale-model:v1 .
docker push <registry>/autoscale-model:v1
\end{lstlisting}

\subsubsection{Kubernetes manifests}
A small deployment, service, and Prometheus-friendly annotations. Save as \texttt{model-deployment.yaml}. The manifest Listing \ref{lst:k8s_deploy} presents the Kubernetes configuration for deploying the autoscaling model service with integrated Prometheus monitoring. The Deployment schedules a single replica with explicit CPU and memory requests to ensure predictable behaviour under autoscaling experiments. Prometheus scrape annotations are included to expose both the \texttt{/predict} and \texttt{/metrics} endpoints for monitoring and model-latency collection. Readiness and liveness probes help maintain service reliability by ensuring that only healthy pods receive traffic. The accompanying Service provides stable in-cluster access through a \texttt{ClusterIP}, enabling operators and other controllers to query the model server consistently during reconciliation cycles.

\begin{lstlisting}[caption={Kubernetes deployment and service configuration for autoscaling model}, label={lst:k8s_deploy}]
apiVersion: v1
kind: Namespace
metadata:
  name: autoscale-system
---
apiVersion: apps/v1
kind: Deployment
metadata:
  name: autoscale-model
  namespace: autoscale-system
spec:
  replicas: 1
  selector:
    matchLabels:
      app: autoscale-model
  template:
    metadata:
      labels:
        app: autoscale-model
      annotations:
        prometheus.io/scrape: "true"
        prometheus.io/port: "5000"
        prometheus.io/path: "/metrics"
    spec:
      containers:
      - name: autoscale-model
        image: <registry>/autoscale-model:v1
        ports:
        - containerPort: 5000
        resources:
          requests:
            cpu: "250m"
            memory: "256Mi"
          limits:
            cpu: "1"
            memory: "1Gi"
        readinessProbe:
          httpGet:
            path: /predict
            port: 5000
          initialDelaySeconds: 5
          periodSeconds: 10
        livenessProbe:
          httpGet:
            path: /metrics
            port: 5000
          initialDelaySeconds: 10
          periodSeconds: 20
---
apiVersion: v1
kind: Service
metadata:
  name: autoscale-model
  namespace: autoscale-system
spec:
  selector:
    app: autoscale-model
  ports:
  - port: 5000
    targetPort: 5000
    protocol: TCP
  type: ClusterIP
\end{lstlisting}

\subsubsection{Prometheus Scraping} If you use the Prometheus operator, a ServiceMonitor is preferred. For a basic Prometheus without operator, add scrape config pointing to the Kubernetes service. Prometheus scraping enables the autoscaling pipeline to collect fine-grained
latency and workload metrics from the model server. When using the Prometheus Operator, a \texttt{ServiceMonitor} is the recommended method because it automatically discovers services based on label selectors and manages scrape configurations declaratively. For setups without the operator, the same service can be scraped by adding a manual entry in Prometheus’ \texttt{scrape\_configs}. As shown in Listing~\ref{lst:servicemonitor}, a minimal \texttt{ServiceMonitor} configuration polls the model’s \texttt{/metrics} endpoint at regular intervals (e.g., every 15 seconds), ensuring consistent observability for forecasting accuracy, model latency, and autoscaling behaviour.
\begin{lstlisting}[caption={Prometheus ServiceMonitor configuration for autoscaling model metrics}, label={lst:servicemonitor}]
# Prometheus ServiceMonitor example (if using Prometheus operator)
apiVersion: monitoring.coreos.com/v1
kind: ServiceMonitor
metadata:
  name: autoscale-model-sm
  namespace: autoscale-system
spec:
  selector:
    matchLabels:
      app: autoscale-model
  endpoints:
  - port: 5000
    path: /metrics
    interval: 15s
\end{lstlisting}

\subsubsection{ConfigMap for model and operator}
A \texttt{ConfigMap} is used to store the runtime parameters like per-pod capacity, prediction horizon, model type and the Resource Removal Strategy (RRS) factor. This configuration will be shared by the model server, adaptation manager and operator, so that the behaviour is consistent across the autoscaling pipeline. These parameters can be injected into pods at runtime, and do not require rebuilding container images for dynamic tuning. Centralisation in a \texttt{ConfigMap} also enables operators to change the scaling sensitivity, model selection and capacity settings via simple configuration changes. The example configuration has key parameters that are used for predictive autoscaling as described in Listing~\ref{lst:configmap}. 

\begin{lstlisting}[caption={ConfigMap for autoscaling configuration parameters}, label={lst:configmap}]
apiVersion: v1
kind: ConfigMap
metadata:
  name: autoscale-config
  namespace: autoscale-system
data:
  POD_CAPACITY: "1000"  # e.g., requests per minute per pod
  RRS_FACTOR: "0.6"
  MODEL_TYPE: "informer"
  HORIZON: "5"
\end{lstlisting}

\subsubsection{PredictiveAutoscaler CRD example}
The desired behaviour of the autoscaling controller is specified in a minimal custom resource (CR) \texttt{PredictiveAutoscaler}. In practice, the CRD schema should be designed and preloaded a priori~\cite{kumar2024optimizing}. The CR is connected to a target Deployment, it defines the forecasting model, it encodes the pod capacity, and it specifies policy constraints like the Resource Removal Strategy (RRS) factor, the replicas bounds etc. At each reconciliation period, the operator reads the CR, fetches metrics from Prometheus, calls the prediction model, calculates the desired replica count and updates both the target Deployment and the CR status. The \texttt{status} field stores the forecast value, the time of the forecasted event, and active replicas for traceability and auditability of scaling decisions. The example below is a minimal configuration for predictive autoscaling, as seen in Listing~\ref{lst:crd}.

\begin{lstlisting}[caption={PredictiveAutoscaler custom resource definition example}, label={lst:crd}]
apiVersion: autoscaling.example.com/v1
kind: PredictiveAutoscaler
metadata:
  name: web-predictive
  namespace: autoscale-system
spec:
  targetRef:
    apiVersion: apps/v1
    kind: Deployment
    name: web-deploy
  model:
    type: informer
    horizon: 5m
  capacity:
    podCapacity: 1000
  policy:
    rrs: 0.6
    minReplicas: 2
    maxReplicas: 50
status:
  lastForecast:
    time: "2025-12-07T10:00:00Z"
    value: 8400
  currentReplicas: 8
\end{lstlisting}

\subsubsection{Operator and Model Control System Integration}
Kubernetes-based predictive autoscaling relies on close coordination between the Operator and the Model Control System (MCS). Two deployment patterns are commonly used. In the first, the forecasting model runs as an independent microservice exposing REST or gRPC endpoints. During reconciliation, the Operator queries this service, retrieves workload forecasts, computes the desired replica count, and updates the target Deployment or StatefulSet. In the second pattern, the model is embedded directly within the Operator, reducing network overhead and latency but increasing binary size, memory usage, and update complexity. Both approaches require a consistent \texttt{PredictiveAutoscaler} CRD schema to ensure uniform interpretation of model type, prediction horizon, and capacity parameters.

Reconciliations usually start when a \texttt{PredictiveAutoscaler} resource is created or updated. The Operator pulls the desired state, collects the metrics from Prometheus or in-cluster sources and pushes them to the MCS for forecasting. It uses the predicted demand to calculate the number of replicas that is required, it uses the Resource Removal Strategy (RRS) to safely scale in the target workload, and it patches the target workload. Forecast values, selected model, replica decisions and timestamps are then added to the CRD status to facilitate traceability. This process is a minimal reconciliation workflow as illustrated in Listing~\ref{lst:reconcile}.

\begin{lstlisting}[caption={Minimal operator reconciliation workflow for predictive autoscaling}, label={lst:reconcile}, language=bash, frame=single, basicstyle=\ttfamily\footnotesize]

# Minimal operator reconciliation logic (illustrative)
1. Watch PredictiveAutoscaler instances
2. Read spec and current status
3. Query Prometheus metrics (CPU_t, Mem_t, etc.)
4. Invoke model: POST /predict { "window": X, "H": 5 }
5. Compute desired replicas: P_des = ceil(forecast / POD_CAP)
6. Apply RRS for safe scale-in
7. Patch Deployment and update CRD status
\end{lstlisting}

\subsubsection{Build and Deployment Workflow}
To deploy the model service and Operator, you need to create images for both the service and the Operator, push them to a registry and apply the manifests for both. The workflow is as natural as it is lightweight while using a local Minikube environment—extending it to production clusters is also straightforward.

\begin{lstlisting}[caption={Build and deployment commands for autoscaling components}, label={lst:build_deploy}, language=bash, frame=single, basicstyle=\ttfamily\footnotesize]
# build and push image
docker build -t <registry>/autoscale-model:v1 .
docker push <registry>/autoscale-model:v1

# apply manifests
kubectl apply -f model-deployment.yaml
kubectl apply -f autoscale-config.yaml
kubectl apply -f predictive-autoscaler-cr.yaml
\end{lstlisting}

When the model service and operator are deployed, as in Listing~\ref{lst:build_deploy}, autoscaling is completely automated and managed by CRDs. All the times the system is monitoring the workload metrics, forecasting and dynamically allocating resources without manual intervention.

\subsubsection{Production and Edge Considerations}
It is important to have a secure, resource-efficient and fault-tolerant design for production deployment. To reduce supply-chain vulnerabilities, container images should be stored in authenticated registries, and pinned to immutable digests. Resources requests and limits should leave room for resource-consuming system processes, especially on resource-constrained edge nodes. PodDisruptionBudgets ensure that the pod does not get evicted when necessary to keep the operation going (e.g. during the FL aggregation); graceful termination means that checkpoints or partial updates are saved before terminating the pod.

Forecasting models need to be optimized for edge environments, either by compressing or quantizing models, pruning or distillation. On-device inference helps to mitigate the need to rely on remote services when connectivity is intermittent. Bursty workloads are supported by asynchronous reconciliation and coroutine-based operators. Structured logging, tracing, and Prometheus metrics should be included in observability, such as prediction latency, model load time, confidence intervals, etc., to aid in drift detection and debugging.

Additional coordination constraints are added by privacy and federated integration. Actively scaling down training pods must be avoided by the operators, and affinity-aware scheduling can help alleviate communication latency for aggregator nodes. If differential privacy (DP) is used, the computational cost needs to be taken into account when planning capacity to avoid “under-provisioning” in high-traffic training periods.

As a whole, this integrated operator-model architecture offers a scalable framework for predictive autoscaling in cloud-native and edge-setups, making it possible to reduce the reliance on human resources and automate and optimise scaling for cloud-native or edge setups.

\subsection{Custom Resource Definitions for Autoscaling} \label{sec4.6}
CRDs extend the Kubernetes API with new resource types, providing a declarative interface for integrating predictive autoscaling directly into the control plane~\cite{yuan2024time,wijesekera2025kubernetes}. Unlike built-in mechanisms such as HPA or external scaling services, CRDs~\cite{kumar2024optimizing} enable users to encode predictive models, parameters, and scaling policies as first-class Kubernetes resources. This aligns with Kubernetes’ declarative design and operator-driven automation paradigm~\cite{kubernetesExtendKubernetes,miracapillo2025enabling}.

A \texttt{PredictiveAutoscaler} CRD allows users to define the forecasting model, prediction horizon, target workload, pod capacity, and scaling policies such as the Resource Removal Strategy (RRS) and replica bounds. The \texttt{spec} field represents the desired autoscaling behaviour, while the \texttt{status} field records runtime outputs, including forecasts and replica decisions. As shown in Listing~\ref{lst:crd_example}, the following example illustrates a minimal predictive autoscaler configuration.

\begin{lstlisting}[caption={PredictiveAutoscaler CRD example for declarative autoscaling}, label={lst:crd_example}]
apiVersion: autoscaling.example.com/v1
kind: PredictiveAutoscaler
metadata:
  name: web-predictive
spec:
  targetRef:
    apiVersion: apps/v1
    kind: Deployment
    name: web-deploy
  model:
    type: informer
    horizon: 5m
  capacity:
    podCapacity: 1000 # requests per minute per pod
  policy:
    rrs: 0.6
    minReplicas: 2
    maxReplicas: 50
status:
  lastForecast:
    time: "2025-12-07T10:00:00Z"
    value: 8400
  currentReplicas: 8
\end{lstlisting}
The central advantage of embedding predictive autoscaling logic in CRDs is that all forecast outputs, scaling plans, and decisions become part of the Kubernetes control-plane state. Operators and controllers can continuously watch these CRD objects and reconcile the actual cluster state with the desired state expressed in the CRD. This enables asynchronous prediction, transparent recording of model decisions, cluster-wide auditability, and clean integration of Informer, LSTM, Transformer, or MV-Transformer models within Kubernetes-native workflows. By using CRDs as the declarative interface, predictive autoscaling becomes fully programmable, versionable, and observable, enabling robust automation across cloud, edge, and federated environments.

\end{document}